%% file: main.tex
\documentclass[journal]{new-aiaa}
\usepackage[utf8]{inputenc}
\usepackage{tikz}
\usetikzlibrary{arrows,automata}
\usetikzlibrary{fit}
\usepackage{graphicx}
\usepackage{subcaption}
\usepackage{import}
\usepackage{amsmath}
\usepackage{mathtools}
\usepackage{bbm}
\usepackage{bm}
\usepackage{rsfso}
\usepackage[version=4]{mhchem}
\usepackage{siunitx}
\usepackage{longtable,tabularx}
\usepackage{lettrine}
\usepackage{color}
\usepackage{xcolor}
\usepackage{transparent}
\usepackage{multicol}
\usepackage{algorithm}
\usepackage{algorithmic}
\usepackage{pgfplotstable}
\usepackage[colorlinks,urlcolor=blue,citecolor=blue,linkcolor=blue,bookmarks=false]{hyperref}
\long\def\comment#1{}

\usetikzlibrary{arrows, calc, chains, positioning, quotes, backgrounds, shapes}
\usetikzlibrary{positioning}
\usepackage[export]{adjustbox}
\usepackage{relsize}
\tikzstyle{block} = [rectangle, draw, fill=white,
    text width=6em, text centered, rounded corners, minimum height=3em]
 \tikzstyle{dot} = [circle, draw, fill=black,
    minimum size=1pt]
\tikzstyle{arrow} = [draw, -latex']
\tikzstyle{virtual} = [coordinate]

\tikzset{CNTRL/.style =
{
gain/.style     = {fill=##1,draw, isosceles triangle,isosceles triangle apex angle=50, text centered,},
gainRot/.style     = {fill=white,draw, isosceles triangle, 								     ,isosceles triangle apex angle=50, shape border 						  rotate=180,},
sumBlock/.style = {draw,circle,minimum size=1pt,},
block/.style = {rectangle, draw, fill=white,
    text width=6em, text centered, rounded corners, minimum height=3em,},
blockLarge/.style = {rectangle, draw, text centered, rounded corners, minimum height=3em,},
gain/.default = white,
integrator/.style = {rectangle, draw, fill=white,
    minimum height=6mm, minimum width=6mm,
    outer sep = 0mm},
rect/.style = {rectangle, draw, fill=white,
    minimum height=6mm, minimum width=8mm,
    outer sep = 0mm},
}
}
\tikzset{
tmp/.style  = {coordinate},
}

\newcommand{\tkzcircsimple}{\tikz{\node[circle,inner sep=0,outer sep=0,draw, line width=1pt]{$\phantom{o}$}}}

\newcounter{nodecount}

\input{macros.tex}

\definecolor{myBlue}{rgb}{0.2235,0.4157,0.6941}
\definecolor{myOrange}{rgb}{0.8549,0.4863,0.1882}
\definecolor{myGreen}{rgb}{0.2431,0.5882,0.3176}
\definecolor{myRed}{rgb}{1,0.4,0.4}

\title{Enhancing Resilience of Airborne Wind Energy Systems Through Upset Condition Avoidance}

\author{Sebastian Rapp\thanks{PhD Researcher, Delft University of Technology, Kluyverweg 1, 2629 HS, Delft, The Netherlands, s.rapp@tudelft.nl.} and Roland Schmehl\thanks{Associate Professor, Delft University of Technology, Kluyverweg 1, 2629 HS, Delft, The Netherlands, r.schmehl@tudelft.nl.}   }
\affil{Delft University of Technology, Faculty of Aerospace Engineering}

\begin{document}

\maketitle

\begin{abstract}
Airborne wind energy (AWE) systems are tethered flying devices that harvest wind resources at higher altitudes which are not accessible to conventional wind turbines. In order to become a viable alternative to other renewable energy technologies, AWE systems are required to fly reliably and autonomously for long periods of time while being exposed to atmospheric turbulence and wind gusts. In this context, the present paper proposes a three-step methodology to improve the resilience of an existing baseline control system towards these environmental disturbances. In the first step, upset conditions are systematically generated that lead to a failure of the control system using the subset simulation method. In the second step, the generated conditions are used to synthesize a surrogate model that can be used to predict upsets beforehand. In the final step an avoidance maneuver is designed which keeps the AWE system operational while minimizing the impact of the maneuver on the average pumping cycle power. The feasibility of the methodology is demonstrated on the example of tether rupture during pumping cycle operation. As an additional contribution a novel transition strategy from retraction to traction phase is presented that can reduce the probability of tether rupture significantly. 

\end{abstract}
\vspace{5mm}

\saythanks

\section{Introduction}

\lettrine{O}{perating} airborne wind energy (AWE) systems requires sophisticated control strategies that try to exploit the full physical capabilities of the system for maximum power generation without compromising safety. The major part of the existing literature about AWE control systems focuses on the former, to maximize the power output using trajectory optimization, see for instance \cite{Horn2013, Zanon2013,Cobb2017, LICITRA2019569}. The only recent publication that analyses reliability and safety of AWE systems is \cite{Salma2019}, which  presents a failure mode and effect analysis along with a fault tree analysis for a flexible wing kite power system. This imbalance between performance optimization and reliability analysis in the AWE literature indicates that more research is necessary to investigate how the resilience and robustness of AWE control systems can be improved, which motivates the present work.

AWE systems need to operate in varying environmental conditions such as slowly varying wind speeds due to the altitude dependent mean wind speed profile but also need to cope with rapid changes due to wind gusts and turbulence. Because of the inherent stochastic nature of the wind conditions it is difficult to explicitly include them in the control design process. In practice, the closed loop system is verified a posteriori for randomly generated  wind conditions, as presented, for instance, in \cite{Rapp2019}. If the controller fails to satisfy all requirements, it either needs to be re-tuned or completely re-designed. To create enough confidence that the controller achieves its objective, a large amount of simulations is necessary. This approach belongs to the direct Monte Carlo simulation methods \cite[p. 83f]{MonteCarlo}. Besides the computational burden of the control system verification process it is also difficult to create enough counter examples where the control system fails. For example a wind gust with a certain shape, that occurs with a probability of $10^\mr{-6}$, requires on average $10^\mr{6}$ simulation runs until it is encountered once. Especially for computationally expensive simulation runs this approach can be practically infeasible if several samples of these rare events need to be generated. Naturally, the more information about the condition that leads to a control system failure is available, the more reliably it can be predicted and prevented in the future. Concretely, if enough data about counter examples is available, a model that runs in parallel to the control system can be constructed that monitors the current flight state. It can then be used to predict how likely it is that the current flight condition leads to an upset and if necessary triggers a maneuver that avoids it.
Creating such a predictor requires a significant amount of data that can not be generated efficiently using the direct Monte Carlo method due to the aforementioned computational burden. Therefore, a different approach is chosen in this work which is based on \textit{subset simulations} (SS). It is an algorithm that has been developed originally to estimate small failure probabilities of high dimensional stochastic systems \cite{AU2001263}. 
Recently SS has already been applied to small failure probability estimation in the context of flight control system verification (see \cite{Lbl2015} and \cite{Wang2019}). In the context of this work the algorithm will not only be used to estimate rare event probabilities but also to generate systematically a knowledge base for external disturbances that lead to a specified control system failure, denoted as an upset. The generated conditions will then be used to train a binary classifier which is either based on a fixed threshold prediction strategy or a \textit{support vector machine} (SVM). These surrogate models are able to predict and eventually prevent the occurrence of a failure beforehand with the overall goal to improve the fail-operational characteristic of the AWE system.

The contribution of this work can be summarized as follows. First, a control system modification to a previously published work of the authors (see \cite{Rapp2019}) is proposed. It is shown that the modification reduces the probability of tether rupture significantly. Second, a generic framework is presented that systematically generates conditions in which the control system fails. Third, two different prediction strategies are presented that are either based on a simple threshold approach or a binary time series classification technique to predict upset conditions.
Fourth, a loss rate function is derived that allows to trade off the prediction performance with respect to the induced economic loss of false positives and false negatives. 
In the last part of this work the framework is applied to predict and prevent tether rupture, a common failure scenario in the context of AWE. A tailored avoidance maneuver is proposed that prevents this specific upset and keeps the system operational. 

Ultimately, the following research questions are answered: How can the transition from retraction to traction phase be shaped in order to damp tether tension peaks during the transients? How can upset conditions in the context of AWE be defined and systematically generated if the probability of encountering one per pumping cycle is low and how can they be predicted? Furthermore, how can the practical impact of different prediction strategies be used to measure classification performance beyond classical metrics? Finally, is it justifiable from an economic point of view to prevent these conditions if that comes at the cost of false positives or is it more reasonable to simply accept them? 

To that end, the paper is structured as follows. In section \ref{sec:clsys} the closed loop system is presented including a brief description of the utilized models and controllers. In addition, a modification to the baseline control architecture is introduced. In section \ref{sec:framework} the novel framework is presented generically and in section \ref{sec:results} it is applied to the specific case of predicting and preventing tether rupture during a pumping cycle. Finally, section \ref{sec:conclusion} concludes the paper.

\section{Closed-Loop System Description}\label{sec:clsys}
In this study a model of a generic AWE system operated in pumping cycle mode is used along a modified version of the control system presented in \cite{Rapp2019} in order to demonstrate the effectiveness of the proposed framework. In the first part of this section the main components of the model will be reviewed. In the second part the key elements of the controller will be presented with a focus on the modification with respect to the baseline architecture as presented in \cite{Rapp2019}.

\begin{figure}[t!]
	\centering
	\def\svgwidth{0.5\textwidth}
	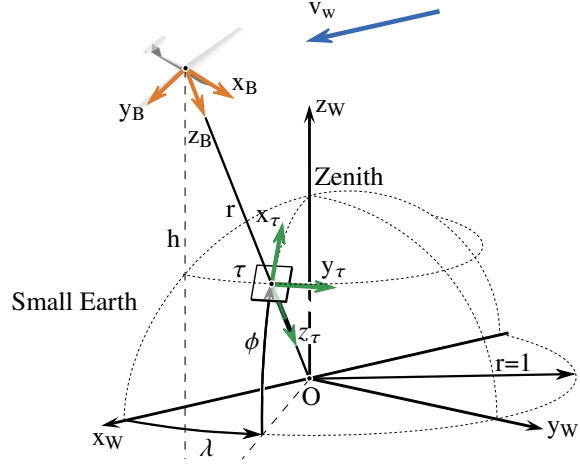
	\caption{Small earth analogy with wind reference frame W, tangential plane frame $\tau$ and body-fixed frame B \cite{Rapp2019}.}
	\label{fig:small_earth}
\end{figure}

\subsection{Aircraft, Ground Station and Wind Models}
The aircraft is modeled as a six degrees of freedom rigid body and its geometric and aerodynamic properties are based on the values in \cite{PhDthesisLicitra:} and \cite{Malz19}. The translational dynamics are given by
\begin{equation}
\left(\dot{\mathbf{v}}_\mr{k}^\mr{G}\right)_\mr{B}=-\left(\bm{\omega}\right)_\mr{B}^\mr{OB}\times \left(\mathbf{v}_\mr{k}^\mr{G}\right)_\mr{B}+\frac{\left(\mathbf{F}_\mr{tot}\right)_\mr{B}}{m_\mr{a}}
\end{equation}
with
\begin{equation}
\left(\mathbf{F}_\mr{tot}\right)_\mr{B} = \left(\mathbf{F}_\mr{a}\right)_\mr{B}+\left(\mathbf{F}_\mr{g}\right)_\mr{B}+\left(\mathbf{F}_\mr{t}\right)_\mr{B}+\left(\mathbf{F}_\mr{p}\right)_\mr{B}
\end{equation}
where $\left(\dot{\mathbf{v}}_\mr{k}^\mr{G}\right)_\mr{B}$, $\left(\mathbf{v}_\mr{k}^\mr{G}\right)_\mr{B}$,  $\left(\bm{\omega}\right)_\mr{B}^\mr{OB}$ and $m_\mr{a}$ represent the kinematic acceleration, the kinematic velocity, the rotational rate of the aircraft with respect to the North-East-Down reference frame and the total mass of the aircraft, respectively. The subscript $B$ indicates vectors given in the conventional body-fixed frame of the aircraft as visualized in Fig.~\ref{fig:small_earth}. The total force acting in the center of mass $G$ of the aircraft consists of the resulting aerodynamic force $\left(\mathbf{F}_\mr{a}\right)_\mr{B}$, the weight $\left(\mathbf{F}_\mr{g}\right)_\mr{B}$, tether force  $\left(\mathbf{F}_\mr{t}\right)_\mr{B}$ as well as propulsion force $\left(\mathbf{F}_\mr{p}\right)_\mr{B}$.
The rotational dynamics are given by
\begin{equation}
\left(\dot{\bm{\omega}}\right)_\mr{B}^\mr{OB}=-\mathbf{J}^\mr{-1}\left(\left(\bm{\omega}\right)_\mr{B}^\mr{OB}\times \mathbf{J}\left(\bm{\omega}\right)_\mr{B}^\mr{OB}-\left(\mathbf{M}_\mr{a}\right)_\mr{B}\right)
\end{equation}
where $\left(\dot{\bm{\omega}}\right)_\mr{B}^\mr{OB}$, $\mathbf{J}$ and $\left(\mathbf{M}_\mr{a}\right)_\mr{B}$ denote the rotational acceleration, inertia tensor and resulting aerodynamic moment acting in the center of mass, respectively. Note, it is assumed that the tether is attached to the center of gravity since information about the exact location of the attachment point is not publicly available. Therefore, the tether does not contribute to the rotational dynamics.
Furthermore, the actuator dynamics for ailerons, elevator and rudder are approximated as first order filters including rate and deflection limits. These values are summarized in Table \ref{tab:actuator}.
\begin{table}[b!]
	\caption{\label{tab:actuator} First order actuator models.}
	\centering
	\begin{tabular}{llllllllllcc} 
		\hline
		\hline
		Parameters & Values & Units \\
		\hline
		Bandwidth aileron $\omega_\mr{a,0}$ & 35 & \SI{}{\radian\per\second} \\
		Aileron deflection limit $\delta_\mr{a,lim}$ & $\pm$20 & \SI{}{\degree} \\
		Aileron rate limit $\dot{\delta}_\mr{a,lim}$ & $\pm$115&\SI{}{\degree\per\second} \\
			Bandwidth elevator $\omega_\mr{e,0}$ & 35 & \SI{}{\radian\per\second} \\
		Elevator deflection limit $\delta_\mr{e,lim}$ & $\pm$20 & \SI{}{\degree} \\
		Elevator rate limit $\dot{\delta}_\mr{e,lim}$ & $\pm$115&\SI{}{\degree\per\second} \\
			Bandwidth rudder $\omega_\mr{r,0}$ & 35 & \SI{}{\radian\per\second} \\
		Rudder deflection limit $\delta_\mr{r,lim}$ & $\pm$30 & \SI{}{\degree} \\
		Rudder rate limit $\dot{\delta}_\mr{r,lim}$ & $\pm$115&\SI{}{\degree\per\second} \\
		\hline
		\hline
	\end{tabular}
\end{table}
 For the post-takeoff phase a simple propeller model is implemented as defined in \cite[p.53f]{Beard:2012:SUA:2462627}. Note, the propeller is only used in the beginning of each simulation to initialize the pumping cycle operation.

\begin{figure}[h]
	\centering
	\def\svgwidth{0.5\textwidth}
	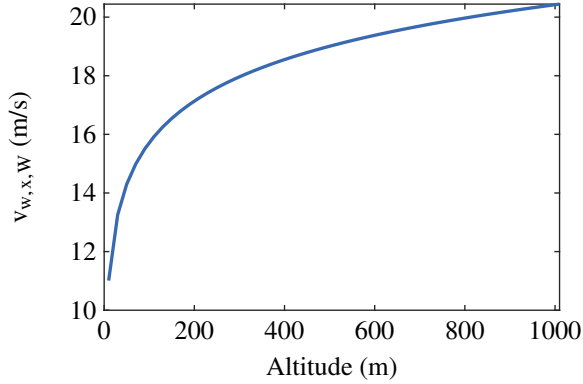
	\caption{Mean wind speed profile as a function of altitude.}
	\label{fig:shear}
\end{figure}

The ground station is modeled as in \cite{Rapp2019} and relevant parameter values are summarized in Table \ref{tab:gs_params}. Furthermore, the discretized tether model of \cite{Fechner2015} is implemented, the utilized values are displayed in Table \ref{tab:t_params}. Wind conditions are simulated using the wind shear model and the discrete Dryden turbulence model provided by the Matlab Aerospace Toolbox \cite{MatlabATB}. The resulting mean wind speed profile in the present work as a function of altitude is depicted in Fig.~\ref{fig:shear}. The turbulence components are superimposed to this wind speed profile.
\begin{table}[b!]
	\caption{\label{tab:gs_params} Ground station parameters.}
	\centering
	\begin{tabular}{llllllllllcc} 
		\hline
		\hline
		Parameters & Values & Units \\
		\hline
		Inertia $J_\mr{W}$ & 0.08 & \SI{}{\kilogram\square\meter} \\
		Viscous friction $\kappa_\mr{W}$ & 0.6 & \SI{}{\kilogram\meter\per\second} \\
		Acceleration limits $a_\mr{W,min/max}$ & $\pm 5$ & \SI{}{\meter\per\square\second} \\
		Maximum speed $v_\mr{W,max}$ & 20 & \SI{}{\meter\per\second} \\
		Minimum speed $v_\mr{W,min}$ & -15 & \SI{}{\meter\per\second} \\
		\hline
		\hline
	\end{tabular}
\end{table}

\begin{table}[b!]
	\caption{\label{tab:t_params} Tether parameters.}
	\centering
	\begin{tabular}{llllllllllcc} 
		\hline
		\hline
		Parameters & Values & Units \\
		\hline
		Particles $n_\mr{T}$ & 5 & - \\
		Mass Density $\rho_\mr{T}$ & 0.0046 & \SI{}{\kilogram\per\cubic\meter} \\
		Diameter $d_\mr{T}$ & 0.0025 & \SI{}{\meter} \\
		Drag coefficient $C_\mr{d,T}$ & 1.2 & - \\
		Stiffness $c_\mr{T}$& 10243 & \SI{}{\newton\per\meter} \\
		Damping $d_\mr{T}$& 7.8833 & \SI{}{\newton\second\per\meter} \\
		\hline
		\hline
	\end{tabular}
\end{table}

\subsection{Control System Description}
\begin{figure}[h]
	\centering
	\def\svgwidth{0.5\textwidth}
	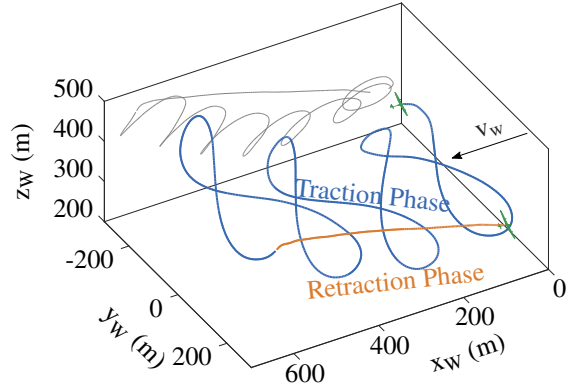
	\caption{Generic pumping cycle trajectory with traction and retraction phase.}
	\label{fig:pc_traj}
\end{figure}

In the following the control objective for an AWE system operated in pumping cycle mode is described. A pumping cycle usually results in a trajectory similar to the one displayed in Fig.~\ref{fig:pc_traj}.
The control objective for such a system can be subdivided into a radial and a tangential direction control task. On the one hand, the controller needs to keep a high tension in the tether during the traction and a low tension during the retraction phase. This radial direction control objective is achieved using the rotational speed of the winch as well as the aircraft angle of attack and bank angle as control variables. On the other hand, the aircraft needs to follow a prescribed flight path, for instance a figure of eight pattern during the traction phase and a straight line glide path during the retraction phase. These two different flight path segments are indicated in Fig.~\ref{fig:pc_traj} by the blue (traction) and orange (retraction) lines. The path following objective will be achieved using only the flight controller. In this work the flight controller is implemented in a cascaded form with three feedback loops. In the first loop the guidance commands in form of course and flight path angle rates are calculated with respect to the path curvature for the figure of eight flight while during the retraction phase command shaping filters are used. These commands are then translated into bank angle and angle of attack commands that are tracked by the attitude controller. In the inner most loop the output of the attitude controller is translated into rate commands that are tracked by the rate controller and eventually allocated to the actuator deflections. Additionally, in each loop pseudo control hedging is implemented to comply with aircraft state and input limits.

In the following paragraphs two modifications to \cite{Rapp2019} are introduced that improve the robustness of the controller in turbulent wind conditions. The first modification consists of an improved guidance strategy for the transition between the retraction and the traction phase. The main challenge here is represented by the rising tether tension i.e. from low tension in the retraction to high tension in the traction phase. Furthermore, since both phases are fundamentally different from a control perspective a transition strategy from straight line following (retraction) to path following on a virtual sphere (traction) needs to be achieved. Subsequently, several possible solutions will be discussed before the final method is presented.

One possible approach is to directly switch into the figure of eight path following mode as soon as the end of the retraction path is reached which does not require any intermediate guidance strategy. Another option is to include a planar circular arc at the end of the retraction phase that defines the turning radius for the transition phase. This delays the activation of the traction mode until the aircraft is steered sufficiently back into the wind which can be defined by a way-point on the arc. The drawback of the first approach is the reduced level of guidance and hence it is difficult to shape the transient behavior. Since the same controller for the transient as for the traction phase is used to avoid unnecessary switching between different controllers, tuning of the controller for better transient behavior would also alter the controller for the traction phase. The downside of the second approach is that it requires additional parameters to be tuned such as the length and curvature of the arc. Modifying the figure of eight would most likely also require to modify the geometry of the transition arc. It can be seen that both approaches are complementary in terms of additional complexity and level of guidance. Note, more sophisticated approaches such as an optimal control strategy that connects the retraction and traction phase are also possible but this requires again an additional controller switch which is not desirable.

The advantages of both approaches can be combined in the following third alternative. Instead of defining a new arc in the horizontal plane the same but rotated figure of eight curve as for the traction phase is used. This is similar to the first approach where the traction phase is directly triggered at the end of the retraction phase. However, instead of directly approaching the traction phase path at a low elevation angle (power zone) a figure eight curve at a high elevation angle (limit is 90 degrees) is used for better guidance in the turning phase (advantage of the second approach). During the transient the curve is rotated towards the desired elevation angle for the traction phase. The time constant that defines the speed with which the path is rotated turns out to be an important parameter which trades of robustness (large value) and performance (small value) since it defines how fast the aircraft will fly into the power zone. In combination with a shaped set point change for the tether force tracking a smooth transition from a straight path with low tether tension to figure of eight flight path following with high tether tension can be achieved. The mathematical implementation of this approach is discussed in the following.

As in \cite{Rapp2019} the figure of eight flight path is parameterized using the definition of a Lemniscate in spherical coordinates on a unit sphere. Concretely, the longitude and latitude of each point on the path is then given by
\begin{equation}
\begin{split}
	\lambda_\mr{p} &= \frac{b \sin(s)}{1+\left(\frac{a}{b}\cos(s) \right)^\mr{2}} \\
    \phi_\mr{p} &= \frac{a \sin(s)\cos(s)}{1+\left(\frac{a}{b}\cos(s) \right)^\mr{2}} \\
\end{split}
\end{equation}
where $a$ and $b$ define the specific shape of the path and $s\in[0,2\pi]$ defines a specific position on the path.
Transforming the path definition from spherical into Cartesian coordinates yields
\begin{equation}
	\left(\mathbf{p}\right)_\mr{P} =\begin{pmatrix}
	 \cos\lambda_\mr{p}\cos\phi_\mr{p}\\
	\sin\lambda_\mr{p}\cos\phi_\mr{p}\\
	\sin\phi_\mr{p}
	\end{pmatrix}
\end{equation}
where the subscript P denotes the path frame. It is essentially defined in the same way as the Wind reference frame W (see \cite{Rapp2019}, and Fig.~\ref{fig:small_earth}) but is tilted by an angle $\phi_\mr{r}$ around the negative $\mr{y}_\mr{W}$ axis. The reference path in the W frame is then defined by
\begin{equation}\label{eq:path_in_W}
	\left(\mathbf{p}\right)_\mr{W} = \begin{pmatrix}
	\cos\phi_\mr{r} & 0 & -\sin\phi_\mr{r}	\\
	0 & 1 & 0 \\
	\sin\phi_\mr{r} & 0 & \cos\phi_\mr{r}
	\end{pmatrix}\left(\mathbf{p}\right)_\mr{P}
\end{equation}
Note that this redefinition of the path requires also a small modification in the algorithm in \cite{Rapp2019} that finds the closest point on the path using Newton's method with respect to the current position. In \cite{Rapp2019} the path is fixed at a certain elevation angle. However, since the rotation matrix in Eq. (\ref{eq:path_in_W}) is constant with respect to $s$ the derivatives are not impacted and only the final result in \cite{Rapp2019} needs to be changed. Concretely, the target on the path as well as the tangent and its derivative with respect to $s$ (see \cite{Rapp2019}) need to be rotated by $\phi_\mr{r}$ using the same rotation matrix as utilized in Eq. (\ref{eq:path_in_W}).
\begin{figure}[t!]
	{\centering
	\setlength\extrarowheight{-7pt}
	\begin{tabular}{r@{ }l r@{ }l  }
		$\color{myBlue}{\bm{-}}$ 		& Flight path &
		$\color{myOrange}{\bm{\ast}}$  &  Start/End retraction phase \\
		&
	\end{tabular}\par}
	\centering
	\def\svgwidth{0.4\textwidth}
	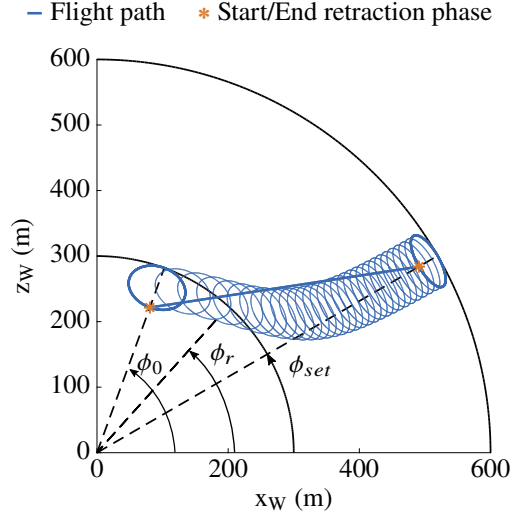
	\caption{Projection of the reference flight path in the xy plane of the W frame.}
	\label{fig:path_rot}
\end{figure}

The transient of the rotation angle $\phi_\mr{r}$ is shaped using a first order filter with bandwidth constant $\omega_\mr{0,r}$ and set point $\phi_\mr{set}$ which corresponds to the reference elevation angle during the traction phase (see Fig.~\ref{fig:path_rot}):
\begin{equation}\label{eq:time_constant_rot}
	\dot{\phi}_\mr{r} =
\begin{cases}
0\quad \text{if} \quad \Delta_\mr{\phi}>\bar{\Delta}_\mr{\phi}\\
 -\omega_\mr{0,r}\phi_\mr{r}+\omega_\mr{0,r}\phi_\mr{set}, \quad \phi_\mr{r}(t=0)=\phi_\mr{0}  \quad \text{else}
	\end{cases}
\end{equation}
In order to avoid that the path is rotated too quickly $\dot{\phi}_\mr{r}$ is set equal to zero as soon as the arc length on the unit sphere $\Delta_\mr{\phi}$  between the path and the current projected position of the aircraft exceeds a certain threshold $\bar{\Delta}_\mr{\phi}$, which is set to one degree for the subsequent simulations.
$\Delta_\mr{\phi}$ is given by
\begin{equation}
\begin{split}
\Delta_\mr{\phi} &= \tilde{\phi}^\mr{G} - \tilde{\phi}^\mr{t}\\
\tilde{\phi}^\mr{G}    &= \arccos\left( \frac{\left( \mathbf{p}_\mr{xy}^\mr{G} \right)_\mr{W}^\top\left( \mathbf{p}^\mr{G} \right)_\mr{W}}{\left\lVert\left( \mathbf{p}_\mr{xy}^\mr{G} \right)_\mr{W}\right\rVert_\mr{2} \left\lVert\left( \mathbf{p}^\mr{G} \right)_\mr{W}        \right\rVert_\mr{2} }\right)\\
\tilde{\phi}^\mr{t} &= \arccos\left( \frac{\left( \mathbf{p}_\mr{xy}^\mr{t} \right)_\mr{W}^\top\left( \mathbf{p}^\mr{t} \right)_\mr{W}}{\left\lVert\left( \mathbf{p}_\mr{xy}^\mr{t} \right)_\mr{W}\right\rVert_\mr{2} \left\lVert\left( \mathbf{p}^\mr{t} \right)_\mr{W}        \right\rVert_\mr{2} }\right)
\end{split}
\end{equation}
where $\left( \mathbf{p}_\mr{xy}^\mr{G} \right)_\mr{W}$ and $\left( \mathbf{p}_\mr{xy}^\mr{t} \right)_\mr{W}$ are the normal projections into the $\mr{x}_\mr{W}\mr{y}_\mr{W}$ plane of the aircraft position $\left( \mathbf{p}^\mr{G} \right)_\mr{W}$ and the target on the path $\left( \mathbf{p}^\mr{t} \right)_\mr{W}$, respectively. All vectors are given in the wind reference frame.
The inverse of $\omega_\mr{0,r}$ represents the time constant $\tau_\mr{r}$ of the transition phase. It is a tuning parameter that defines how quickly the path is rotated into the power zone. In the limit, as $\tau_\mr{r}$ goes to zero the transition scenario without guidance is reached. On the contrary, for large time constants the aircraft will fly most of the time at high elevation angles which will reduce the power output. Hence, the parameter value reflects the trade-off between robustness (large $\tau_\mr{r}$) and maximum power output (small $\tau_\mr{r}$). The impact of the time constant on the robustness is addressed in more detail later in the paper.  
Note, the initial condition is usually chosen smaller than 90 degrees (between 70 and 80 degrees) otherwise this would cause the aircraft to overfly the ground station. The filter is reset at the beginning of the transition phase.

For the retraction phase the straight glide path is defined as the connecting line of the point at which the retraction mode got triggered and a waypoint on the rotated reference path defined by $s=\{\frac{\pi}{2},\frac{3\pi}{2}\}$ and  $\phi_\mr{0}$. The $s$ value is chosen depending in which part of the figure of eight (positive or negative $\mr{y}_\mr{W}$ coordinate) the traction phase got triggered. The retraction phase is triggered if two conditions are met. First, a specified tether length needs to be reached, second, the aircraft needs to pass the point on the path specified by $s_\mr{1}=\frac{\pi}{2}$ or $s_\mr{2}=\frac{3\pi}{2}$. As opposed to directly triggering the retraction phase if the maximum tether length is reached, this approach reduces the possible retraction points on the path to two, which is more convenient for robustness analysis. The downside of this approach is that the maximum length of the tether can vary in one pumping cycle since the increment in tether length per half figure of eight flight varies with the reeling speed. To mitigate this effect the increment in tether length for each half figure of eight flight is predicted based on the previous increment. If the aircraft reaches one of the two possible retraction points the increase in tether length until the other retraction point is reached will be estimated. If the estimated tether length is higher than the maximum allowable tether length the retraction phase will be already triggered at the current retraction point. This feature ensures that the maximum allowable tether length is never exceeded.

Finally, a minor modification of the winch controller is presented. In contrast to \cite{Rapp2019} the winch controller was simplified since the feed-forward part turned out to be too aggressive in highly turbulent wind conditions leading to instabilities due to the acceleration limits of the winch. Instead, a simple PI controller is implemented that calculates a reference torque based on the difference between the tether force set point and the measured tether force on the ground. Based on the reference torque the winch will adapt the reeling speed. This strategy works for traction and retraction phase and requires only different set points.

\section{Upset Condition Generation, Prediction and Avoidance Framework}\label{sec:framework}

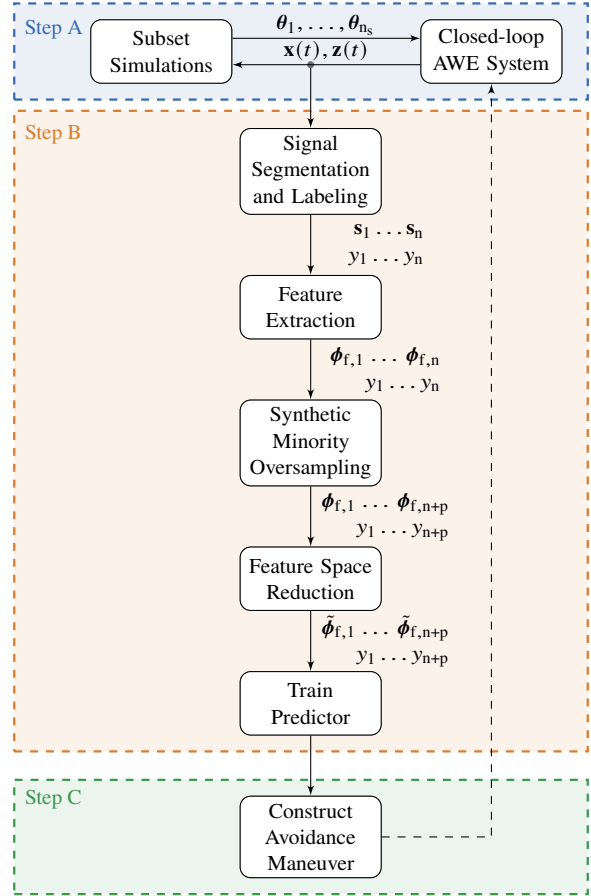
\begin{figure}[t]
	\begin{adjustbox}{max totalsize={1\textwidth}{1\textheight},center}
		\fontsize{8pt}{10pt}\selectfont
		\begin{tikzpicture}
		\node (SS) [block] {Subset Simulations};
		\node (AWES) [block] [right=2.5cm of SS] {Closed-loop AWE System};
	    \node (crossing1)[dot,scale=0.3] [right=1cm of SS,yshift=-16.6667pt]{};
		\node (SigSeg) [block] [below=0.8cm of crossing1] {Signal Segmentation and Labeling};
		\node (FE) [block] [below=0.8cm of SigSeg] {Feature Extraction  };
		\node (GFS) [block] [below=0.8cm of FE] {Synthetic Minority Oversampling};
		\node (SMOTE) [block] [below=0.8cm of GFS] {Feature Space Reduction};
		\node (Pred) [block] [below=0.8cm of SMOTE] {Train Predictor};
		\node (Avoid) [block] [below=0.8cm of Pred] {Construct Avoidance Maneuver};
		
		\node [virtual, below of=SS, xshift=9pt] (node_northwest)  {};
		\node [virtual, below of=AWES,  xshift=-9pt] (node_northeast)  {};
		\node [virtual, right of=AWES,  xshift=2pt] (upright)  {asdf};
		\node [virtual, below of=upright] (midright)  {asdf};
		\node [virtual, below of=midright, yshift=-9cm] (bottomright)  {asdf};
		
		\node [virtual, left= 0.8cm of SS] (upleft)  {};
	    \node [virtual, below= of upleft] (midleft)  {};
		\node [virtual, below= of upleft, yshift=-9cm] (bottomleft)  {};
		
		\node [fit=(upleft) (SS) (upright),draw,dashed,line width=0.3mm,myBlue,inner sep=6pt, fill=myBlue!20,fill opacity=.5] {};
		\node [] [left=of SS,xshift=1cm, yshift=0.3cm]{ \color{myBlue}{Step A} };

		\node [fit=(midleft) (midright) (Pred),draw,dashed,myOrange,line width=0.3mm,inner sep=6pt, fill=myOrange!20,fill opacity=.5] {};
		\node [] [left=of SigSeg,xshift=-1cm, yshift=0.5cm]{ \color{myOrange}{Step B} };
				
	    \node [fit= (bottomleft) (Avoid) (bottomright),draw,dashed,myGreen,line width=0.3mm,inner sep=6pt, fill=myGreen!20,fill opacity=.5] {};
	    \node [] [left=of Avoid,xshift=-1cm, yshift=0.5cm]{ \color{myGreen}{Step C} };
		
		\node (SS2) [block] {Subset Simulations};
		\node (AWES2) [block] [right=2.5cm of SS] {Closed-loop AWE System};
		\node (SigSeg2) [block] [below=0.8cm of crossing1] {Signal Segmentation and Labeling};
		\node (FE2) [block] [below=0.8cm of SigSeg] {Feature Extraction  };
		\node (GFS2) [block] [below=0.8cm of FE] {Synthetic Minority Oversampling};
		\node (SMOTE2) [block] [below=0.8cm of GFS] {Feature Space Reduction};
		\node (Pred2) [block] [below=0.8cm of SMOTE] {Train Predictor};
		\node (Avoid2) [block] [below=0.8cm of Pred] {Construct Avoidance Maneuver};

		\draw[arrow] ([yshift=5pt]SS.east)  --  ([yshift=5pt]AWES.west) node [above,pos=0.5,yshift=-3pt] {\begin{tabular}{c} $\bm{\theta}_\mr{1},\dots, \bm{\theta}_\mr{n_\mr{s}}$\end{tabular}};
		
		\draw[arrow] ([yshift=-5pt]AWES.west)  --  ([yshift=-5pt]SS.east) node [above,yshift=-3pt,pos=0.5] {\begin{tabular}{c} $\mathbf{x}(t), \mathbf{z}(t)$\end{tabular}};
		\draw[arrow] (crossing1) -- (SigSeg);
		\draw[arrow, dashed] (Avoid) -| (AWES);
		
		\draw[arrow] (SigSeg) edge node[align=right, xshift=1cm]{$\mathbf{s}_\mr{1}\dots\mathbf{s}_\mr{n}$\\ $y_\mr{1}\dots y_\mr{n}$ } (FE);
		
		\draw[arrow] (FE) edge node[align=right,xshift=1cm]{ $\bm{\phi}_\mr{f,1}$ $\dots$ $\bm{\phi}_\mr{f,n}$\\ $y_\mr{1}\dots y_\mr{n}$} (GFS);
		
		\draw[arrow] (GFS)  edge node[align=right,xshift=1cm]{ ${\bm{\phi}}_\mr{f,1}$ $\dots$ ${\bm{\phi}}_\mr{f,n+p}$\\
			$y_\mr{1}\dots y_\mr{n+p}$} (SMOTE);
		
		\draw[arrow] (SMOTE) edge node[align=right,xshift=1cm]{ $\tilde{\bm{\phi}}_\mr{f,1}$ $\dots$ $\tilde{\bm{\phi}}_\mr{f,n+p}$\\ 
			$y_\mr{1}\dots y_\mr{n+p}$}(Pred);
		
		
			\draw[arrow] (Pred) edge node[align=right,xshift=1cm]{} (Avoid);
		\end{tikzpicture}
	\end{adjustbox}
	\captionsetup{justification=centering}
	\caption{Workflow of the proposed framework.}
	\label{fig:framework}
\end{figure}

The framework consists of three steps denoted with A, B and C. The different steps can be designed to a large extent independently, which allows to improve the framework in the future in a modular manner. In step A (Upset Condition Generation) the subset simulation (SS) algorithm is utilized to systematically generate samples that lead to a specific upset condition. In step B (Upset Condition Prediction) the prediction model is designed based on the created samples from step A in order to learn to distinguish between upset and nominal conditions. Finally, in step C (Upset Condition Avoidance) the avoidance maneuver is designed. The complete framework is visualized in Fig.~\ref{fig:framework} where the highlighted rectangles enclose the tasks associated to every individual step. The three steps are discussed generically and in more detail in the subsequent paragraphs. In section \ref{sec:results} the framework is applied to generate, predict and prevent tether rupture during pumping cycle operation.

\subsection{Upset Condition Generation}\label{subsec:ucg}
In this work upset conditions are generated using the SS algorithm. The introduction of SS in this section follows \cite{Zuev2021}. Further details about SS and mathematical proofs can also be found in \cite{AU2001263}. 
In general, SS is a popular algorithm to estimate small event probabilities for high dimensional systems \cite{AU2001263}. An event, or failure, probability is a function of a multidimensional random variable $\bm{\Theta}$ . As a function of its probability density function $f_\mr{\bm{\Theta}}$ a failure can be written as a multidimensional integral: 
\begin{equation}\label{eq:failer_prob}
p_\mr{f} = \int_\mr{\Theta} \mathbbm{1}_\mr{F}\left(\bm{\theta}\right)f_\mr{\bm{\Theta}}\left({\bm{\theta}}\right)d\bm{\theta}
\end{equation}
where $\Theta$ represents the entire space of $\bm{\theta}$, $\mathbbm{1}_\mr{F}\left(\bm{\theta}\right)$ is the indicator function that is either one if a certain realization  $\bm{\theta}$ leads to a failure or zero otherwise. Furthermore, in the context of SS it is usually assumed that the random variables are identically and independently (iid) distributed hence
\begin{equation}
f_\mr{\bm{\Theta}}\left({\bm{\theta}}\right) = \prod_{k=1}^{d} f_\mr{\Theta_\mr{k}}\left(\theta_\mr{k}\right)
\end{equation}
In addition, it is assumed that the random variables are transformed such that the transformed variables are iid standard normal random variables with probability density function $f'$.
Directly evaluating the integral in Eq.~(\ref{eq:failer_prob}) analytically or even numerically is not feasible for complex high dimensional systems due to the curse of dimensionality \cite{AU2001263}. One approach to calculate this integral is using direct Monte Carlo methods that randomly sample from the parameter marginal distributions, evaluating the indicator function by simulation and using eventually the sample average to approximate the failure probability. If $p_\mr{f}$ is small this can require an unfeasible amount of simulation runs which is especially critical if one simulation run is time consuming i.e. several minutes or more. Contrarily, in the context of SS $p_\mr{f}$ is written as a product of conditional probabilities which involves the definition of intermediate failure domains. The main idea behind this strategy is that transitioning from one intermediate failure domain to the next has a higher chance than directly transitioning from nominal conditions into the failure domain. The failure probability can then equivalently be expressed as a product of conditional probabilities
 \begin{equation}
p_\mr{f} = \Pr\left(F_\mr{1}\right) \prod_{i=1}^{m_\mr{s}-1} \Pr\left(F_\mr{i+1} \;\middle\vert\; F_\mr{i}\right)
\end{equation}
The first intermediate failure probability $\Pr\left(F_\mr{1}\right)$ is obtained via a direct Monte Carlo approach where $n_\mr{s}$ samples are generated at random. Next, a limit function $g$ that characterizes how close the current sample is to the failure is evaluated for each sample. The limit function is defined such that a higher value indicates a sample that is closer to the actual failure defined by $g^\mr{*}$. The current intermediate failure domain is defined by a threshold $\mathcal{T}$ for the limit function which is in general smaller than the critical value $g^\mr{*}$. Given the current sample set of size $n_\mr{s}$ the threshold $\mathcal{T}$ is calculated by splitting the $n_\mr{s}$ samples into two subsets. One that contains the $n_\mr{s} \cdot p_\mr{s}$ samples with the highest limit function values and one that contains the remaining $n_\mr{s}-n_\mr{s} \cdot p_\mr{s}$ samples. Then $\mathcal{T}$ is the average of the limit function values that separate the two sets. Concretely, if the threshold values of the limit function are arranged in descending order then the intermediate threshold value is defined by
\begin{equation}\label{eq:int_th}
\mathcal{T} = \frac{g^\mr{n_\mr{s} \cdot p_\mr{s}}+g^\mr{n_\mr{s} \cdot p_\mr{s} + 1}}{2}
\end{equation}
where $g^\mr{n_\mr{s} \cdot p_\mr{s}}$ and $g^\mr{n_\mr{s} \cdot p_\mr{s}+1}$ denote the $n_\mr{s} \cdot p_\mr{s}$th and $(n_\mr{s} \cdot p_\mr{s}+1)$-th largest samples with respect to their limit function values in the current sample set.
In that case, the transition probabilities $\Pr\left(F_\mr{i+1} \;\middle\vert\; F_\mr{i}\right)$ are by definition equal to $p_\mr{s}$ which is usually set to 0.1 \cite{AU2001263}. 
In order to populate a intermediate failure domain with new samples a Markov chain Monte Carlo method, such as the modified Metropolis algorithms (Algorithm \ref{alg:metropolis}), is used. The algorithm is is briefly introduced in the following based on \cite{AU2001263}. 
In the context of SS the task of the Metropolis algorithm is to populate an intermediate failure domain with samples that also belong to the current intermediate failure domain i.e. $\tilde{\bm{\theta}}\in F_\mr{i}$. This means that $\tilde{\bm{\theta}}$ leads to a limit function value that is larger than the current intermediate threshold $\mathcal{T}$. As soon as $n_\mr{s}$ samples are contained in the current domain $F_\mr{i}$ the subsequent intermediate failure domain $F_\mr{i+1}$ will be defined. First, the new threshold $\mathcal{T}$ using Eq. (\ref{eq:int_th}) is calculated and afterwards new Markov chains are created to populate $F_\mr{i+1}$.
\begin{algorithm}
	\caption{Modified Metropolis Algorithm}
	\begin{algorithmic}[1]
		\STATE Pick $\bm{\theta}\in F_\mr{i}$
		\FOR{each coordinate  $k=1...d$ in $\bm{\theta}$}
		\STATE  Sample $\tilde{\theta}_\mr{k} \sim \tilde{f}\left( \cdot \middle| \theta_\mr{k} \right)$
		\STATE Compute $\alpha=\frac{f'(\tilde{\theta}_\mr{k})}{f'(\theta_\mr{k})}$
		\STATE  Accept $\tilde{\theta}_\mr{k}$ if $\alpha>1$ or if $\alpha>u$ with $u\sim\mathcal{U}(0,1)$
		\ENDFOR
		\STATE 	Accept $\tilde{\bm{\theta}}$ if $\tilde{\bm{\theta}}\in F_\mr{i}$ o.w. set $\tilde{\bm{\theta}}=\bm{\theta}$
		\RETURN $\tilde{\bm{\theta}}$
	\end{algorithmic}
	\label{alg:metropolis}
\end{algorithm}
New samples, conditioned on an existing sample $\bm{\theta}$ in an intermediate failure domain $F_\mr{i}$, are created by centering a symmetric proposal function $\tilde{f}$ around each coordinate $\theta_\mr{k}$ of $\bm{\theta}$. In this work a Gaussian proposal function is used. Its variance can be calculated adaptively as described in \cite{PAPAIOANNOU201589}.
This results in $n_\mr{s}\cdot p_\mr{s}$ Markov chains with $\frac{1}{p_\mr{s}}-1$ elements. An accept/reject strategy, as defined in line 5 of the algorithm, leads to a non-greedy random walk around the previous state in the Markov chain. Since the intermediate thresholds are selected adaptively with respect to the most promising samples (higher limit function value) and new samples are only accepted if they are contained in the current intermediate failure domain (line 7) the algorithm will return at every stage inputs that drive the system more towards an upset condition (critical limit function value).
This procedure is repeated until more than $n_\mr{s}\cdot p_\mr{s}$ samples lie in the actual failure domain. The actual failure probability can then be approximated by 
\begin{equation}
p_\mr{f} \approx p_\mr{s}^\mr{m_\mr{s}-1}\frac{n_\mr{f}}{n_\mr{s}}
\end{equation}
where $n_\mr{f}>n_\mr{s} \cdot p_\mr{s}$ is the number of samples that lie in the actual failure domain and $m_\mr{s}$ are the number of epochs in the SS run.

Applying the SS algorithm in the context of this work requires to define the upset conditions formally in form of a scalar limit function. All samples that lead to a limit function value that is beyond a defined  threshold value are considered as upset conditions.
The crucial part in modeling an upset condition is the allocation of the upset condition to a reasonable signal value or a combination of different signal values. For instance, if the analyzed upset is stall, the angle of attack represents the obvious choice as a limit function. Since this framework is mostly applicable to control system failure, finding the right limit function is usually done by taking the complement of the control objective. For instance, as described in section \ref{sec:clsys}, the control objective for AWE systems operated in pumping cycle mode can be decomposed into a path-following problem (tangential direction control) and a tether force tracking problem (radial direction control). Hence, the limit function should be able to describe a failure in the tangential or radial direction control objective. The performance of the tangential direction control objective is reflected by the path-following tracking error, which suggest to choose this signal as a limit function to generate conditions in which the controller is not able to keep the aircraft close enough to the flight path. Similarly, in the radial direction the controller needs to track a high tension in the tether for maximum power production while keeping the tether force below the maximum tensile force that the aircraft and the tether itself can still support. An upset condition in this case can then be defined as a condition where the tension in the tether exceeds this critical value. The latter example is investigated in depth in section \ref{sec:results}. Depending on the model fidelity, more complex upset conditions such as too high wing bending or vibrations with certain amplitudes in a certain frequency range can be analyzed, where the external excitation is generated using SS. Ultimately, a wide range of different upset conditions can be converted into a scalar function $g(\bm{\theta})$ with a threshold value beyond which the upset occurs. Note that the choice of this limit function is not limited to a specific functional form. It can be represented by an arbitrary nonlinear scalar function that just needs to be tailored to the specific upset condition. The only constraint is that the function needs to be monotonous such that maximizing the functional value indeed drives the system towards the considered upset condition. In most cases it is dependent on the aircraft states and outputs (e.g. angle of attack, airspeed, wing bending, tether tension,...).
Having defined the limit function the SS algorithm can be applied to sample $\bm{\theta}$s in order to drive $g$ into the specific upset $g>g^*$.



\subsection{Upset Condition Prediction}\label{sec:upset_pred}
In this section two prediction approaches are presented. Since it is assumed that an upset can be defined by the value of the corresponding limit function, a first intuitive prediction approach is to predict an upset solely based on the current functional value of $g$. Due to the stochastic nature of the system the values of $g$ will fluctuate according to the joint distribution of the uncertainties. Threshold values can then be selected based on the distribution of the maximum $g$ values obtained from Monte Carlo simulations. 
 For the sole purpose of classification it is obvious that selecting a threshold value arbitrarily close to the maximum limit function value will yield the highest prediction accuracy (least conservative). However, due to the inertia of the system as well as time delays this will in most cases not allow to avoid the upset condition. Contrarily, if the threshold value decreases the false positive rate will grow (more conservative). For this reason different threshold values need to be tested and a benchmark strategy as presented at the end of this section can be used to identify the best threshold.

In addition to the fixed threshold approach an alternative strategy based on binary time series classification is proposed. The main motivation for this approach is that also the time history of certain states and outputs contains information which can be exploited for prediction. This approach is especially beneficial in the context of controls since critical disturbances often cause an oscillatory behavior prior to the actual upset. Obviously, oscillations can only be detected by analyzing the time and frequency content of a signal in a certain time window which is not possible if the simple threshold approach is utilized. In this work the time series classifier is realized as a \textit{support vector machine} (SVM) which is optimized based on the generated samples in step A of the framework. In the following a concise description of the SVM algorithm is given which is based on \cite[p.383-387]{barberBRML2012}. More details about SVMs can also be found in \cite{cristianini_shawe-taylor_2000}. 

The goal of the SVM algorithm is to find a hyperplane for each class such that the margin between the two planes is maximized. The two spaces defined by the hyperplanes can be defined as
\begin{equation}
\mathbf{w}^{\top}\bm{\phi}_\mr{f,i}+b
\begin{cases}
\geq 1\quad \text{if} \quad \bm{\phi}_\mr{f,i} \quad \text{belongs to class 1}\\
\leq-1 \quad \text{otherwise}
\end{cases}
\end{equation}
where $\mathbf{w}$ is the normal vector of both hyperplanes and $b$ is the bias term.
The distance between the two hyperplanes is given by $\frac{2}{\sqrt{\mathbf{w}^{\top}\mathbf{w}}}$. In order to maximize the distance between the two planes the scalar product $\mathbf{w}^{\top}\mathbf{w}$ needs to be minimized which leads to the following quadratic programming problem \cite[p.384]{barberBRML2012}:
\begin{equation}
\begin{split}
\text{minimise} & \quad \frac{1}{2}\mathbf{w}^{\top} \mathbf{w} \\
\text{subject to} & \quad y_\mr{i}\left( \mathbf{w}^{\top}\bm{\phi}_\mr{f,i}+b \right)\geq 1, \quad i=1,...,n
\end{split}
\end{equation}
with $y_\mr{i}\in\{-1,1\}$. 
The optimization problem can be rewritten in terms of its Lagrangian as defined in \cite[p.386]{barberBRML2012}. It will contain the input vector only as the scalar product $\bm{\phi}_\mr{f,i}^{\top}\bm{\phi}_\mr{f,i}$ which allows to apply the kernel trick. The kernel function essentially maps the input parameter into a higher dimensional space in which both classes are linearly separable \cite[p.382]{barberBRML2012}. A common kernel is the radial basis function, or Gaussian kernel, which is also used in this work. Ultimately, the SVM is used to solve a binary classification problem where a given data set $\mathcal{D}=\{(\bm{\phi}_\mr{f,i}, y_\mr{i}),\text{i}=1,...,\text{n}\}$ is used to construct a model that can predict if a certain input vector belongs to class -1 or 1. The predictor equation that is eventually added to the control system is given by
\begin{equation}
\hat{f} = \sum_{j=1}^{m}\alpha_\mr{j}y_\mr{j}e^\mr{-\left( \frac{  \left( \bm{\phi}_\mr{f,i}- \bm{\phi}_\mr{j}  \right)^\mr{\top} \left( \bm{\phi}_\mr{f,i}- \bm{\phi}_\mr{j}  \right)  }{\sigma^\mr{2}}  \right)}
\end{equation}
where the $\alpha_\mr{j}$'s are the $m$ non-zero Lagrange multipliers of the corresponding support vectors $\bm{\phi}_\mr{j}$ as well as their class labels $y_\mr{j}$, and $\sigma^\mr{2}$ is the variance of the Gaussian kernel which is a hyperparameter that needs to be tuned. $\bm{\phi}_\mr{f,i}$ corresponds to the current feature vector.
The class label is then determined based on the condition
\begin{equation}
\hat{y} = \begin{cases}
1\quad \text{if} \quad \hat{f}\ge 0\\
-1 \quad \text{else}
\end{cases}
\end{equation}
 In this work an upset is defined by $y=-1$ and a nominal condition by $y=1$. Estimated quantities are indicated by the "hat" operator.

The inputs to the SVM based predictor will be specific estimations of aircraft states and wind conditions. Note, it will be assumed that the utilized signal values can be measured at a specific rate, no state estimation is performed. The approach can however be extended by including a state estimator in between the predictor and the sensor outputs.

\begin{figure}[h]
	{\centering
  		\setlength\extrarowheight{-7pt}
		\centering
		\begin{tabular}{r@{ }l r@{ }l r@{ }l }
		$\color{myBlue}{\bm{-}}$ 		& Generic signal &
		$\color{myOrange}{\bm{--}}$ 	& Limit &
		$\color{myOrange}{\bm{\ast}}$  &  Point of failure  \\
		& 
	\end{tabular} 
	\par}
	
	\centering
	\def\svgwidth{0.5\textwidth}
	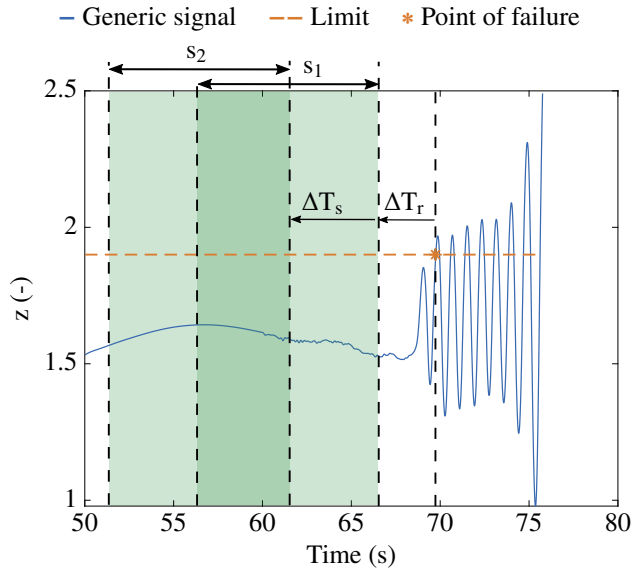
	\caption{Training example with reaction time definition for a hypothetical signal.}
	\label{fig:training_ex2}
\end{figure}

Instead of capturing the complete time history of each signal, specific signal statistics are extracted and collected in a finite dimensional feature vector. Therefore, each signal is cut into smaller segments according to a chosen time window size. For instance, the highlighted green area in Fig.~\ref{fig:training_ex2} indicates a time window with length $\SI{10}{\second}$ of an arbitrary signal denoted here with z which has a hypothetical maximum value $\mathrm{z}_\mr{max}$ of 1.8 (orange, dashed line). At around  $\SI{67}{\second}$ the signal content between $\SI{57}{\second}$ and  $\SI{67}{\second}$, denoted with $s_\mr{1}$ is translated into a feature vector. To create the training examples the time window will be moved from either the final logged data point to the first data point, or if the complete signal contains an upset i.e. $g(\bm{\theta})>g*$, in this example $g(\bm{\theta}) = \mathrm{z}>\mathrm{z}_\mr{max}$,
the segmentation starts where the first upset occurs minus a shift $\Delta \mathrm{T}_\mr{r}$ as depicted in Fig.~\ref{fig:training_ex2}. The additional shift is required otherwise the predictor might fail to forecast an upset prior to its occurrence. The signal segmentation contains overlaps between the segments, hence the first time window is only shifted by $\Delta \mathrm{T}_\mr{s}$ and not by the window length. Note, only the first segment in Fig.~\ref{fig:training_ex2} would be labeled as an upset  i.e. $y=-1$, the following segments starting with $s_\mr{2}$ belong all to the non-upset class and are labeled with $y=1$. Note, the time shifts are hyperparameters that need to be tuned to improve the classifier performance.

If a binary classifier is trained based on the generated data, the prediction accuracy can be improved by balancing the training data set. Although the SS algorithm will systematically generate upset conditions, the segmentation of the logged signals within a pumping cycle will always lead to more non-upset than upset conditions and hence to an extremely imbalanced data set. In fact, most of the simulated pumping cycles will not contain a single upset. One approach, which belongs to the data-level methods of learning from imbalanced data (see \cite{Krawczyk2016}), suggests to use a similar amount of samples from both classes. In this case randomly chosen non-upset samples are removed from the training data set (undersampling). This has the disadvantage that samples are thrown away. Another more sophisticated approach is to synthetically create more samples of the minority class.
This can be achieved using the so called SMOTE algorithm (see \cite{SMOTE}). The algorithm randomly pics a sample from the minority class, determines its k-nearest neighbors, picks one of the k neighbors at random and interpolates again randomly between the two samples to synthesize a new minority class sample. Since the variances vary strongly between the different features the k-nearest neighbors are determined based on the Mahalanobis distance which normalizes the Euclidean distance between two samples using the sample covariance matrix of the training set. This process is repeated until a specified amount of minority class samples has been created. Note, SMOTE can also be applied to the uncertainty vector $\bm{\theta}$. In that case the synthesized input vectors can be tested by simulation if they indeed lead to an upset and hence belong to the minority class. If a synthesized input vector is not leading to an upset it is discarded. This is an advantage over the approach where SMOTE is used to synthesize new  feature vectors. In this case it is not guaranteed with certainty that a new feature vector indeed belongs to the upset class. A drawback of applying SMOTE to the input vectors is that it requires significantly more time to create more samples of the minority class since every synthesized sample requires an additional simulation run. In this work SMOTE is applied directly to the feature space to save training time.

Based on the balanced training set a greedy forward feature selection algorithm as described in \cite{Fulcher2014} is proposed to identify the most relevant features. The relevance of a feature is determined using 10-fold cross-validation and as a metric the average Matthews correlation coefficient (MCC) is used to measure classification performance. The MCC is defined as
\begin{equation}
	\mathrm{MCC} = \frac{n_\mr{TP} \cdot n_\mr{TN} - n_\mr{FP} \cdot n_\mr{FN}}{\sqrt{(n_\mr{TP}+n_\mr{FP})(n_\mr{TP}+n_\mr{FN})(n_\mr{TN}+n_\mr{FP})(n_\mr{TN}+n_\mr{FN})}}
\end{equation}
where $n_\mr{TP},n_\mr{TN},n_\mr{FP}$ and $n_\mr{FN}$ denote the number of true positives, true negatives, false positives and false negatives, respectively. The MCC is the preferred performance measure in binary classification problems since it condenses information of all four quadrants of the confusion matrix in one single number. This is not the case if other measures are used such as accuracy or F1 score which is discussed in detail in \cite{Chicco2020}.
Ultimately, each continuous time series segment is condensed in a $\mathbb{R}^\mr{m}$ dimensional vector $\tilde{\bm{\phi}}_\mr{f,i}$ and the predictor is optimized based on the relationship
\begin{equation}
\begin{pmatrix}
\tilde{\bm{\phi}}_\mr{f,1}	\\
\tilde{\bm{\phi}}_\mr{f,2}  \\
\vdots \\
\tilde{\bm{\phi}}_\mr{f,n+p} 
\end{pmatrix} \rightarrow 	\begin{pmatrix}
y_\mr{1} \\
y_\mr{2} \\
\vdots \\
y_\mr{n+p}
\end{pmatrix}
\end{equation} 
where $y_\mr{i}\in\{-1,1\}$. Note, the "tilde" operator indicates the reduced feature vector, n indicates the amount of samples generated by the SS algorithm and p is the amount of additionally synthesized samples using SMOTE.  In this work the SVM is trained using the Matlab Statistics and Machine Learning Toolbox \cite{MatlabStat}.

In the previous paragraphs two prediction strategies are presented. On the one hand, a simple threshold based predictor and on the other hand a time series classification prediction strategy. 
One open question to be answered is how the performance of the prediction methods can be compared to each other in the context of upset condition prediction for an AWE system. Besides the classical metrics such as accuracy, F1 score or MCC it is beneficial to associate weights to false positives and false negatives that reflect the practical impact on the  system performance.
Since in practice false positives and false negatives have in general a different impact, ranking predictors simply based on their prediction accuracy is not a recommended approach. In the context of AWE a solution is to weight both terms proportionally to the resulting economic loss. In case of a false positive this loss equals the energy loss due to the triggered emergency maneuver $E_\mr{FP}=E_\mr{em}$. The loss stemming from a false negative $E_\mr{FN}$ is more difficult  to estimate since it requires a cost model that is able to predict the energy loss due to system downtime, repair costs and material costs in case the upset damaged the system. In order to combine the impact of false negatives and false positives in a single number, an economic loss rate is introduced which is defined as the weighted linear combination:
\begin{equation}\label{eq:eco_loss}
L = w_\mr{1} E_\mr{FP} +  w_\mr{2} E_\mr{FN}
\end{equation}
where $E_\mr{em}$ and $E_\mr{FN}$ are the associated energy losses in kWh due to false predictions. 
In this work, the weights $w_\mr{1}$ and $w_\mr{2}$ are derived based on the probabilities of false predictions.
Mathematically, the occurrence of either a FP or a FN is modeled as a Poisson process. The Poisson process that models the arrivals of FPs runs until the first arrival time within the Poisson process that models the arrival of a FN. The expected value of the arrival time of a FN allows then to estimate the amount of FPs until that point in time and hence the resulting energy loss. The rate for the process that models the occurrence of a FN is given by
\begin{equation}
\begin{split}
\lambda_\mr{FN}&=\Pr\left( \hat{y}=1, y=-1\right) \\  
&= \Pr\left(\hat{y}=1 \;\middle\vert\;  y=-1 \right)
\Pr\left(y=-1\right)\\
&= \frac{n_\mr{FN}}{n_\mr{FN}+n_\mr{TP}}p_\mr{f}
\end{split}
\end{equation}
The conditional probability is simply given by the false positive rate of the prediction strategy, the probability that $y=-1$ is the upset condition probability which is independent of the prediction approach.
Estimating the conditional probability is done by re-simulating upset conditions for each of the different predictors using the results from the SS run in step A. Note, it is paramount here to use a different SS run than the one used to train the predictor.
With the estimated FN rate the number of pumping cycles until the first expected FN occurs is then given by
\begin{equation}
n_\mr{pc}=\frac{1}{\lambda_\mr{FN}}
\end{equation}
The expected number of FP up to the first FN is given by the expected value of the corresponding Poisson process defined by
\begin{equation}
\begin{split}
n_\mr{FP}&=\lambda_\mr{FP} n_\mr{pc} \\
									 &=\Pr\left(\hat{y}=-1,  y=1 \right)  n_\mr{pc}
\end{split}
\end{equation}
With the SVM based predictor the probability of encountering a false positive per pumping cycle can be estimated by counting falsely predicted upsets in a separate Monte Carlo simulation run. With a fixed threshold predictor this probability can be directly calculated using Eq.~(\ref{eq:pr_false_pos}).
\begin{equation}\label{eq:pr_false_pos}
\Pr(\hat{y}=-1,y=1)=1-F_\mr{\bar{g}(\bm{\theta)}}\left(q^*\right)-p_\mr{f}
\end{equation}
where $q^*$ represents the chosen threshold value and $F_\mr{\bar{g}(\bm{\theta})}$ is the cumulative distribution function of the maximum values of ${g}(\bm{\theta})$ which are calculated for each simulation run. Note, the treshold value represents a quantile of the distribution of $\bar{g}(\bm{\theta)}$ and hence the false positive probability is the corresponding area under the PDF right from the threshold minus the upset probability. 

If a FN occurs the system will not be operational for a specific amount of time $\Delta T_\mr{nop}$. It reflects the required time to conduct a possible emergency landing, maintenance and relaunching. This mainly leads to a power loss in terms of missed pumping cycles. Assuming an average  pumping cycle time of $t_\mr{pc}$ the number of missed pumping cycles is
\begin{equation}
	 n_\mr{mpc}=\frac{\Delta T_\mr{nop}}{t_\mr{pc}}
\end{equation}
The expected energy loss per pumping cycle due to predictions errors is eventually given by
\begin{equation}
\begin{split}
L &= w_\mr{1} E_\mr{FP} +  w_\mr{2} E_\mr{FN} \\
  &= \frac{ 1}{n_\mr{pc}+n_\mr{mpc}} \left( n_\mr{FP} P_\mr{em}t_\mr{pc} + n_\mr{mpc}P_\mr{pc}t_\mr{pc}+E_\mr{misc}\right)
\\\end{split}
\end{equation}
This expression can be normalized by the average energy $E_\mr{pc}=P_\mr{pc}t_\mr{pc}$ converted in one pumping cycle which yields 
\begin{equation}\label{eq:loss}
\frac{L}{E_\mr{pc}} =\frac{1}{n_\mr{pc}+n_\mr{mpc}} \left( n_\mr{FP} \frac{P_\mr{em}}{P_\mr{pc}} + n_\mr{mpc}+\frac{E_\mr{misc}}{E_\mr{pc}}\right)
\end{equation}
$E_\mr{misc}$ combines all additional losses involved with a FN such as replacement costs of damaged parts. 
Equation (\ref{eq:loss}) allows to rank different predictors with respect to their expected energy loss relative to the average converted energy in one pumping cycle. This metric is better suited to assess prediction performance because it associates  weights to false positives and negatives that have a practical meaning. This is not the case if standard metrics for prediction performance are used. Note, at this stage only guesses about the average downtime $\Delta T_\mr{nop}$ as well as the additional involved costs, summarized in $E_\mr{misc}$, can be made. Furthermore, due to the lack of a comprehensive cost model for AWE systems, Eq.~(\ref{eq:loss}) is only an approximation of the monetary loss that might be encountered in reality. In the future, and as soon as more data becomes available, a more accurate cost model should replace the simple model defined in Eq.~(\ref{eq:loss}).

Moreover, note that the features used by the SVM based predictor are selected with respect to the achieved MCC. One could also directly choose Eq.~(\ref{eq:loss}) to rank the performance of feature combinations. However, this requires to rank the features as a function of the parameter values in the loss function for which only rough estimations are available at the moment. 
For this reason the MCC is used to optimize the SVM and Eq.~(\ref{eq:loss}) is only used to compare different prediction strategies after the design phase. Based on these results the best predictor can be chosen and deployed on the real system. 

\subsection{Upset Condition Avoidance}
In step C of the framework the avoidance maneuver is defined. Due to the possibility of false positives it is desirable that the impact of the maneuver on the pumping cycle operation is minimized. One generic approach for upset avoidance during the pumping cycle is to abort the current traction or retraction phase and use the onboard propulsion system to either land the aircraft or to go into a loiter mode from which the normal operation can again be initiated. Both approaches however reduce the average power output of the system significantly. A more efficient upset avoidance strategy needs to be tailored to the upset condition itself, which is demonstrated in the next section for the case of tether rupture. 


\section{Application of the Framework to Generate, Predict and Avoid Tether Rupture}\label{sec:results}

\subsection{Setup}\label{sec:setup}
In this section the the three steps A, B and C of the framework  are applied to the case of tether rupture which is an important upset condition in the field of AWE. Due to the complex interaction between ground station and flight control system, wind, tether as well as the aircraft dynamics it is basically impossible to analytically derive conditions that lead to this critical event.  Furthermore, assuming that a reliable control system is implemented tether rupture has a low probability of occurrence which makes it a suitable example to demonstrate the methodology proposed in this work.
Additionally, since this event has a high relevance for the AWE community, the cause for tether rupture based on the obtained results is investigated in depths. 

In step A the SS algorithm is used to generate systematically conditions that drive the tether force peak within a pumping cycle beyond its maximum allowable value. The limit function is in this case given by
\begin{equation}
g\left(\bm{\theta}\right)=F_\mr{t}
\end{equation}
In the present example the stochastic excitation is limited to the uncertainties in the wind conditions. It is arguably also the highest uncertainty that makes AWE systems  difficult to control. Of course, the framework can be easily extended in order to include model parameter uncertainties, sensor noise or hardware failures as well but this is left for future work. The wind conditions in the simulations are generated using the Dryden Turbulence model that has as input standard Gaussian distributed random variables $\theta_\mr{k}$ that are filtered to recover the Dryden turbulence spectrum. In total $d=T_\mr{sim}f_\mr{s}$ random variables are sampled per run where $T_\mr{sim}$ is the simulation run-time and $f_\mr{s}$ the sampling frequency which is set to $10$ Hz. Further possible variations in the wind field such as discrete gusts or changes in the wind speed profile and the mean wind direction are not considered in this work and are also left for future research.
In general, upset conditions for a complete pumping cycle, or even several pumping cycles in a row can be generated with the proposed framework. However, since the dimension of the joint probability density function from which the wind condition is sampled grows linearly with simulation time all the results are generated for only one pumping cycle per sample. For this specific example the SS algorithm created 4221 tether ruptures and around $7.1\cdot 10^\mr{5}$ segments without tether rupture are extracted. The results are created with one SS run that included in total $3\cdot 10^\mr{4}$ pumping cycle simulations. The selected time window size for one segment is $\SI{5}{\second}$ and the reaction time shift $\Delta T_\mr{r}$ is set to $\SI{0.2}{\second}$.

In step B reasonable state and output variables are selected to predict the upset and the predictor is designed based on the results of step A. In this example only signals that are available at the aircraft are considered in order to avoid communication delays between the ground station and the aircraft. Concretely, the following signals are chosen:
\begin{itemize}
	\item wind speed components $v_{w,x,W}$,  $v_{w,y,W}$, and $v_{w,z,W}$
	\item aircraft acceleration in radial direction $a_\mr{z,\tau}$
	\item Tether force $F_\mr{t}$
	\item angle of attack  $\alpha$
	\item path following error  $e_\mr{p}$
\end{itemize}
Following section \ref{sec:framework}.\ref{sec:upset_pred}, each signal is discretized into smaller overlapping time windows and statistical properties in the time and frequency domain are calculated. The utilized features that are calculated for each of the signals in the time domain are: mean, median, rms-value, variance, maximum, minimum, maximum peak-to-peak ratio, skewness, kurtosis, crest factor, median absolute deviation, range of the cumulative sum, the time-reversal asymmetry statistic given by Eq. (\ref{eq:shift_stat}) and the maximum signal slope.
The time-reversal asymmetry statistic is defined by \cite{Fulcher2014}
\begin{equation}\label{eq:shift_stat}
p = \frac{\mathbb{E}\left( \left( s(t) - s(t-\tau)\right)^\mr{3}\right)}{\left( \mathbb{E}\left( \left( s(t) - s(t-\tau)\right)^\mr{2} \right)   \right)^\mr{\frac{3}{2}}}
\end{equation}
where different values for $\tau$ are considered.
In the frequency domain the following characteristics are calculated: median and maximum amplitude, and additionally the maximum amplitude above 1 Hz using a fast Fourier transform. Note, this set of features is created heuristically. The individual features are chosen because they are computationally cheap to evaluate and easy to comprehend or because they turned out to be useful features in other applications (e.g. the time-reversal statistic in \cite{Fulcher2014}). It will be shown later in the paper that  an optimized subset, which is derived from the initial feature pool, leads to an acceptable prediction performance. Note, it is always possible to add more features to the initial feature pool for instance if the initial prediction performance is poor. However, the larger the initial feature pool, the longer it takes to optimize the smaller subset. Therefore, it is recommended to only gradually increase the size of the feature pool and always check if an acceptable classification performance can be achieved before new features are added. 

\begin{figure}[h]
	\centering
	\def\svgwidth{0.5\textwidth}
	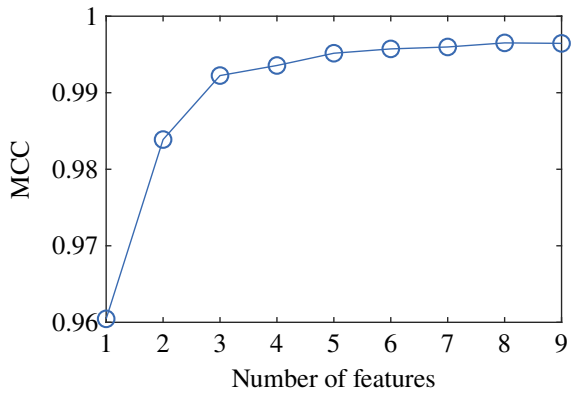
	\caption{Maximization of MCC using greedy feature selection.}
	\label{fig:feature_selection_conv}
\end{figure}

In order to balance the data set additional feature vectors are created using the SMOTE algorithm and afterwards the feature selection algorithm is applied to reduce the dimension of the original feature space. With SMOTE around $6\cdot 10^\mr{5}$ feature vectors are synthesized from the 4221 original samples created by the SS algorithm in step A. Note, in contrast to the original feature vectors it cannot be guaranteed that the synthesized feature vectors indeed belong to the set of feature vectors that lead to tether rupture. One reason is that the set corresponding to a tether rupture is not necessarily convex. Hence, an interpolated feature vector can also end up outside the non-convex set. The optimized subset of features is displayed in Table \ref{tab:featue_table} (ordered according to significance). Note, the MCC value in the second column is the cumulative MCC value. In Fig.~\ref{fig:feature_selection_conv} the convergence of the selection process is displayed. Convergence is defined as the point where the relative change in the MCC after adding a feature to the list is smaller than $10^\mr{-4}$. This convergence criteria is also proposed in \cite{Fulcher2014}.
\begin{table*}[t]
	\caption{\label{tab:featue_table} Ordered feature list. }
	\centering
	\begin{tabular}{llllllllllcc} 
		\hline
		\hline
		Feature  & MCC \\
		\hline
		Crest factor $F_t$  & 9.60\\
		Time-reversal asymmetry statistic for the path tracking error $e_\mr{p}$ Eq. (\ref{eq:shift_stat}) with  $\tau=\SI{1}{\second}$ & 9.83\\
		Maximum $F_\mr{t}$ slope &  9.92\\
		Mean of  $F_\mr{t}$  &  9.935 \\
		Maximum amplitude above 1 Hz of $\alpha$ &  9.951\\
		Median amplitude of $F_\mr{t}$ & 9.957\\
		Minimum $\alpha$ &  9.959\\
		Variance $a_\mr{z,\tau}$ &  9.965\\
		Variance of $F_\mr{t}$ &   9.965\\
		\hline
		\hline
	\end{tabular}
\end{table*}
With the optimized feature list the SVM predictor is trained as explained in section \ref{sec:framework}.\ref{sec:upset_pred}. 
Additionally, fixed thresholds are selected based on the estimated distribution of the tether force peaks using the results of the first stage of the SS run (direct Monte Carlo run). In this case the thresholds are selected with respect to the tether force set point in the traction phase. Concretely, the thresholds $\mathrm{F}_\mr{t,set}+8\%$, $\mathrm{F}_\mr{t,set}+10\%$,  $\mathrm{F}_\mr{t,set}+12\%$, $\mathrm{F}_\mr{t,set}+14\%$ and $\mathrm{F}_\mr{t,set}+16\%$ are considered which are all larger than the 0.99-quantile of the tether force peak distribution which corresponds to $\mathrm{F}_\mr{t,set}+7\%$. The set point $F_\mr{t,set}$ itself is chosen to be $-20\%$ of the maximum allowable tether tension which is set to $\SI{2}{\kilo\newton}$. 

In step C of the framework the avoidance maneuver is designed. In case of a predicted tether rupture the contingency maneuver must reduce the current tension in the tether as quickly as possible. It turns out that with the underlying control system this can be achieved with a set point change for the tether force. On the one hand, the tension in the tether is tracked by the winch controller via the reeling out/in speed and on the other hand by the flight path controller through the angle of attack and bank angle. Therefore, changing the set point for the tether force leads to an adaptation of the winch reeling speed but also of the angle of attack and bank angle commands $\alpha_\mr{set}$ and $\mu_\mr{a,set}$, respectively. As derived in \cite{Rapp2019} both attitude commands are determined by inverting the flight path dynamics which yields
\begin{equation}\label{eq:bank_set}
\begin{split}
f_\mr{y,m} &= m_\mr{a} \nu_{\chi_\mr{k}} \cos\gamma_\mr{k} v_\mr{k} - f_\mr{t,y,K }\\
f_\mr{z,m} &= m_\mr{a} \nu_{\gamma_\mr{k}} v_\mr{k} + \cos\gamma_\mr{k} m_\mr{a} g +  f_\mr{t,z,K }
\end{split}
\end{equation}
\begin{equation}\label{eq:alpha_set}
\begin{split}
\mu_\mr{a,set} \approx \mu_\mr{k,set} &= \arctan\left(\frac{f_\mr{y,m}}{f_\mr{z,m}} \right) \\
C_\mr{L,set}(\alpha) &= \frac{\sqrt{f_\mr{y,m}^2 + f_\mr{z,m}^2}}{0.5\rho v_a^2 S_\mr{w}}\\
\alpha_\mr{set} &= C_\mr{L,set}^{-1}\left( \dots \right)
\end{split}
\end{equation}
where the tether force set point components $f_\mr{t,y,K}$ and $f_\mr{t,z,K}$ are obtained by
\begin{equation}\label{eq:ftxyz}
\begin{pmatrix}
f_\mr{t,x,K} \\ f_\mr{t,y,K} \\f_\mr{t,z,K} \end{pmatrix} = -\mathbf{M}_\mr{KO}(\chi_\mr{k},\gamma_\mr{k})\frac{\left(\mathbf{p}^G\right)_\mr{O}}{\left\lVert \left(\mathbf{p}^G\right)_\mr{O}\right\rVert_2}F_\mr{t,set}
\end{equation}
$\left(\mathbf{p}^\mr{G}\right)_\mr{O}$ is the position of the aircraft in the Nort-East-Down frame O and $\mathbf{M}_\mr{KO}(\chi_\mr{k},\gamma_\mr{k})$ transforms a vector from the O frame into the kinematic frame $K$ (see \cite{Rapp2019}).
Essentially, Eq.~(\ref{eq:bank_set}-\ref{eq:ftxyz}) calculate the angle of attack and bank angle commands based on the desired path curvature represented by the pseudo-control inputs $\nu_{\chi_\mr{k}}$ and $\nu_{\gamma_\mr{k}}$ as well as the tether force set point $F_\mr{t,set}$. Note, for consistency the kinematic bank angle should in general be converted into the aerodynamic bank angle (banking around the aerodynamic instead of the kinematic velocity vector) but the effect is negligible here which leads to the approximation $\mu_\mr{a,set} \approx \mu_\mr{k,set}$. The angle of attack command reflects the required lift magnitude which is estimated by the required forces $f_\mr{y,m}, f_\mr{z,m}$ and involves the inversion of the lift coefficient as shown in Eq. (\ref{eq:alpha_set}). Furthermore, $\chi_\mr{k}$ is the aircraft course angle, $\gamma_\mr{k}$ is the flight path angle, $g$ denotes gravity, $v_\mr{a}$ is the airspeed, $\rho$ is the air density and $S_\mr{w}$ is the wing reference area.
Note, in order to track the tether force the angle of attack and bank angle commands are calculated using the tether force set point $F_\mr{t,set}$ and not the measured tether force currently acting on the aircraft. During nominal operation this allows to effectively keep the tether under the desired tension. As soon as an upset is predicted this set point will be reduced to a low value ($F_\mr{t,set}=\SI{10}{\newton}$). As a result the aircraft will correct the current bank and angle of attack commands accordingly. This allows to reliably reduce the tension in the tether quickly even if the winch is currently saturating which is discussed in the next section.
As soon as the tether force drops below a second threshold value i.e. $ F_\mr{t}\leq c\cdot\SI{10}{\newton}$ with for instance $c=1.2$ and the predictor output switches from $\hat{y}=-1$ (upset) back to $\hat{y}=1$ (no upset) the force set point is increased again to the original traction phase set point. The set point change is shaped smoothly using a first order filter.


\subsection{Results}

\begin{figure}[h]
	\centering
	\def\svgwidth{0.5\textwidth}
	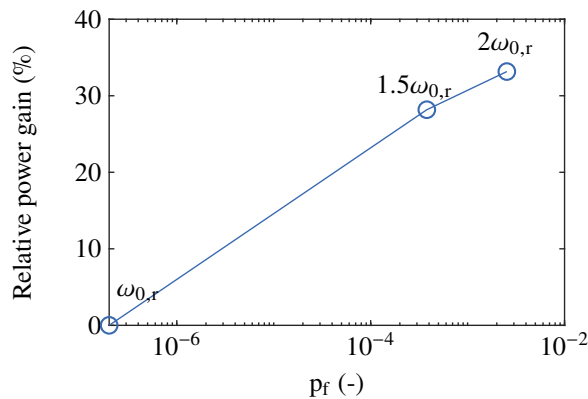
	\caption{Relative gain in pumping cycle power over tether rupture probability as a function of multiples of the path rotation constant $\omega_\mathrm{0,r}$.}
	\label{fig:power_gain_over_reliab_loss}
\end{figure}

The introduced control system modification in section \ref{sec:clsys} has an important impact on the overall probability of tether rupture and on the power output which is why it is also included in this paper. Concretely, the choice of the bandwidth  $\omega_\mathrm{0,r}$ that defines how quickly the path is rotated from higher to lower elevation angles during the transition from retraction to traction phase allows to trade off robustness against performance.  In Fig.~\ref{fig:power_gain_over_reliab_loss} the results of three independent subset simulations are presented with three different choices for $\omega_\mathrm{0,r}$.
As a reference $\omega_\mathrm{0,r}=0.05$ is chosen which leads to an approximate tether rupture probability during a pumping cycle of $\mathrm{p}_\mr{f}\approx2\cdot 10^\mr{-7}$. Increasing the reference value by a factor of $1.5$ and $2$ increases the power output by $28\%$ and  $33\%$  but also leads to a significant increase of the tether rupture probability by a factor of approximately $1.9\cdot 10^\mr{3}$ and $1.3\cdot 10^\mr{4}$, respectively. Since there is no external standard that defines the allowable tether rupture probability the conservative value of $\omega_\mathrm{0,r}=0.05$ is chosen to generate the subsequent results. The corresponding low probability also justifies the use of SS to generate this type of upset condition in the first place, whereas the other two controller settings defined by $1.5\omega_\mathrm{0,r}$ and $2\omega_\mathrm{0,r}$ lead to tether rupture probabilities that might be analyzed with simple Monte Carlo simulations. Given a desired level of reliability $\omega_\mathrm{0,r}$ can be adapted in the future accordingly.

The performance of the different prediction strategies  (SVM and thresholds) is tested on a separately generated data set that is not used to construct the predictors. The test data set is generated in the same manner as the training data set using the SS algorithm.
In the first part of this section the effectiveness of the avoidance maneuver is analyzed. Subsequently, the prediction and prevention performance among the different predictors is assessed using Eq.~(\ref{eq:loss}).

\begin{figure}[h]
	{\centering
		\begin{tabular}{r@{ }l r@{ }l  }
			$\color{myBlue}{\bm{-}}$ 		& With avoidance &
			$\color{myOrange}{\bm{--}}$ 	& Without avoidance \\
			$\color{myGreen}{\bm{\times}}$  & Tether break point &
			$\color{red}{\bigstar}$ & Start point \\
			$\color{red}{\tkzcircsimple}$ 	& End point &
		\end{tabular}\par}
	\centering
	\def\svgwidth{0.5\textwidth}
	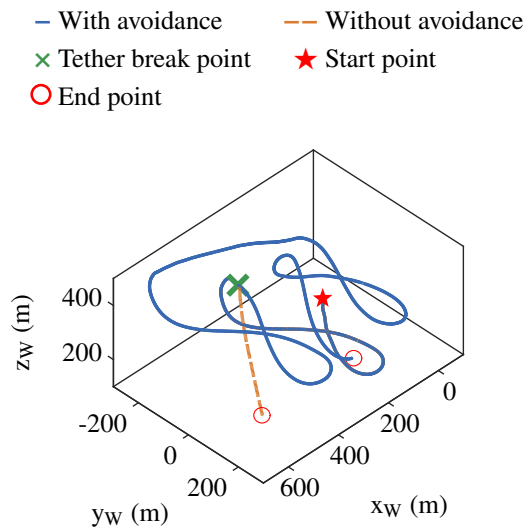
	\caption{Three dimensional flight paths with and without avoidance strategy.}
	\label{fig:path_3d_rupture_no_rupture}
\end{figure}

\begin{figure}[h]
	{\centering
		\begin{tabular}{r@{ }l r@{ }l  }
			$\color{myBlue}{\bm{-}}$ 		& With avoidance &
			$\color{myOrange}{\bm{--}}$ 	& Without avoidance \\
			$\color{myGreen}{\bm{\times}}$  & Tether break point &
			$\color{red}{\bigstar}$ & Start point \\
			$\color{red}{\tkzcircsimple}$ 	& End point &
		\end{tabular}\par}
	\centering
	\def\svgwidth{0.5\textwidth}
	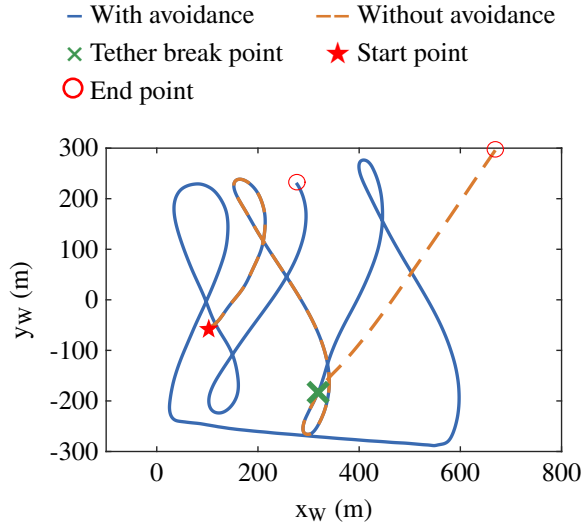
	\caption{Projected flight paths with and without avoidance strategy.}
	\label{fig:path_xy_rupture_no_rupture}
\end{figure}

\begin{figure}[h]
	{\centering
		\begin{tabular}{r@{ }l r@{ }l }
			$\color{myBlue}{\bm{-}}$ 		& With avoidance &
			$\color{myOrange}{\bm{--}}$ 	& Without avoidance \\
			$\color{myGreen}{\bm{--}}$  & Avoidance trigger point 
		\end{tabular}\par}
	\centering
	\def\svgwidth{0.5\textwidth}
	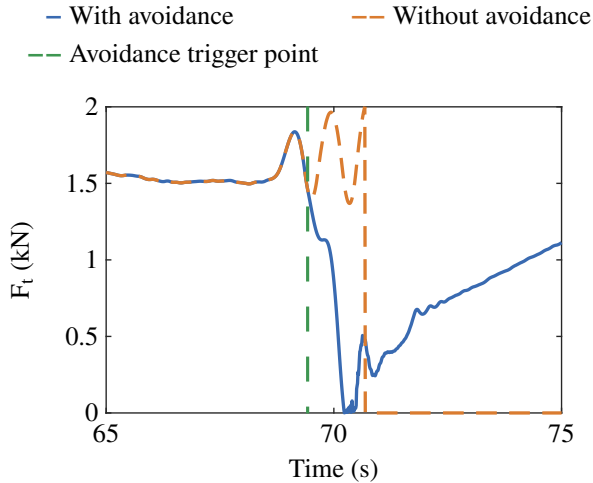
	\caption{Evolution of the tether tension with and without prediction and avoidance strategy.}
	\label{fig:tether_rupture_zoom}
\end{figure}

The effectiveness of the prediction and avoidance strategy is demonstrated and explained in detail using the results of one sample of the test set that contains a tether rupture. To that end, the same simulation is carried out twice once with prediction and avoidance method and once without. To limit the scope of the result section only the results using the SVM prediction strategy are displayed and analyzed. The resulting flight path of both scenarios is displayed in Fig.~\ref{fig:path_3d_rupture_no_rupture}, the corresponding projection in the $\mathrm{x}_\mr{W}\mathrm{y}_\mr{W}$ plane is depicted in Fig.~\ref{fig:path_xy_rupture_no_rupture}. The blue path shows one complete pumping cycle where the tether rupture is prevented using the proposed avoidance maneuver. It can be observed that the system is able to continue its operation and the avoidance maneuver has no visible impact on the path following performance. The green cross indicates the point where the tether breaks and as a result, in the scenario without avoidance strategy, the aircraft is ejected from the reference flight path and is no longer able to continue the pumping cycle. 
The evolution of the tether force in both scenarios is depicted in Fig.~\ref{fig:tether_rupture_zoom}. At around $\SI{69}{\second}$ the avoidance maneuver is triggered which leads to a significant tether tension reduction as indicated by the blue solid line. In contrast to that, without the avoidance maneuver the tether tension continues to oscillate and at around $\SI{70.5}{\second}$ the tether breaks.

\begin{figure}[h]
	{\centering
	\begin{tabular}{r@{ }l r@{ }l r@{ }l  }
		$\color{myBlue}{\bm{-}}$ 		& $\mu_\mr{a,set}$ &
		$\color{myOrange}{\bm{--}}$ 	& $\mu_\mr{a,ref}$ &
		$\color{myGreen}{\bm{\cdots}}$  & $\mu_\mr{a}$ 
	\end{tabular}\par}
	\centering
	\def\svgwidth{0.5\textwidth}
	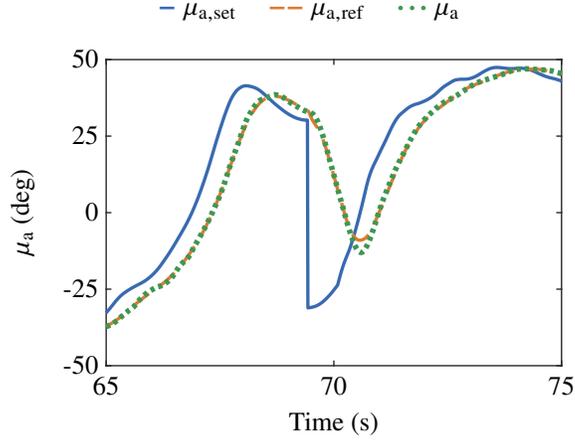
	\caption{Evolution of the aerodynamic bank angle. Set point $\mu_\mr{a,set}$, reference $\mu_\mr{a,ref}$ and achieved bank angle $\mu_\mr{a}$.}
	\label{fig:mu_a_ref_wpred}
\end{figure}

The control system performance during the avoidance maneuver is analyzed more in detail using the evolution of the aerodynamic bank angle and the angle of attack. In Fig.~\ref{fig:mu_a_ref_wpred} the impact of the tether tension set point change is clearly visible in the evolution of the bank angle set point $\mu_\mr{a,set}$ (blue, solid line). At around $\SI{69}{\second}$ the set point drops to around $\SI{-30}{\degree}$. Since the controller uses a dynamic inversion based control strategy the set point is filtered (orange, dashed line) and the actual bank angle is controlled such that it follows the corresponding reference model (green, dotted line). A similar behavior results for the angle of attack $\alpha_\mr{set}$. Also in this case the tether tension set point change leads to a drop in the angle of attack set point. The actual angle of attack follows the corresponding reference model with an overshoot of approximately $\SI{2.3}{\degree}$.
\begin{figure}[h]
	{\centering
		\begin{tabular}{r@{ }l r@{ }l r@{ }l  }
			$\color{myBlue}{\bm{-}}$ 		& $\alpha_\mr{set}$ &
			$\color{myOrange}{\bm{--}}$ 	& $\alpha_\mr{ref}$ &
			$\color{myGreen}{\bm{\cdots}}$  & $\alpha$ 
		\end{tabular}\par}
	\centering
	\def\svgwidth{0.5\textwidth}
	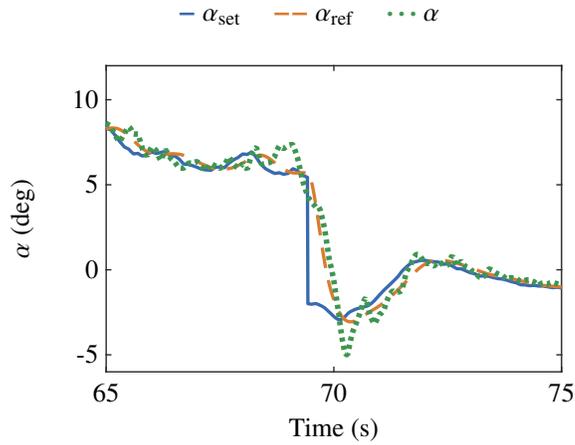
	\caption{Evolution of the angle of attack with set point $\alpha_\mr{set}$, reference filter state $\alpha_\mr{ref}$ and achieved angle of attack $\alpha$ .}
	\label{fig:alpha_a_ref_wpred}
\end{figure}
 
\begin{figure}[h]
	 {\centering
		\begin{tabular}{r@{ }l r@{ }l  }
			$\color{myBlue}{\bm{-}}$ 		& With avoidance &
			$\color{myOrange}{\bm{--}}$ 	& Without avoidance \\  
			$\color{myGreen}{\bm{--}}$  	& Avoidance trigger point &
			$\color{myRed}{\bm{\cdots}}$ 	& Tether break point
		\end{tabular}\par}
	\centering
	\def\svgwidth{0.5\textwidth}
	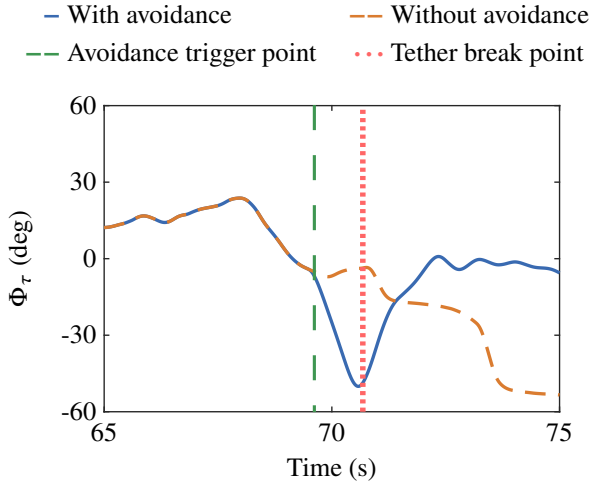
	\caption{Roll angle with and without avoidance maneuver.}
	\label{fig:phi_tau_upset}
\end{figure}

\begin{figure}[h]
	{\centering
		\begin{tabular}{r@{ }l r@{ }l  }
			$\color{myBlue}{\bm{-}}$ 		& With avoidance &
			$\color{myOrange}{\bm{--}}$ 	& Without avoidance \\  
			$\color{myGreen}{\bm{--}}$  	& Avoidance trigger point &
			$\color{myRed}{\bm{\cdots}}$ 	& Tether break point
		\end{tabular}\par}
	\centering
	\def\svgwidth{0.5\textwidth}
	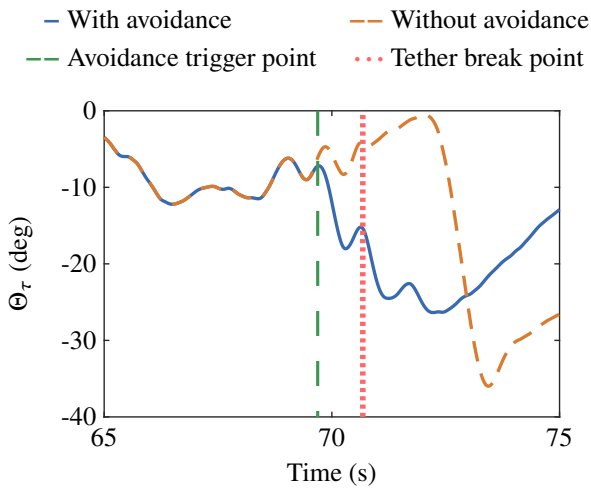
	\caption{Pitch angle with and without avoidance maneuver.}
	\label{fig:theta_tau_upset}
\end{figure}

\begin{figure}[h]
	\centering
	\def\svgwidth{0.55\textwidth}
	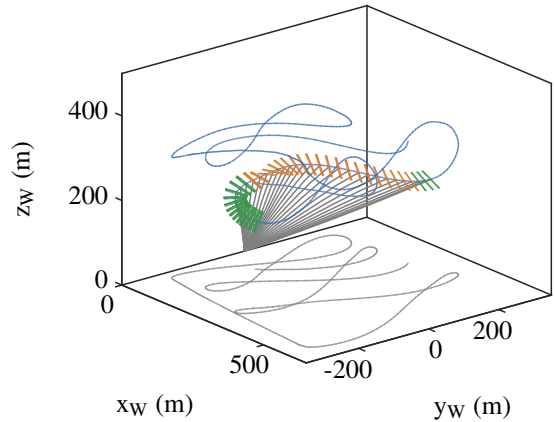
	\caption{Flight path of entire pumping cycle (blue) with the same aircraft attitude visualization as in Fig.~\ref{fig:upset_aircraft_visual_magnified}}
	\label{fig:upset_aircraft_visual}%
\end{figure}

\begin{figure}
	\centering
	\def\svgwidth{0.55\textwidth}
	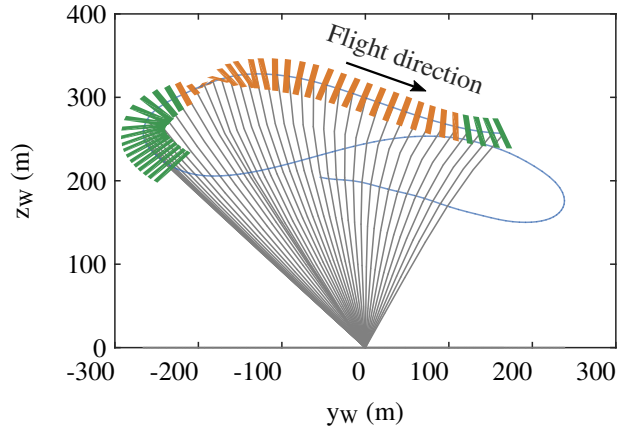
	\captionsetup{margin=0.2cm}
	\caption{Flight path (blue) with aircraft attitude visualization. Green indicates the pre- and post-avoidance maneuver state, orange indicates the avoidance maneuver state.}
	\label{fig:upset_aircraft_visual_magnified}%
\end{figure}

The adaption of the bank angle and the angle of attack leads to an adaption of the aircraft attitude with respect to the tangential plane. Therefore, besides the change in lift magnitude through the adaption of the angle of attack also the rotation of the lift force leads to a tether force reduction.  If the aircraft is flying in the tangential plane most of the lift force is pointing in the radial direction. Increasing the attitude angles (absolute value) with respect to the tangential plane by a simultaneous roll and pitch maneuver can reduce the tension in the tether since the component of the lift vector perpendicular to the tether direction increases. This behavior can be observed in Fig.~\ref{fig:phi_tau_upset} and Fig.~\ref{fig:theta_tau_upset}. 
Both plots demonstrate that due to the tether force set point change the aircraft is indeed rotated into the tangential plane (blue, solid line). The roll angle $\Phi_\mr{\tau}$ is reduced from a nearly horizontal attitude (with respect to the tangential plane) to $\SI{-50}{\degree}$, the pitch angle $\Theta_\mr{\tau}$ is reduced from around $\SI{-5}{\degree}$ to $\SI{-25}{\degree}$. Without the avoidance maneuver (orange, dashed line) the roll angle stays nearly constant and the pitch angle starts to oscillate and to increase. The green dotted line indicates the point where the tether breaks. 
For the case with tether rupture avoidance the resulting trajectory is again displayed in Fig.~\ref{fig:upset_aircraft_visual}. The blue line represents the flight path of one pumping cycle, of which only the part around the prevented tether rupture is displayed in Fig.~\ref{fig:upset_aircraft_visual_magnified}.  The tether is shown as a solid gray line connecting the aircraft with the ground station. Additionally, a simple aircraft visualization (colored rectangle) is added to the figure which represents the orientation of the aircraft wing. The aircraft visualization color changes from green to orange as soon as the avoidance maneuver is triggered. The resulting attitude change is visible in the beginning of the maneuver where the aircraft rolls negatively, with respect to the body-fixed frame x-axis, into the tangential plane. The color changes back to green as soon as the avoidance maneuver is finished. In this case the end of the avoidance maneuver is defined as the first time the tether force set point reaches again $90\%$ of the original traction phase tether tension set point. A visible drawback of the avoidance maneuver is that the aircraft flies about a quarter of the figure of eight at low tether tension which results in a power loss. Hence, the amount of falsely predicted upsets (i.e. false positives) needs to be traded off against the power loss associated with a tether rupture. On the other hand, a visible deviation from the flight path cannot be observed and the aircraft continues the traction phase without interruption which is an advantage over strategies that need to abort the current operational mode in order to prevent tether rupture. 

%

\begin{figure}[h]
		{\centering
		\begin{tabular}{r@{ }l r@{ }l }
			$\color{myBlue}{\bm{-}}$ 		& $\mr{a}_\mr{winch}$ &
			$\color{myOrange}{\bm{--}}$ 	& Tether break point 
		\end{tabular}\par}
	\centering
	\def\svgwidth{0.5\textwidth}
	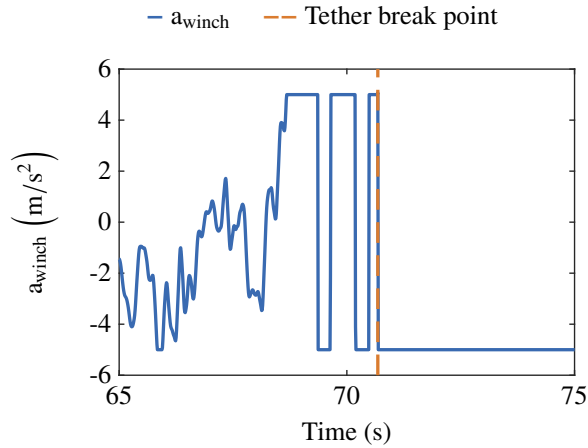
	\caption{Evolution of the winch acceleration without tether rupture avoidance.}
	\label{fig:w_acc_without_pred}
\end{figure}

\begin{figure}[h]
		{\centering
		\begin{tabular}{r@{ }l r@{ }l }
			$\color{myBlue}{\bm{-}}$ 		& $\mr{a}_\mr{winch}$ &
			$\color{myGreen}{\bm{--}}$ 	& Avoidance trigger point 
		\end{tabular}\par}
	\centering
	\def\svgwidth{0.5\textwidth}
	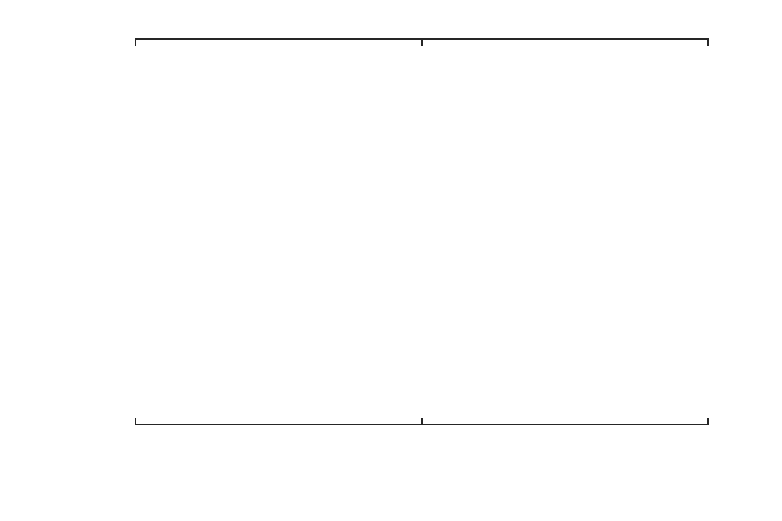
	\caption{Evolution of the winch acceleration with tether rupture avoidance.}
	\label{fig:w_acc_with_pred}
\end{figure}

\begin{figure}[h]
	{\centering
		\begin{tabular}{r@{ }l r@{ }l }
			$\color{myBlue}{\bm{-}}$ 		& $\mr{v}_\mr{winch}$ &
			$\color{myOrange}{\bm{--}}$ 	& Tether break point 
		\end{tabular}\par}
	\centering
	\def\svgwidth{0.5\textwidth}
	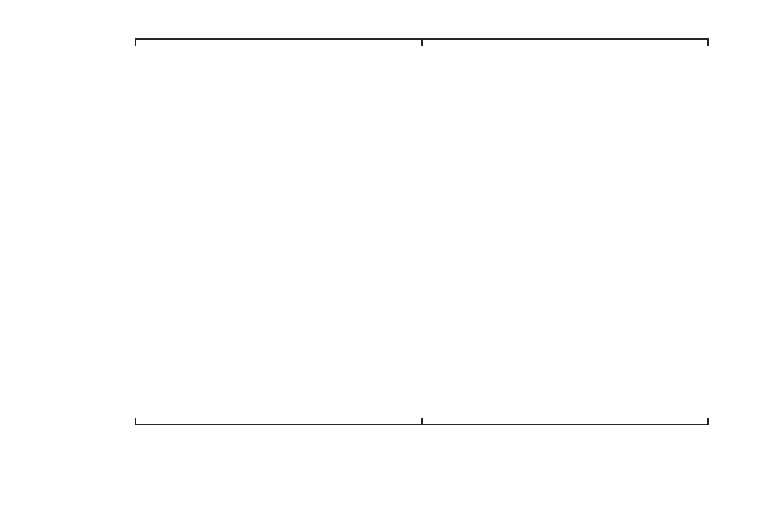
	\caption{Evolution of the winch speed without tether rupture avoidance.}
	\label{fig:w_v_ro_without_pred}
\end{figure}

\begin{figure}[h]
	{\centering
		\begin{tabular}{r@{ }l r@{ }l }
			$\color{myBlue}{\bm{-}}$ 		& $\mr{v}_\mr{winch}$ &
			$\color{myGreen}{\bm{--}}$ 	& Avoidance trigger point 
		\end{tabular}\par}
	\centering
	\def\svgwidth{0.5\textwidth}
	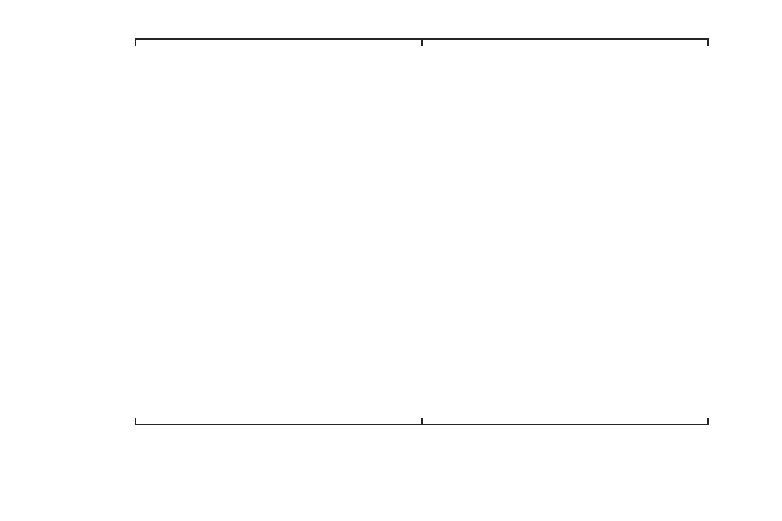
	\caption{Evolution of the winch speed with tether rupture avoidance.}
	\label{fig:w_v_ro_with_pred}
\end{figure}

The actual reason for the tether break can be found by looking at the evolution of the tether, or winch, acceleration measured on the ground. In Fig.~\ref{fig:w_acc_without_pred} the winch acceleration for the scenario without avoidance maneuver is displayed. The dashed orange line indicates the point where the tether breaks. Before the tether breaks the winch acceleration saturates at around $\SI{68}{\second}$ and starts to jump between the maximum and the minimum acceleration limit with increasing frequency until the maximum supported tension is exceeded and the tether breaks. In contrast, Fig.~\ref{fig:w_acc_with_pred} shows the winch acceleration for the scenario with avoidance maneuver. In this case, the oscillation is prevented and instead the winch stays in the upper saturation limit leading to a fast reeling out of the tether. The start of the avoidance maneuver is indicated by the dashed green line.
The corresponding winch speed for the flight with tether rupture is displayed in Fig.~\ref{fig:w_v_ro_without_pred}. It can be observed that the winch speed itself is not saturating but also starts to oscillate due to the saturated acceleration. In contrast, with the avoidance maneuver the reeling out speed continues to increase after a small kink at the prediction point (see Fig.~\ref{fig:w_v_ro_with_pred}) and  tether rupture is prevented. As soon as the avoidance maneuver is completed the winch starts reeling out slower according to the increasing tether force set point at around $\SI{70.2}{\second}$ (see Fig.~\ref{fig:tether_rupture_zoom}).

\begin{figure}[h]
	{\centering
		\begin{tabular}{r@{ }l r@{ }l }
			$\color{myBlue}{\bm{-}}$ 		& $\mr{v}_\mr{w,x,W}$ &
			$\color{myOrange}{\bm{--}}$ 	& Tether break point 
		\end{tabular}\par}
	\centering
	\def\svgwidth{0.5\textwidth}
	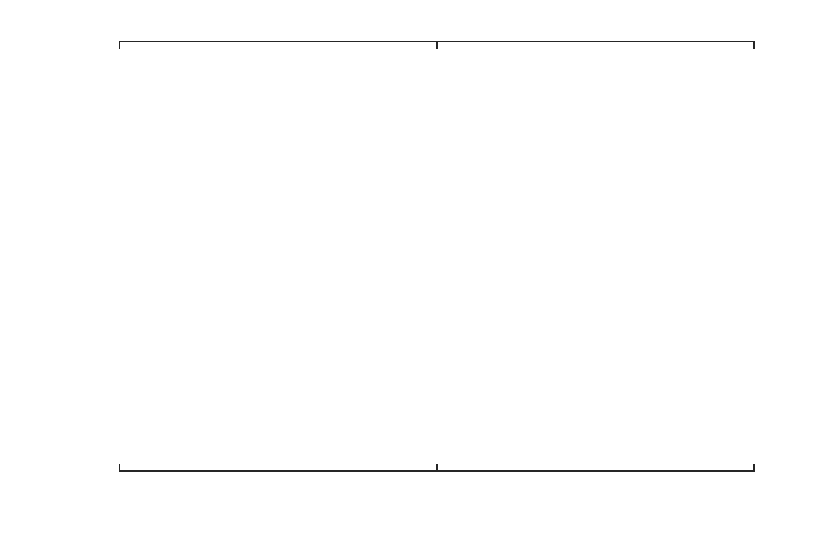
	\caption{Evolution of wind speed in mean wind direction without avoidance maneuver. }
	\label{fig:v_w_x_rupture_zoom_noavoid}
\end{figure}

\begin{figure}[h]
	{\centering
		\begin{tabular}{r@{ }l r@{ }l }
			$\color{myBlue}{\bm{-}}$ 		& $\mr{v}_\mr{w,x,W}$ &
			$\color{myGreen}{\bm{--}}$ 	& Avoidance trigger point 
		\end{tabular}\par}
	\centering
	\def\svgwidth{0.5\textwidth}
	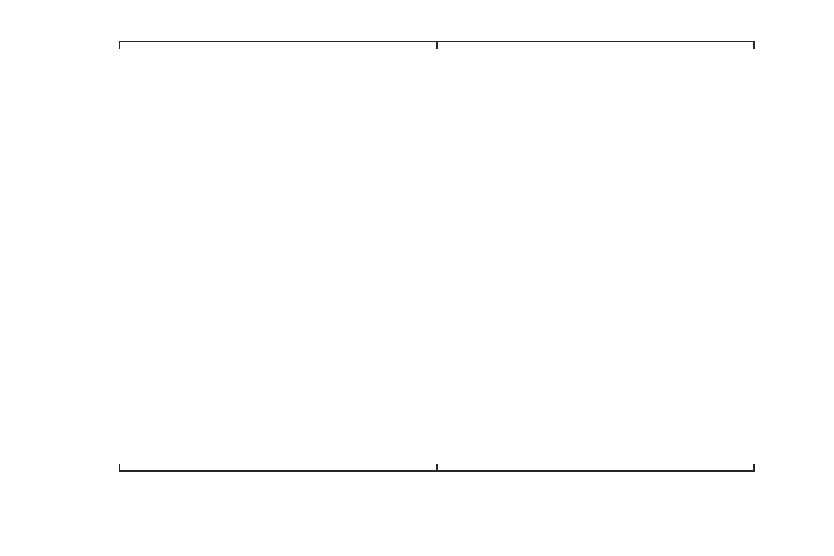
	\caption{Evolution of wind speed in mean wind direction with avoidance maneuver. }
	\label{fig:v_w_x_rupture_zoom}
\end{figure}

For completeness the mean wind speed at the aircraft around the point in time at which the avoidance maneuver is triggered is displayed in Fig.~\ref{fig:v_w_x_rupture_zoom} and the wind speed evolution of the flight without avoidance maneuver in the same time window is displayed in Fig.~\ref{fig:v_w_x_rupture_zoom_noavoid}. The evolution of the wind speed before the avoidance maneuver is triggered or before the tether ruptures does not show any visible changes compared to wind speed after the avoidance maneuver and after tether rupture. This indicates that the tether rupture is not caused by an easy to comprehend change in the wind conditions at the aircraft but rather is a result of the complex interaction between aircraft and winch dynamics, as well as the control system and the wind conditions. This is consistent with the results displayed in Table \ref{tab:featue_table} where the wind speed is not among the selected features.

\begin{figure}[h]
	\centering
	\def\svgwidth{0.5\textwidth}
	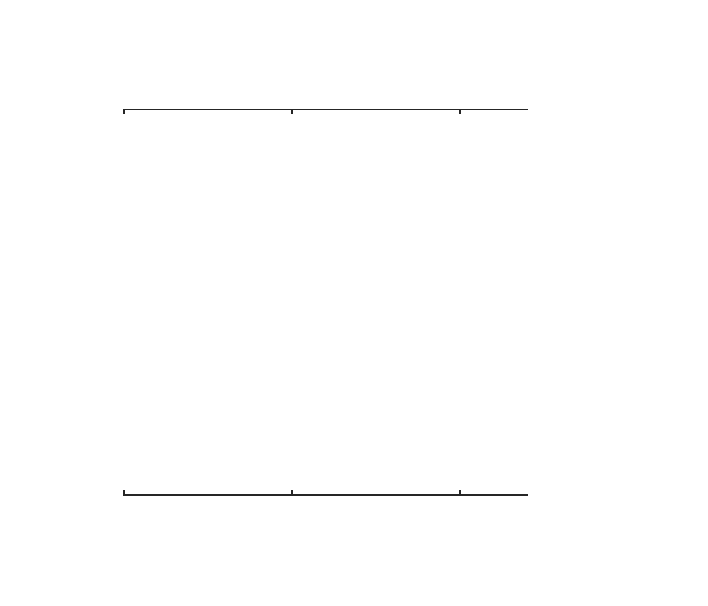
	\caption{Distribution of airspeed $\mathrm{v}_\mr{a}$ and angle of attack  $\alpha_\mr{a}$ pairs at tether rupture.}
	\label{fig:va_over_alpha_upset}
\end{figure}

\begin{figure}[h]
	\centering
	\def\svgwidth{0.5\textwidth}
	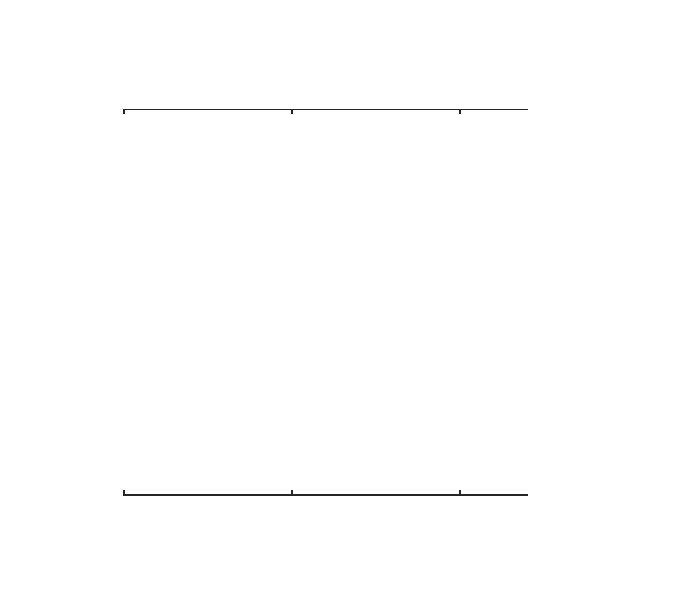
	\caption{Distribution of of airspeed $\mathrm{v}_\mr{a}$ and angle of attack  $\alpha_\mr{a}$ pairs at randomly selected times during traction phases without tether rupture.}
	\label{fig:va_over_alpha_noupset}
\end{figure}

In the previous paragraph the winch acceleration limits are identified as one cause for the tether rupture. However, a second factor represented by a specific airspeed and angle of attack combination can be identified. This can be shown by analyzing the distribution of airspeed and angle of attack pairs for simulation runs with and without tether rupture. Note, for the upset case the values at the upset are taken and for the nominal flight the values are picked at randomly chosen positions on the flight path. The results are displayed in Fig.~\ref{fig:va_over_alpha_upset} and Fig.~\ref{fig:va_over_alpha_noupset}. In total 22200 simulations without tether rupture and 5749 simulations with tether rupture from three different subset simulation runs are used to approximate the distributions. The red solid line in both figures represents the same optimized separation boundary. The distributions itself are plotted in two different figures for visualization purposes. It can be observed that most of the samples above the boundary are simulation runs where the tether ruptured whereas most of the simulation runs below the boundary are flight without tether rupture. The color gradient represents the conditional probability of a specific airspeed and angle of attack combination given a tether rupture or given no tether rupture. The results show that a perfect separation between the two distributions is not possible. This is however to be expected since using only airspeed and angle of attack to distinguish tether rupture conditions from nominal flights reduces the dimension of the problem significantly. However, except some minor overlap in the tails the two modes of the distributions are indeed distinguishable (brighter color). Furthermore, above an airspeed of $\SI{37}{\meter\per\second}$ there is a high chance that a sample belongs to the upset class independently of the angle of attack value. Similarly, below $\SI{30}{\meter\per\second}$ and independent of the angle of attack no tether rupture will occur with a high probability. This result suggests an additional strategy to avoid tether rupture by limiting the angle of attack set point as a function of airspeed according to the plotted linear decision boundary. However, this is not investigated in this work further and is left for future research. 

\begin{figure}
		{\centering
		\begin{tabular}{r@{ }l r@{ }l }
			\color{myBlue}{\tkzcircsimple} 	& Thresholds &
			$\color{myOrange}{\bm{*}}$ 	& SVM 
		\end{tabular}\par}
	\centering
	\def\svgwidth{0.5\textwidth}
	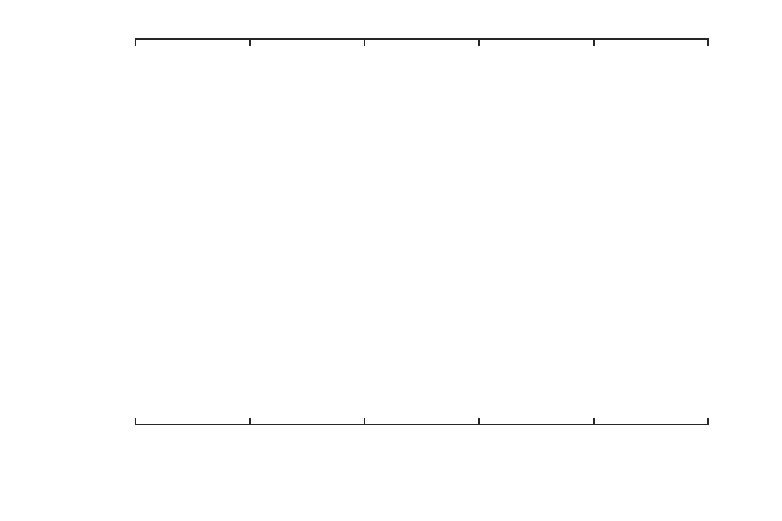
	\captionsetup{margin=0.2cm}
	\caption{Conditional probability of not preventing tether rupture given a tether rupture for different prediction thresholds and the SVM predictor.}
	\label{fig:fp_fn_curve}%
\end{figure}

In the following the performance of different predictors will be investigated with respect to classical performance metrics as well as the introduced economic loss rate given by Eq.~(\ref{eq:loss}). In total, five different threshold based predictors and one SVM based predictor are compared to each other. The results are visualized in Fig.~\ref{fig:fp_fn_curve} where the conditional probabilities of not detecting and preventing an occurring tether rupture is plotted over the probability of a false positive. The blue, solid line with circular markers connects the performance pairs of the five threshold strategies. The performance of the SVM predictor is represented by the orange asterisk. The numerical values are listed in Table \ref{tab:up_det_p}. The false negative rates are estimated based on  764 flights with tether rupture and the false positive probability is estimated based on 20803 samples without tether rupture.  
\begin{table}[b!]
	\caption{\label{tab:up_det_p} Upset detection performance.}
	\centering
	\begin{tabular}{llllccc} 
		\hline
		\hline
		Method & $\Pr\left(\textrm{FP}\right)$  & $\Pr\left(\hat{y}=1 \;\middle\vert\;  y=-1 \right)$  \\
		\hline
		$F_\mr{t,set}+8\%$ & 0.48\% & 0\%       \\
		$F_\mr{t,set}+10\%$ & 0.13 \% & 0\%    \\
		$F_\mr{t,set}+12\%$ & 0.04\% & 0.26\%  \\
		$F_\mr{t,set}+14\%$ & 0.01\% & 1.44\%  \\
		$F_\mr{t,set}+16\%$ & 0\% & 7.20\%    \\
		SVM & 0\% & 0.79\%  \\
		\hline
		\hline
	\end{tabular}
\end{table}
For the fixed thresholds the probability of a FP is negatively correlated with the probability of a not correctly identified/prevented tether rupture, as expected. The closer the threshold value is selected  to the critical value the more likely it is that the tether rupture cannot be prevented due to the inertia of the system. In contrast to that, the more conservative the threshold is chosen, i.e. closer to the set point, the more likely it is to prevent a tether rupture but at the cost of an increasing false positive rate. Based on the numerical values and also based on Fig.~\ref{fig:fp_fn_curve} it is difficult to decide which is the best prediction strategy. The SVM achieves the best performance with respect to false positives together with the highest threshold which however has a almost 10 times higher probability of not avoiding a tether rupture. Between the thresholds $F_\mr{t,set}+14\%$ and $F_\mr{t,set}+12\%$ the largest threshold value can be found that outperforms the SVM in terms of the false negative rate. However, this predictor and predictors below this threshold lead to higher false positive rates. Therefore, selecting the right prediction strategy solely based on these results is difficult because no reasonable acceptable FP and FN rate can be defined a priori and both metrics are not equally important from a practical point of view.

The energy loss rate defined in Eq.~(\ref{eq:loss}) tries to solve the aforementioned issue by assigning weights to the FP and FN rate proportionally to the associated performance loss. Since no reasonable estimation of the term $\mathrm{E}_\mathrm{misc}$ can be made at this stage of the research, $\mathrm{E}_\mathrm{misc}$ is set to zero in the following analysis. Note, in this case a false negative impacts the performance loss only through the power loss due to the number of missed pumping cycles during the downtime of the system. The expected downtime after a tether rupture is not available either but a reasonable range of values can be defined and the losses can be plotted over the selected range. All other parameter values can be estimated using sample averages from the Monte Carlo simulations. The numerical values are listed in Table \ref{tab:param_ranges}. 
 \begin{table}[b!]
 	\caption{\label{tab:param_ranges} Average parameter values.}
 	\centering
 	\begin{tabular}{llllccc} 
 		\hline
 		\hline
 		Parameter & Value  & Unit \\
 		\hline
 		$\mathrm{P}_\mr{em}$ & $0.4$ & $\SI{}{\kilo\watt}$ \\
 		$\mathrm{P}_\mr{pc}$ & $3.9$ & $\SI{}{\kilo\watt}$ \\
 		$\mathrm{t}_\mr{pc}$ & $2.5$  & minutes\\
 	    $\mathrm{p}_\mr{f}$ & $2\times 10^{-7}$  & -\\
 		\hline
 		\hline
 	\end{tabular}
 \end{table}

\begin{figure}[h]
	{\centering
	\begin{tabular}{r@{ }l r@{ }l r@{ }l}
	 $\color{myOrange}{\bm{\ast}}$ &  $F_\mr{t,set}+8\%$ & $\color{myGreen}{\bm{\triangledown}}$ & $F_\mr{t,set}+10\%$ & $\color{myRed}{\bm{\times}}$ & $F_\mr{t,set}+12\%$ \\
		$\color{black}{\bm{\cdots}}$ &  $F_\mr{t,set}+14\%$ & 	$\color{myBlue}{\tkzcircsimple}$ & SVM &   $\color{myBlue}{\bm{--}}$ & no predictor 
	\end{tabular}\par}
	\centering
	\def\svgwidth{0.5\textwidth}
	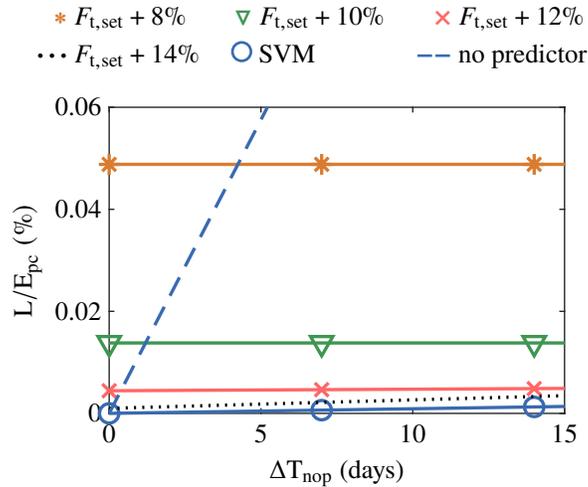
	\caption{Comparison of loss rates across different predictors.}
	\label{fig:loss_over_downtime}
\end{figure}

In Fig.~\ref{fig:loss_over_downtime} the performance loss rate $L$ per average converted energy per pumping cycle $E_\mr{pc}$ is displayed as a function of system downtime using Eq.~(\ref{eq:loss}). The results show that without prevention strategy (dashed line) the relative performance loss grows quickly with increasing system downtime even for the estimated low probability of tether rupture. For the thresholds $F_\mr{t,set}+8\%$ and $F_\mr{t,set}+10\%$ the loss rate remains constant since their false negative rate is zero and hence the downtime has no impact on the loss. The other thresholds and the SVM predictor loss rates remain nearly constant as well due to the overall small probability of false negatives among the predictors. The SVM leads to the lowest loss rate among the predictors in the considered time window. 

\subsection{Discussion Model Validity and Future Work}
The presented framework uses models of the AWE system as well as the wind to create, predict and prevent upset conditions. The accuracy of the models is critical in order to be able to project the results to reality. The aircraft model has been validated to some extent as described in \cite{PhDthesisLicitra:} and \cite{Malz19} but especially for quick changes in the wind conditions the aerodynamic model is probably too aggressive since changes in the local flow immediately change the resulting lift force. It is expected that with a more realistic aerodynamic model an additional time delay between changes in the local flow field around the aircraft and the resulting change of the tether force is present which might alter the presented results in the previous section. The present model can hence be regarded as conservative and it is expected that the prediction accuracy can be further improved with a more realistic model. Testing the framework with a more realistic aircraft model is therefore regarded as the main suggestion for future work. This will also allow to investigate further upset conditions related to the structural and aerodynamic integrity of the aircraft. For instance, wind conditions that lead to critical wing bending or severe vibrations can be generated using the SS algorithm and a data-driven predictor such as the SVM predictor can be used to trigger a load and/or vibration alleviation strategy if necessary. Finally, it needs to be emphasized that the presented results are strongly dependent on the specific controller. In order to investigate how well the results generalize it is recommended to apply the presented methodology to a different closed loop system model in the future. Moreover, additional data-driven methods for the predictor can be tested to further decrease both false positive and negative rates and hence the economic loss. Finally, if more information about the terms in Eq.~(\ref{eq:loss}) is available the SVM should be optimized with respect to the loss and not the MCC.

\section{Conclusion}\label{sec:conclusion}
The contribution of the present work consists of two major parts. First, a modification to an existing pumping cycle control system is presented. It improves the transition from retraction to traction phase on the guidance level by a controlled rotation of the figure of eight flight path from high to low elevation angles. Second, a framework to generate, predict and prevent upset conditions that jeopardize the long term reliability of Airborne Wind Energy (AWE) systems is presented. The feasibility of the framework is demonstrated with the example of tether rupture. The presented results in the paper allow to draw the following conclusions.

First, the introduced control modification can reduce the tether rupture probability significantly if a large time constant for the transition phase is chosen. However, for increasing transition times the aircraft flies longer at higher elevation angles which decreases the average pumping cycle power. Hence, the choice of the time constant involves a trade off between robustness and average power output. 

Second, a conservative implementation of the control modification leads to a low tether rupture probability. This makes it impractical to create knowledge about the conditions in which the tether breaks using a direct Monte Carlo simulation approach. The subset simulation (SS) algorithm can achieve this task more efficiently. The samples generated by the SS algorithm can be used to design and optimize a prediction model that is able to detect tether rupture before it occurs. In order to trade off false positive and false negative rates a cost function is introduced that is better suited to rank predictor performances than conventional classification measures. It allows to associate an average energy loss rate to each of the two prediction error types and hence weights prediction errors according to their practical impact. The support vector machine based predictor achieves the lowest loss in the investigated scenario but more accurate information about the involved parameters in the cost function such as system downtime, repair costs, maintenance costs and so on is required in the future to improve the validity of the loss rate function.  

Third, the proposed upset avoidance maneuver can reliably prevent tether rupture while keeping the system operational. The avoidance maneuver does not require to abort the current pumping cycle or even land the aircraft which lowers the impact of false positives on the average power output. 

Fourth, the analysis of the flights with tether rupture shows that the upset is a result of the winch acceleration saturation as well as a specific combination of airspeed and angle of attack during the traction phase of the pumping cycle. No visible patterns in the wind conditions could be identified which leads to the conclusion that this specific upset is due to the complex interaction between the dynamics of the subsystems in addition to the atmospheric turbulence. 

Finally, for well tuned control systems upset conditions occur with low probability which poses the question if accepting upsets is better than preventing them and therefore avoiding any prediction error induced costs. However, the results in this work show that in the long run also rare upset conditions can have an impact on the average power output, hence augmenting AWE baseline control systems with an upset tailored prediction and prevention strategy, such as the one presented in this work, is recommended.

\section*{Acknowledgments}

This research has been supported financially by the project AWESCO (H2020-ITN-642682), funded by the European Union’s Horizon 2020 research and innovation program under the Marie Skłodowska-Curie grant agreement No. 642682.

\clearpage

\bibliography{./bibliography}

\end{document}

%% file: macros.tex
\newcommand{\mr}[1]{{\mathrm{#1}}}

\newcommand{\zse}{{z_\mr{\tau}}}



%% file: small_earth.pdf_tex
\begingroup%
  \makeatletter%
  \providecommand\color[2][]{%
    \errmessage{(Inkscape) Color is used for the text in Inkscape, but the package 'color.sty' is not loaded}%
    \renewcommand\color[2][]{}%
  }%
  \providecommand\transparent[1]{%
    \errmessage{(Inkscape) Transparency is used (non-zero) for the text in Inkscape, but the package 'transparent.sty' is not loaded}%
    \renewcommand\transparent[1]{}%
  }%
  \providecommand\rotatebox[2]{#2}%
  \newcommand*\fsize{\dimexpr\f@size pt\relax}%
  \newcommand*\lineheight[1]{\fontsize{\fsize}{#1\fsize}\selectfont}%
  \ifx\svgwidth\undefined%
    \setlength{\unitlength}{308.19735718bp}%
    \ifx\svgscale\undefined%
      \relax%
    \else%
      \setlength{\unitlength}{\unitlength * \real{\svgscale}}%
    \fi%
  \else%
    \setlength{\unitlength}{\svgwidth}%
  \fi%
  \global\let\svgwidth\undefined%
  \global\let\svgscale\undefined%
  \makeatother%
  \begin{picture}(1,0.73611933)%
    \lineheight{1}%
    \setlength\tabcolsep{0pt}%
    \put(0,0){\includegraphics[width=\unitlength,page=1]{small_earth.pdf}}%
    \put(0.35160383,0.59337417){\color[rgb]{0,0,0}\makebox(0,0)[lt]{\lineheight{0}\smash{\begin{tabular}[t]{l}$\mr{x}_\mr{B}$\end{tabular}}}}%
    \put(0.16956762,0.55253436){\color[rgb]{0,0,0}\makebox(0,0)[lt]{\lineheight{0}\smash{\begin{tabular}[t]{l}$\mr{y}_\mr{B}$\end{tabular}}}}%
    \put(0.28047125,0.50831964){\color[rgb]{0,0,0}\makebox(0,0)[lt]{\lineheight{0}\smash{\begin{tabular}[t]{l}$\mr{z}_\mr{B}$\end{tabular}}}}%
    \put(0.37148253,0.17630441){\color[rgb]{0,0,0}\makebox(0,0)[lt]{\lineheight{0}\smash{\begin{tabular}[t]{l}$\phi$\end{tabular}}}}%
    \put(0.4954236,0.29948244){\color[rgb]{0,0,0}\makebox(0,0)[lt]{\lineheight{0}\smash{\begin{tabular}[t]{l}$\mr{y}_\tau$\end{tabular}}}}%
    \put(0.39015369,0.38182435){\color[rgb]{0,0,0}\makebox(0,0)[lt]{\lineheight{0}\smash{\begin{tabular}[t]{l}$\mr{x}_\tau$\end{tabular}}}}%
    \put(0.45712749,0.19336249){\color[rgb]{0,0,0}\makebox(0,0)[lt]{\lineheight{0}\smash{\begin{tabular}[t]{l}$\zse$\end{tabular}}}}%
    \put(0.12647301,0.02245232){\color[rgb]{0,0,0}\makebox(0,0)[lt]{\lineheight{0}\smash{\begin{tabular}[t]{l}$\mr{x}_\mr{W}$\end{tabular}}}}%
    \put(0.30000036,0.00370731){\color[rgb]{0,0,0}\makebox(0,0)[lt]{\lineheight{0}\smash{\begin{tabular}[t]{l}$\lambda$\end{tabular}}}}%
    \put(0.4642947,0.08695715){\color[rgb]{0,0,0}\makebox(0,0)[lt]{\lineheight{0}\smash{\begin{tabular}[t]{l}O\end{tabular}}}}%
    \put(-0.00304188,0.23929432){\color[rgb]{0,0,0}\makebox(0,0)[lt]{\lineheight{0}\smash{\begin{tabular}[t]{l}Small Earth\end{tabular}}}}%
    \put(0.77628192,0.14096767){\color[rgb]{0,0,0}\makebox(0,0)[lt]{\lineheight{0}\smash{\begin{tabular}[t]{l}r=1\end{tabular}}}}%
    \put(0.85919415,0.04195593){\color[rgb]{0,0,0}\makebox(0,0)[lt]{\lineheight{0}\smash{\begin{tabular}[t]{l}$\mr{y}_\mr{W}$\end{tabular}}}}%
    \put(0.48634801,0.55888971){\color[rgb]{0,0,0}\makebox(0,0)[lt]{\lineheight{0}\smash{\begin{tabular}[t]{l}$\mr{z}_\mr{W}$\end{tabular}}}}%
    \put(0.48498099,0.44071817){\color[rgb]{0,0,0}\makebox(0,0)[lt]{\lineheight{0}\smash{\begin{tabular}[t]{l}Zenith\end{tabular}}}}%
    \put(0,0){\includegraphics[width=\unitlength,page=2]{small_earth.pdf}}%
    \put(0.47516486,0.71612548){\color[rgb]{0,0,0}\makebox(0,0)[lt]{\lineheight{0}\smash{\begin{tabular}[t]{l}$\mr{v}_\mr{w}$\end{tabular}}}}%
    \put(0.34247375,0.38984828){\color[rgb]{0,0,0}\makebox(0,0)[lt]{\lineheight{0}\smash{\begin{tabular}[t]{l}r\end{tabular}}}}%
    \put(0.24728056,0.34072504){\color[rgb]{0,0,0}\makebox(0,0)[lt]{\lineheight{0}\smash{\begin{tabular}[t]{l}h\end{tabular}}}}%
    \put(0.35362187,0.29333354){\color[rgb]{0,0,0}\makebox(0,0)[lt]{\lineheight{0}\smash{\begin{tabular}[t]{l}$\tau$\end{tabular}}}}%
  \end{picture}%
\endgroup%

%% file: shear.pdf_tex
\begingroup%
  \makeatletter%
  \providecommand\color[2][]{%
    \errmessage{(Inkscape) Color is used for the text in Inkscape, but the package 'color.sty' is not loaded}%
    \renewcommand\color[2][]{}%
  }%
  \providecommand\transparent[1]{%
    \errmessage{(Inkscape) Transparency is used (non-zero) for the text in Inkscape, but the package 'transparent.sty' is not loaded}%
    \renewcommand\transparent[1]{}%
  }%
  \providecommand\rotatebox[2]{#2}%
  \newcommand*\fsize{\dimexpr\f@size pt\relax}%
  \newcommand*\lineheight[1]{\fontsize{\fsize}{#1\fsize}\selectfont}%
  \ifx\svgwidth\undefined%
    \setlength{\unitlength}{225bp}%
    \ifx\svgscale\undefined%
      \relax%
    \else%
      \setlength{\unitlength}{\unitlength * \real{\svgscale}}%
    \fi%
  \else%
    \setlength{\unitlength}{\svgwidth}%
  \fi%
  \global\let\svgwidth\undefined%
  \global\let\svgscale\undefined%
  \makeatother%
  \begin{picture}(1,0.66666667)%
    \lineheight{1}%
    \setlength\tabcolsep{0pt}%
    \put(0,0){\includegraphics[width=\unitlength,page=1]{shear.pdf}}%
    \put(0.17333333,0.07622233){\makebox(0,0)[t]{\lineheight{1.25}\smash{\begin{tabular}[t]{c}0\end{tabular}}}}%
    \put(0.318548,0.07622233){\makebox(0,0)[t]{\lineheight{1.25}\smash{\begin{tabular}[t]{c}200\end{tabular}}}}%
    \put(0.46376233,0.07622233){\makebox(0,0)[t]{\lineheight{1.25}\smash{\begin{tabular}[t]{c}400\end{tabular}}}}%
    \put(0.608977,0.07622233){\makebox(0,0)[t]{\lineheight{1.25}\smash{\begin{tabular}[t]{c}600\end{tabular}}}}%
    \put(0.75419133,0.07622233){\makebox(0,0)[t]{\lineheight{1.25}\smash{\begin{tabular}[t]{c}800\end{tabular}}}}%
    \put(0.899406,0.07622233){\makebox(0,0)[t]{\lineheight{1.25}\smash{\begin{tabular}[t]{c}1000\end{tabular}}}}%
    \put(0.54000033,0.01933333){\makebox(0,0)[t]{\lineheight{1.25}\smash{\begin{tabular}[t]{c}Altitude (m)\end{tabular}}}}%
    \put(0,0){\includegraphics[width=\unitlength,page=2]{shear.pdf}}%
    \put(0.160889,0.11){\makebox(0,0)[rt]{\lineheight{1.25}\smash{\begin{tabular}[t]{r}10\end{tabular}}}}%
    \put(0.160889,0.20447267){\makebox(0,0)[rt]{\lineheight{1.25}\smash{\begin{tabular}[t]{r}12\end{tabular}}}}%
    \put(0.160889,0.298945){\makebox(0,0)[rt]{\lineheight{1.25}\smash{\begin{tabular}[t]{r}14\end{tabular}}}}%
    \put(0.160889,0.39341767){\makebox(0,0)[rt]{\lineheight{1.25}\smash{\begin{tabular}[t]{r}16\end{tabular}}}}%
    \put(0.160889,0.48789){\makebox(0,0)[rt]{\lineheight{1.25}\smash{\begin{tabular}[t]{r}18\end{tabular}}}}%
    \put(0.160889,0.58236267){\makebox(0,0)[rt]{\lineheight{1.25}\smash{\begin{tabular}[t]{r}20\end{tabular}}}}%
    \put(0.04866667,0.37000033){\rotatebox{90}{\makebox(0,0)[t]{\lineheight{1.25}\smash{\begin{tabular}[t]{c}$\mathrm{v}_\mr{w,x,W}$ (m/s)\end{tabular}}}}}%
    \put(0,0){\includegraphics[width=\unitlength,page=3]{shear.pdf}}%
  \end{picture}%
\endgroup%

%% file: traj_3d.pdf_tex
\begingroup%
  \makeatletter%
  \providecommand\color[2][]{%
    \errmessage{(Inkscape) Color is used for the text in Inkscape, but the package 'color.sty' is not loaded}%
    \renewcommand\color[2][]{}%
  }%
  \providecommand\transparent[1]{%
    \errmessage{(Inkscape) Transparency is used (non-zero) for the text in Inkscape, but the package 'transparent.sty' is not loaded}%
    \renewcommand\transparent[1]{}%
  }%
  \providecommand\rotatebox[2]{#2}%
  \newcommand*\fsize{\dimexpr\f@size pt\relax}%
  \newcommand*\lineheight[1]{\fontsize{\fsize}{#1\fsize}\selectfont}%
  \ifx\svgwidth\undefined%
    \setlength{\unitlength}{308.28414917bp}%
    \ifx\svgscale\undefined%
      \relax%
    \else%
      \setlength{\unitlength}{\unitlength * \real{\svgscale}}%
    \fi%
  \else%
    \setlength{\unitlength}{\svgwidth}%
  \fi%
  \global\let\svgwidth\undefined%
  \global\let\svgscale\undefined%
  \makeatother%
  \begin{picture}(1,0.66272063)%
    \lineheight{1}%
    \setlength\tabcolsep{0pt}%
    \put(0,0){\includegraphics[width=\unitlength,page=1]{traj_3d.pdf}}%
    \put(0.21510835,0.29327247){\makebox(0,0)[rt]{\lineheight{1.25}\smash{\begin{tabular}[t]{r}200\end{tabular}}}}%
    \put(0,0){\includegraphics[width=\unitlength,page=2]{traj_3d.pdf}}%
    \put(0.21510835,0.35792565){\makebox(0,0)[rt]{\lineheight{1.25}\smash{\begin{tabular}[t]{r}300\end{tabular}}}}%
    \put(0,0){\includegraphics[width=\unitlength,page=3]{traj_3d.pdf}}%
    \put(0.21510835,0.42257858){\makebox(0,0)[rt]{\lineheight{1.25}\smash{\begin{tabular}[t]{r}400\end{tabular}}}}%
    \put(0,0){\includegraphics[width=\unitlength,page=4]{traj_3d.pdf}}%
    \put(0.24843677,0.21145842){\makebox(0,0)[rt]{\lineheight{1.25}\smash{\begin{tabular}[t]{r}-200\end{tabular}}}}%
    \put(0,0){\includegraphics[width=\unitlength,page=5]{traj_3d.pdf}}%
    \put(0.11231025,0.38293139){\rotatebox{90}{\makebox(0,0)[t]{\lineheight{1.25}\smash{\begin{tabular}[t]{c}$\textrm{z}_\textrm{W}$ (m)\end{tabular}}}}}%
    \put(0,0){\includegraphics[width=\unitlength,page=6]{traj_3d.pdf}}%
    \put(0.21510835,0.48723176){\makebox(0,0)[rt]{\lineheight{1.25}\smash{\begin{tabular}[t]{r}500\end{tabular}}}}%
    \put(0,0){\includegraphics[width=\unitlength,page=7]{traj_3d.pdf}}%
    \put(0.96664528,0.16389288){\makebox(0,0)[t]{\lineheight{1.25}\smash{\begin{tabular}[t]{c}0\end{tabular}}}}%
    \put(0,0){\includegraphics[width=\unitlength,page=8]{traj_3d.pdf}}%
    \put(0.27350853,0.09118153){\rotatebox{-44.9}{\makebox(0,0)[t]{\lineheight{1.25}\smash{\begin{tabular}[t]{c}$\textrm{y}_\textrm{W}$ (m)\end{tabular}}}}}%
    \put(0,0){\includegraphics[width=\unitlength,page=9]{traj_3d.pdf}}%
    \put(0.32736355,0.1330474){\makebox(0,0)[rt]{\lineheight{1.25}\smash{\begin{tabular}[t]{r}0\end{tabular}}}}%
    \put(0,0){\includegraphics[width=\unitlength,page=10]{traj_3d.pdf}}%
    \put(0.82993999,0.11862223){\makebox(0,0)[t]{\lineheight{1.25}\smash{\begin{tabular}[t]{c}200\end{tabular}}}}%
    \put(0,0){\includegraphics[width=\unitlength,page=11]{traj_3d.pdf}}%
    \put(0.82222925,0.0626023){\rotatebox{18}{\makebox(0,0)[t]{\lineheight{1.25}\smash{\begin{tabular}[t]{c}$\textrm{x}_\textrm{W}$ (m)\end{tabular}}}}}%
    \put(0,0){\includegraphics[width=\unitlength,page=12]{traj_3d.pdf}}%
    \put(0.6932347,0.07335158){\makebox(0,0)[t]{\lineheight{1.25}\smash{\begin{tabular}[t]{c}400\end{tabular}}}}%
    \put(0,0){\includegraphics[width=\unitlength,page=13]{traj_3d.pdf}}%
    \put(0.40629057,0.05463662){\makebox(0,0)[rt]{\lineheight{1.25}\smash{\begin{tabular}[t]{r}200\end{tabular}}}}%
    \put(0,0){\includegraphics[width=\unitlength,page=14]{traj_3d.pdf}}%
    \put(0.5565294,0.02808117){\makebox(0,0)[t]{\lineheight{1.25}\smash{\begin{tabular}[t]{c}600\end{tabular}}}}%
    \put(0,0){\includegraphics[width=\unitlength,page=15]{traj_3d.pdf}}%
    \put(0.85684627,0.42832029){\color[rgb]{0.14901961,0.14901961,0.14901961}\rotatebox{18}{\makebox(0,0)[t]{\lineheight{1.25}\smash{\begin{tabular}[t]{c}$\textrm{v}_\textrm{w}$ \end{tabular}}}}}%
    \put(0.66412298,0.31485406){\color[rgb]{0.22352941,0.41568627,0.69411765}\rotatebox{-6}{\makebox(0,0)[t]{\lineheight{1.25}\smash{\begin{tabular}[t]{c}Traction Phase\end{tabular}}}}}%
    \put(0.70398524,0.17663747){\color[rgb]{0.85490196,0.48627451,0.18823529}\rotatebox{6}{\makebox(0,0)[t]{\lineheight{1.25}\smash{\begin{tabular}[t]{c}Retraction Phase\end{tabular}}}}}%
  \end{picture}%
\endgroup%

%% file: path_rotated.pdf_tex
\begingroup%
  \makeatletter%
  \providecommand\color[2][]{%
    \errmessage{(Inkscape) Color is used for the text in Inkscape, but the package 'color.sty' is not loaded}%
    \renewcommand\color[2][]{}%
  }%
  \providecommand\transparent[1]{%
    \errmessage{(Inkscape) Transparency is used (non-zero) for the text in Inkscape, but the package 'transparent.sty' is not loaded}%
    \renewcommand\transparent[1]{}%
  }%
  \providecommand\rotatebox[2]{#2}%
  \newcommand*\fsize{\dimexpr\f@size pt\relax}%
  \newcommand*\lineheight[1]{\fontsize{\fsize}{#1\fsize}\selectfont}%
  \ifx\svgwidth\undefined%
    \setlength{\unitlength}{316.18359375bp}%
    \ifx\svgscale\undefined%
      \relax%
    \else%
      \setlength{\unitlength}{\unitlength * \real{\svgscale}}%
    \fi%
  \else%
    \setlength{\unitlength}{\svgwidth}%
  \fi%
  \global\let\svgwidth\undefined%
  \global\let\svgscale\undefined%
  \makeatother%
  \begin{picture}(1,0.93117654)%
    \lineheight{1}%
    \setlength\tabcolsep{0pt}%
    \put(0,0){\includegraphics[width=\unitlength,page=1]{path_rotated.pdf}}%
    \put(0.17384919,0.07080072){\makebox(0,0)[t]{\lineheight{1.25}\smash{\begin{tabular}[t]{c}0\end{tabular}}}}%
    \put(0.43793619,0.07080072){\makebox(0,0)[t]{\lineheight{1.25}\smash{\begin{tabular}[t]{c}200\end{tabular}}}}%
    \put(0.70202342,0.07080072){\makebox(0,0)[t]{\lineheight{1.25}\smash{\begin{tabular}[t]{c}400\end{tabular}}}}%
    \put(0.96611041,0.07080072){\makebox(0,0)[t]{\lineheight{1.25}\smash{\begin{tabular}[t]{c}600\end{tabular}}}}%
    \put(0.56998031,0.01126253){\makebox(0,0)[t]{\lineheight{1.25}\smash{\begin{tabular}[t]{c}$\mathrm{x}_\mr{W}$ (m)\end{tabular}}}}%
    \put(0.60467311,0.28155741){\color[rgb]{0,0,0}\makebox(0,0)[t]{\lineheight{1.25}\smash{\begin{tabular}[t]{c}$\phi_{set}$\end{tabular}}}}%
    \put(0.28061672,0.28650635){\color[rgb]{0,0,0}\makebox(0,0)[t]{\lineheight{1.25}\smash{\begin{tabular}[t]{c}$\phi_{0}$\end{tabular}}}}%
    \put(0.42840246,0.30499708){\color[rgb]{0,0,0}\makebox(0,0)[t]{\lineheight{1.25}\smash{\begin{tabular}[t]{c}$\phi_{r}$\end{tabular}}}}%
    \put(0,0){\includegraphics[width=\unitlength,page=2]{path_rotated.pdf}}%
    \put(0.15961695,0.10685572){\makebox(0,0)[rt]{\lineheight{1.25}\smash{\begin{tabular}[t]{r}0\end{tabular}}}}%
    \put(0.15961695,0.23889934){\makebox(0,0)[rt]{\lineheight{1.25}\smash{\begin{tabular}[t]{r}100\end{tabular}}}}%
    \put(0.15961695,0.37094272){\makebox(0,0)[rt]{\lineheight{1.25}\smash{\begin{tabular}[t]{r}200\end{tabular}}}}%
    \put(0.15961695,0.50298633){\makebox(0,0)[rt]{\lineheight{1.25}\smash{\begin{tabular}[t]{r}300\end{tabular}}}}%
    \put(0.15961695,0.63502995){\makebox(0,0)[rt]{\lineheight{1.25}\smash{\begin{tabular}[t]{r}400\end{tabular}}}}%
    \put(0.15961695,0.76707333){\makebox(0,0)[rt]{\lineheight{1.25}\smash{\begin{tabular}[t]{r}500\end{tabular}}}}%
    \put(0.15961695,0.89911694){\makebox(0,0)[rt]{\lineheight{1.25}\smash{\begin{tabular}[t]{r}600\end{tabular}}}}%
    \put(0.03745692,0.51532141){\rotatebox{90}{\makebox(0,0)[t]{\lineheight{1.25}\smash{\begin{tabular}[t]{c}$\mathrm{z}_\mr{W}$ (m)\end{tabular}}}}}%
    \put(0,0){\includegraphics[width=\unitlength,page=3]{path_rotated.pdf}}%
  \end{picture}%
\endgroup%

%% file: training_example.pdf_tex
\begingroup%
  \makeatletter%
  \providecommand\color[2][]{%
    \errmessage{(Inkscape) Color is used for the text in Inkscape, but the package 'color.sty' is not loaded}%
    \renewcommand\color[2][]{}%
  }%
  \providecommand\transparent[1]{%
    \errmessage{(Inkscape) Transparency is used (non-zero) for the text in Inkscape, but the package 'transparent.sty' is not loaded}%
    \renewcommand\transparent[1]{}%
  }%
  \providecommand\rotatebox[2]{#2}%
  \newcommand*\fsize{\dimexpr\f@size pt\relax}%
  \newcommand*\lineheight[1]{\fontsize{\fsize}{#1\fsize}\selectfont}%
  \ifx\svgwidth\undefined%
    \setlength{\unitlength}{374.30859375bp}%
    \ifx\svgscale\undefined%
      \relax%
    \else%
      \setlength{\unitlength}{\unitlength * \real{\svgscale}}%
    \fi%
  \else%
    \setlength{\unitlength}{\svgwidth}%
  \fi%
  \global\let\svgwidth\undefined%
  \global\let\svgscale\undefined%
  \makeatother%
  \begin{picture}(1,0.82810066)%
    \lineheight{1}%
    \setlength\tabcolsep{0pt}%
    \put(0,0){\includegraphics[width=\unitlength,page=1]{training_example.pdf}}%
    \put(0.12280846,0.05972029){\makebox(0,0)[t]{\lineheight{1.25}\smash{\begin{tabular}[t]{c}50\end{tabular}}}}%
    \put(0.2659057,0.05972029){\makebox(0,0)[t]{\lineheight{1.25}\smash{\begin{tabular}[t]{c}55\end{tabular}}}}%
    \put(0.40900273,0.05972029){\makebox(0,0)[t]{\lineheight{1.25}\smash{\begin{tabular}[t]{c}60\end{tabular}}}}%
    \put(0.55209997,0.05972029){\makebox(0,0)[t]{\lineheight{1.25}\smash{\begin{tabular}[t]{c}65\end{tabular}}}}%
    \put(0.6951972,0.05972029){\makebox(0,0)[t]{\lineheight{1.25}\smash{\begin{tabular}[t]{c}70\end{tabular}}}}%
    \put(0.83829424,0.05972029){\makebox(0,0)[t]{\lineheight{1.25}\smash{\begin{tabular}[t]{c}75\end{tabular}}}}%
    \put(0.98139147,0.05972029){\makebox(0,0)[t]{\lineheight{1.25}\smash{\begin{tabular}[t]{c}80\end{tabular}}}}%
    \put(0.55210037,0.00942756){\makebox(0,0)[t]{\lineheight{1.25}\smash{\begin{tabular}[t]{c}Time (s)\end{tabular}}}}%
    \put(0,0){\includegraphics[width=\unitlength,page=2]{training_example.pdf}}%
    \put(0.11078629,0.09960863){\makebox(0,0)[rt]{\lineheight{1.25}\smash{\begin{tabular}[t]{r}1\end{tabular}}}}%
    \put(0.11078629,0.31954254){\makebox(0,0)[rt]{\lineheight{1.25}\smash{\begin{tabular}[t]{r}1.5\end{tabular}}}}%
    \put(0.11078629,0.53947644){\makebox(0,0)[rt]{\lineheight{1.25}\smash{\begin{tabular}[t]{r}2\end{tabular}}}}%
    \put(0.11078629,0.75941034){\makebox(0,0)[rt]{\lineheight{1.25}\smash{\begin{tabular}[t]{r}2.5\end{tabular}}}}%
    \put(0.03164037,0.435213){\rotatebox{90}{\makebox(0,0)[t]{\lineheight{1.25}\smash{\begin{tabular}[t]{c} z (-) \end{tabular}}}}}%
    \put(0,0){\includegraphics[width=\unitlength,page=3]{training_example.pdf}}%
    \put(0.60475791,0.5777495){\color[rgb]{0,0,0}\makebox(0,0)[lt]{\lineheight{1.25}\smash{\begin{tabular}[t]{l}$\Delta \mathrm{T}_\mathrm{r}$\end{tabular}}}}%
    \put(0,0){\includegraphics[width=\unitlength,page=4]{training_example.pdf}}%
    \put(0.47247914,0.57810275){\color[rgb]{0,0,0}\makebox(0,0)[lt]{\lineheight{1.25}\smash{\begin{tabular}[t]{l}$\Delta \mathrm{T}_\mathrm{s}$\end{tabular}}}}%
    \put(0,0){\includegraphics[width=\unitlength,page=5]{training_example.pdf}}%
    \put(0.47450726,0.79878316){\color[rgb]{0,0,0}\makebox(0,0)[lt]{\lineheight{1.25}\smash{\begin{tabular}[t]{l}$\mathrm{s}_\mathrm{1}$\end{tabular}}}}%
    \put(0.28613272,0.82212403){\color[rgb]{0,0,0}\makebox(0,0)[lt]{\lineheight{1.25}\smash{\begin{tabular}[t]{l}$\mathrm{s}_\mathrm{2}$\end{tabular}}}}%
  \end{picture}%
\endgroup%

%% file: feature_selection_conv.pdf_tex
\begingroup%
  \makeatletter%
  \providecommand\color[2][]{%
    \errmessage{(Inkscape) Color is used for the text in Inkscape, but the package 'color.sty' is not loaded}%
    \renewcommand\color[2][]{}%
  }%
  \providecommand\transparent[1]{%
    \errmessage{(Inkscape) Transparency is used (non-zero) for the text in Inkscape, but the package 'transparent.sty' is not loaded}%
    \renewcommand\transparent[1]{}%
  }%
  \providecommand\rotatebox[2]{#2}%
  \newcommand*\fsize{\dimexpr\f@size pt\relax}%
  \newcommand*\lineheight[1]{\fontsize{\fsize}{#1\fsize}\selectfont}%
  \ifx\svgwidth\undefined%
    \setlength{\unitlength}{225bp}%
    \ifx\svgscale\undefined%
      \relax%
    \else%
      \setlength{\unitlength}{\unitlength * \real{\svgscale}}%
    \fi%
  \else%
    \setlength{\unitlength}{\svgwidth}%
  \fi%
  \global\let\svgwidth\undefined%
  \global\let\svgscale\undefined%
  \makeatother%
  \begin{picture}(1,0.66666667)%
    \lineheight{1}%
    \setlength\tabcolsep{0pt}%
    \put(0,0){\includegraphics[width=\unitlength,page=1]{feature_selection_conv.pdf}}%
    \put(0.17333333,0.07622233){\makebox(0,0)[t]{\lineheight{1.25}\smash{\begin{tabular}[t]{c}1\end{tabular}}}}%
    \put(0.265,0.07622233){\makebox(0,0)[t]{\lineheight{1.25}\smash{\begin{tabular}[t]{c}2\end{tabular}}}}%
    \put(0.35666667,0.07622233){\makebox(0,0)[t]{\lineheight{1.25}\smash{\begin{tabular}[t]{c}3\end{tabular}}}}%
    \put(0.44833333,0.07622233){\makebox(0,0)[t]{\lineheight{1.25}\smash{\begin{tabular}[t]{c}4\end{tabular}}}}%
    \put(0.54,0.07622233){\makebox(0,0)[t]{\lineheight{1.25}\smash{\begin{tabular}[t]{c}5\end{tabular}}}}%
    \put(0.63166667,0.07622233){\makebox(0,0)[t]{\lineheight{1.25}\smash{\begin{tabular}[t]{c}6\end{tabular}}}}%
    \put(0.72333333,0.07622233){\makebox(0,0)[t]{\lineheight{1.25}\smash{\begin{tabular}[t]{c}7\end{tabular}}}}%
    \put(0.815,0.07622233){\makebox(0,0)[t]{\lineheight{1.25}\smash{\begin{tabular}[t]{c}8\end{tabular}}}}%
    \put(0.90666667,0.07622233){\makebox(0,0)[t]{\lineheight{1.25}\smash{\begin{tabular}[t]{c}9\end{tabular}}}}%
    \put(0.54000033,0.01933367){\makebox(0,0)[t]{\lineheight{1.25}\smash{\begin{tabular}[t]{c}Number of features\end{tabular}}}}%
    \put(0,0){\includegraphics[width=\unitlength,page=2]{feature_selection_conv.pdf}}%
    \put(0.160889,0.10999967){\makebox(0,0)[rt]{\lineheight{1.25}\smash{\begin{tabular}[t]{r}0.96\end{tabular}}}}%
    \put(0.160889,0.23333367){\makebox(0,0)[rt]{\lineheight{1.25}\smash{\begin{tabular}[t]{r}0.97\end{tabular}}}}%
    \put(0.160889,0.356667){\makebox(0,0)[rt]{\lineheight{1.25}\smash{\begin{tabular}[t]{r}0.98\end{tabular}}}}%
    \put(0.160889,0.48){\makebox(0,0)[rt]{\lineheight{1.25}\smash{\begin{tabular}[t]{r}0.99\end{tabular}}}}%
    \put(0.160889,0.60333333){\makebox(0,0)[rt]{\lineheight{1.25}\smash{\begin{tabular}[t]{r}1\end{tabular}}}}%
    \put(0.04866667,0.37000033){\rotatebox{90}{\makebox(0,0)[t]{\lineheight{1.25}\smash{\begin{tabular}[t]{c}MCC\end{tabular}}}}}%
    \put(0,0){\includegraphics[width=\unitlength,page=3]{feature_selection_conv.pdf}}%
  \end{picture}%
\endgroup%

%% file: power_gain_over_reliab_loss.pdf_tex
\begingroup%
  \makeatletter%
  \providecommand\color[2][]{%
    \errmessage{(Inkscape) Color is used for the text in Inkscape, but the package 'color.sty' is not loaded}%
    \renewcommand\color[2][]{}%
  }%
  \providecommand\transparent[1]{%
    \errmessage{(Inkscape) Transparency is used (non-zero) for the text in Inkscape, but the package 'transparent.sty' is not loaded}%
    \renewcommand\transparent[1]{}%
  }%
  \providecommand\rotatebox[2]{#2}%
  \newcommand*\fsize{\dimexpr\f@size pt\relax}%
  \newcommand*\lineheight[1]{\fontsize{\fsize}{#1\fsize}\selectfont}%
  \ifx\svgwidth\undefined%
    \setlength{\unitlength}{225bp}%
    \ifx\svgscale\undefined%
      \relax%
    \else%
      \setlength{\unitlength}{\unitlength * \real{\svgscale}}%
    \fi%
  \else%
    \setlength{\unitlength}{\svgwidth}%
  \fi%
  \global\let\svgwidth\undefined%
  \global\let\svgscale\undefined%
  \makeatother%
  \begin{picture}(1,0.66666667)%
    \lineheight{1}%
    \setlength\tabcolsep{0pt}%
    \put(0,0){\includegraphics[width=\unitlength,page=1]{power_gain_over_reliab_loss.pdf}}%
    \put(0.28224367,0.07622233){\makebox(0,0)[t]{\lineheight{1.25}\smash{\begin{tabular}[t]{c}$10^{-6}$\end{tabular}}}}%
    \put(0.594455,0.07622233){\makebox(0,0)[t]{\lineheight{1.25}\smash{\begin{tabular}[t]{c}$10^{-4}$\end{tabular}}}}%
    \put(0.90666667,0.07622233){\makebox(0,0)[t]{\lineheight{1.25}\smash{\begin{tabular}[t]{c}$10^{-2}$\end{tabular}}}}%
    \put(0.54000033,0.018){\makebox(0,0)[t]{\lineheight{1.25}\smash{\begin{tabular}[t]{c}$\mathrm{p}_\mathrm{f}$ (-)\end{tabular}}}}%
    \put(0,0){\includegraphics[width=\unitlength,page=2]{power_gain_over_reliab_loss.pdf}}%
    \put(0.160889,0.11){\makebox(0,0)[rt]{\lineheight{1.25}\smash{\begin{tabular}[t]{r}0\end{tabular}}}}%
    \put(0.160889,0.23333333){\makebox(0,0)[rt]{\lineheight{1.25}\smash{\begin{tabular}[t]{r}10\end{tabular}}}}%
    \put(0.160889,0.35666667){\makebox(0,0)[rt]{\lineheight{1.25}\smash{\begin{tabular}[t]{r}20\end{tabular}}}}%
    \put(0.160889,0.48){\makebox(0,0)[rt]{\lineheight{1.25}\smash{\begin{tabular}[t]{r}30\end{tabular}}}}%
    \put(0.160889,0.60333333){\makebox(0,0)[rt]{\lineheight{1.25}\smash{\begin{tabular}[t]{r}40\end{tabular}}}}%
    \put(0.04733333,0.37000033){\rotatebox{90}{\makebox(0,0)[t]{\lineheight{1.25}\smash{\begin{tabular}[t]{c}Relative power gain (\%)\end{tabular}}}}}%
    \put(0,0){\includegraphics[width=\unitlength,page=3]{power_gain_over_reliab_loss.pdf}}%
    \put(0.185694,0.17033333){\makebox(0,0)[lt]{\lineheight{1.25}\smash{\begin{tabular}[t]{l}$\omega_\mathrm{0,r}$\end{tabular}}}}%
    \put(0.603231,0.49786967){\makebox(0,0)[lt]{\lineheight{1.25}\smash{\begin{tabular}[t]{l}$1.5\omega_\mathrm{0,r}$\end{tabular}}}}%
    \put(0.76676533,0.56671967){\makebox(0,0)[lt]{\lineheight{1.25}\smash{\begin{tabular}[t]{l}$2\omega_\mathrm{0,r}$\end{tabular}}}}%
  \end{picture}%
\endgroup%

%% file: path_3d_rupture_no_rupture.pdf_tex
\begingroup%
  \makeatletter%
  \providecommand\color[2][]{%
    \errmessage{(Inkscape) Color is used for the text in Inkscape, but the package 'color.sty' is not loaded}%
    \renewcommand\color[2][]{}%
  }%
  \providecommand\transparent[1]{%
    \errmessage{(Inkscape) Transparency is used (non-zero) for the text in Inkscape, but the package 'transparent.sty' is not loaded}%
    \renewcommand\transparent[1]{}%
  }%
  \providecommand\rotatebox[2]{#2}%
  \newcommand*\fsize{\dimexpr\f@size pt\relax}%
  \newcommand*\lineheight[1]{\fontsize{\fsize}{#1\fsize}\selectfont}%
  \ifx\svgwidth\undefined%
    \setlength{\unitlength}{225bp}%
    \ifx\svgscale\undefined%
      \relax%
    \else%
      \setlength{\unitlength}{\unitlength * \real{\svgscale}}%
    \fi%
  \else%
    \setlength{\unitlength}{\svgwidth}%
  \fi%
  \global\let\svgwidth\undefined%
  \global\let\svgscale\undefined%
  \makeatother%
  \begin{picture}(1,0.66666667)%
    \lineheight{1}%
    \setlength\tabcolsep{0pt}%
    \put(0,0){\includegraphics[width=\unitlength,page=1]{path_3d_rupture_no_rupture.pdf}}%
    \put(0.77836867,0.21454733){\makebox(0,0)[t]{\lineheight{1.25}\smash{\begin{tabular}[t]{c}0\end{tabular}}}}%
    \put(0,0){\includegraphics[width=\unitlength,page=2]{path_3d_rupture_no_rupture.pdf}}%
    \put(0.21804533,0.277422){\makebox(0,0)[rt]{\lineheight{1.25}\smash{\begin{tabular}[t]{r}200\end{tabular}}}}%
    \put(0,0){\includegraphics[width=\unitlength,page=3]{path_3d_rupture_no_rupture.pdf}}%
    \put(0.258298,0.16279933){\makebox(0,0)[rt]{\lineheight{1.25}\smash{\begin{tabular}[t]{r}-200\end{tabular}}}}%
    \put(0.697863,0.16279933){\makebox(0,0)[t]{\lineheight{1.25}\smash{\begin{tabular}[t]{c}200\end{tabular}}}}%
    \put(0,0){\includegraphics[width=\unitlength,page=4]{path_3d_rupture_no_rupture.pdf}}%
    \put(0.712823,0.031511){\makebox(0,0)[t]{\lineheight{1.25}\smash{\begin{tabular}[t]{c}$\mathrm{x}_\mathrm{W}$ (m)\end{tabular}}}}%
    \put(0,0){\includegraphics[width=\unitlength,page=5]{path_3d_rupture_no_rupture.pdf}}%
    \put(0.100994,0.32245933){\rotatebox{90}{\makebox(0,0)[t]{\lineheight{1.25}\smash{\begin{tabular}[t]{c}$\mathrm{z}_\mathrm{W}$ (m)\end{tabular}}}}}%
    \put(0,0){\includegraphics[width=\unitlength,page=6]{path_3d_rupture_no_rupture.pdf}}%
    \put(0.26346467,0.018574){\makebox(0,0)[t]{\lineheight{1.25}\smash{\begin{tabular}[t]{c}$\mathrm{y}_\mathrm{W}$ (m)\end{tabular}}}}%
    \put(0,0){\includegraphics[width=\unitlength,page=7]{path_3d_rupture_no_rupture.pdf}}%
    \put(0.21804533,0.36463767){\makebox(0,0)[rt]{\lineheight{1.25}\smash{\begin{tabular}[t]{r}400\end{tabular}}}}%
    \put(0,0){\includegraphics[width=\unitlength,page=8]{path_3d_rupture_no_rupture.pdf}}%
    \put(0.33880367,0.11105167){\makebox(0,0)[rt]{\lineheight{1.25}\smash{\begin{tabular}[t]{r}0\end{tabular}}}}%
    \put(0.61735767,0.11105167){\makebox(0,0)[t]{\lineheight{1.25}\smash{\begin{tabular}[t]{c}400\end{tabular}}}}%
    \put(0,0){\includegraphics[width=\unitlength,page=9]{path_3d_rupture_no_rupture.pdf}}%
    \put(0.53685233,0.05930367){\makebox(0,0)[t]{\lineheight{1.25}\smash{\begin{tabular}[t]{c}600\end{tabular}}}}%
    \put(0.419309,0.05930367){\makebox(0,0)[rt]{\lineheight{1.25}\smash{\begin{tabular}[t]{r}200\end{tabular}}}}%
    \put(0,0){\includegraphics[width=\unitlength,page=10]{path_3d_rupture_no_rupture.pdf}}%
  \end{picture}%
\endgroup%

%% file: path_xy_rupture_no_rupture.pdf_tex
\begingroup%
  \makeatletter%
  \providecommand\color[2][]{%
    \errmessage{(Inkscape) Color is used for the text in Inkscape, but the package 'color.sty' is not loaded}%
    \renewcommand\color[2][]{}%
  }%
  \providecommand\transparent[1]{%
    \errmessage{(Inkscape) Transparency is used (non-zero) for the text in Inkscape, but the package 'transparent.sty' is not loaded}%
    \renewcommand\transparent[1]{}%
  }%
  \providecommand\rotatebox[2]{#2}%
  \newcommand*\fsize{\dimexpr\f@size pt\relax}%
  \newcommand*\lineheight[1]{\fontsize{\fsize}{#1\fsize}\selectfont}%
  \ifx\svgwidth\undefined%
    \setlength{\unitlength}{225bp}%
    \ifx\svgscale\undefined%
      \relax%
    \else%
      \setlength{\unitlength}{\unitlength * \real{\svgscale}}%
    \fi%
  \else%
    \setlength{\unitlength}{\svgwidth}%
  \fi%
  \global\let\svgwidth\undefined%
  \global\let\svgscale\undefined%
  \makeatother%
  \begin{picture}(1,0.66666667)%
    \lineheight{1}%
    \setlength\tabcolsep{0pt}%
    \put(0,0){\includegraphics[width=\unitlength,page=1]{path_xy_rupture_no_rupture.pdf}}%
    \put(0.25481467,0.07344433){\makebox(0,0)[t]{\lineheight{1.25}\smash{\begin{tabular}[t]{c}0\end{tabular}}}}%
    \put(0.41777767,0.07344433){\makebox(0,0)[t]{\lineheight{1.25}\smash{\begin{tabular}[t]{c}200\end{tabular}}}}%
    \put(0.58074067,0.07344433){\makebox(0,0)[t]{\lineheight{1.25}\smash{\begin{tabular}[t]{c}400\end{tabular}}}}%
    \put(0.74370367,0.07344433){\makebox(0,0)[t]{\lineheight{1.25}\smash{\begin{tabular}[t]{c}600\end{tabular}}}}%
    \put(0.90666667,0.07344433){\makebox(0,0)[t]{\lineheight{1.25}\smash{\begin{tabular}[t]{c}800\end{tabular}}}}%
    \put(0.54000033,0.01522233){\makebox(0,0)[t]{\lineheight{1.25}\smash{\begin{tabular}[t]{c}$\mathrm{x}_\mathrm{W}$ (m)\end{tabular}}}}%
    \put(0,0){\includegraphics[width=\unitlength,page=2]{path_xy_rupture_no_rupture.pdf}}%
    \put(0.160889,0.10722233){\makebox(0,0)[rt]{\lineheight{1.25}\smash{\begin{tabular}[t]{r}-300\end{tabular}}}}%
    \put(0.160889,0.18870367){\makebox(0,0)[rt]{\lineheight{1.25}\smash{\begin{tabular}[t]{r}-200\end{tabular}}}}%
    \put(0.160889,0.27018533){\makebox(0,0)[rt]{\lineheight{1.25}\smash{\begin{tabular}[t]{r}-100\end{tabular}}}}%
    \put(0.160889,0.35166667){\makebox(0,0)[rt]{\lineheight{1.25}\smash{\begin{tabular}[t]{r}0\end{tabular}}}}%
    \put(0.160889,0.433148){\makebox(0,0)[rt]{\lineheight{1.25}\smash{\begin{tabular}[t]{r}100\end{tabular}}}}%
    \put(0.160889,0.51462967){\makebox(0,0)[rt]{\lineheight{1.25}\smash{\begin{tabular}[t]{r}200\end{tabular}}}}%
    \put(0.160889,0.596111){\makebox(0,0)[rt]{\lineheight{1.25}\smash{\begin{tabular}[t]{r}300\end{tabular}}}}%
    \put(0.04733333,0.36500033){\rotatebox{90}{\makebox(0,0)[t]{\lineheight{1.25}\smash{\begin{tabular}[t]{c}$\mathrm{y}_\mathrm{W}$ (m)\end{tabular}}}}}%
    \put(0,0){\includegraphics[width=\unitlength,page=3]{path_xy_rupture_no_rupture.pdf}}%
  \end{picture}%
\endgroup%

%% file: tether_rupture_zoom.pdf_tex
\begingroup%
  \makeatletter%
  \providecommand\color[2][]{%
    \errmessage{(Inkscape) Color is used for the text in Inkscape, but the package 'color.sty' is not loaded}%
    \renewcommand\color[2][]{}%
  }%
  \providecommand\transparent[1]{%
    \errmessage{(Inkscape) Transparency is used (non-zero) for the text in Inkscape, but the package 'transparent.sty' is not loaded}%
    \renewcommand\transparent[1]{}%
  }%
  \providecommand\rotatebox[2]{#2}%
  \newcommand*\fsize{\dimexpr\f@size pt\relax}%
  \newcommand*\lineheight[1]{\fontsize{\fsize}{#1\fsize}\selectfont}%
  \ifx\svgwidth\undefined%
    \setlength{\unitlength}{225bp}%
    \ifx\svgscale\undefined%
      \relax%
    \else%
      \setlength{\unitlength}{\unitlength * \real{\svgscale}}%
    \fi%
  \else%
    \setlength{\unitlength}{\svgwidth}%
  \fi%
  \global\let\svgwidth\undefined%
  \global\let\svgscale\undefined%
  \makeatother%
  \begin{picture}(1,0.66666667)%
    \lineheight{1}%
    \setlength\tabcolsep{0pt}%
    \put(0,0){\includegraphics[width=\unitlength,page=1]{tether_rupture_zoom.pdf}}%
    \put(0.17333333,0.07622233){\makebox(0,0)[t]{\lineheight{1.25}\smash{\begin{tabular}[t]{c}65\end{tabular}}}}%
    \put(0.54,0.07622233){\makebox(0,0)[t]{\lineheight{1.25}\smash{\begin{tabular}[t]{c}70\end{tabular}}}}%
    \put(0.90666667,0.07622233){\makebox(0,0)[t]{\lineheight{1.25}\smash{\begin{tabular}[t]{c}75\end{tabular}}}}%
    \put(0.54000067,0.01933333){\makebox(0,0)[t]{\lineheight{1.25}\smash{\begin{tabular}[t]{c}Time (s)\end{tabular}}}}%
    \put(0,0){\includegraphics[width=\unitlength,page=2]{tether_rupture_zoom.pdf}}%
    \put(0.16088867,0.11){\makebox(0,0)[rt]{\lineheight{1.25}\smash{\begin{tabular}[t]{r}0\end{tabular}}}}%
    \put(0.16088867,0.23333333){\makebox(0,0)[rt]{\lineheight{1.25}\smash{\begin{tabular}[t]{r}0.5\end{tabular}}}}%
    \put(0.16088867,0.35666667){\makebox(0,0)[rt]{\lineheight{1.25}\smash{\begin{tabular}[t]{r}1\end{tabular}}}}%
    \put(0.16088867,0.48){\makebox(0,0)[rt]{\lineheight{1.25}\smash{\begin{tabular}[t]{r}1.5\end{tabular}}}}%
    \put(0.16088867,0.60333333){\makebox(0,0)[rt]{\lineheight{1.25}\smash{\begin{tabular}[t]{r}2\end{tabular}}}}%
    \put(0.04866667,0.37000033){\rotatebox{90}{\makebox(0,0)[t]{\lineheight{1.25}\smash{\begin{tabular}[t]{c}$\mathrm{F}_\mathrm{t}$ (kN)\end{tabular}}}}}%
    \put(0,0){\includegraphics[width=\unitlength,page=3]{tether_rupture_zoom.pdf}}%
  \end{picture}%
\endgroup%

%% file: mu_a_ref_wpred.pdf_tex
\begingroup%
  \makeatletter%
  \providecommand\color[2][]{%
    \errmessage{(Inkscape) Color is used for the text in Inkscape, but the package 'color.sty' is not loaded}%
    \renewcommand\color[2][]{}%
  }%
  \providecommand\transparent[1]{%
    \errmessage{(Inkscape) Transparency is used (non-zero) for the text in Inkscape, but the package 'transparent.sty' is not loaded}%
    \renewcommand\transparent[1]{}%
  }%
  \providecommand\rotatebox[2]{#2}%
  \newcommand*\fsize{\dimexpr\f@size pt\relax}%
  \newcommand*\lineheight[1]{\fontsize{\fsize}{#1\fsize}\selectfont}%
  \ifx\svgwidth\undefined%
    \setlength{\unitlength}{225bp}%
    \ifx\svgscale\undefined%
      \relax%
    \else%
      \setlength{\unitlength}{\unitlength * \real{\svgscale}}%
    \fi%
  \else%
    \setlength{\unitlength}{\svgwidth}%
  \fi%
  \global\let\svgwidth\undefined%
  \global\let\svgscale\undefined%
  \makeatother%
  \begin{picture}(1,0.66666667)%
    \lineheight{1}%
    \setlength\tabcolsep{0pt}%
    \put(0,0){\includegraphics[width=\unitlength,page=1]{mu_a_ref_wpred.pdf}}%
    \put(0.17333333,0.07622233){\makebox(0,0)[t]{\lineheight{1.25}\smash{\begin{tabular}[t]{c}65\end{tabular}}}}%
    \put(0.54,0.07622233){\makebox(0,0)[t]{\lineheight{1.25}\smash{\begin{tabular}[t]{c}70\end{tabular}}}}%
    \put(0.90666667,0.07622233){\makebox(0,0)[t]{\lineheight{1.25}\smash{\begin{tabular}[t]{c}75\end{tabular}}}}%
    \put(0.54000067,0.01933333){\makebox(0,0)[t]{\lineheight{1.25}\smash{\begin{tabular}[t]{c}Time (s)\end{tabular}}}}%
    \put(0,0){\includegraphics[width=\unitlength,page=2]{mu_a_ref_wpred.pdf}}%
    \put(0.16088867,0.11){\makebox(0,0)[rt]{\lineheight{1.25}\smash{\begin{tabular}[t]{r}-50\end{tabular}}}}%
    \put(0.16088867,0.23333333){\makebox(0,0)[rt]{\lineheight{1.25}\smash{\begin{tabular}[t]{r}-25\end{tabular}}}}%
    \put(0.16088867,0.35666667){\makebox(0,0)[rt]{\lineheight{1.25}\smash{\begin{tabular}[t]{r}0\end{tabular}}}}%
    \put(0.16088867,0.48){\makebox(0,0)[rt]{\lineheight{1.25}\smash{\begin{tabular}[t]{r}25\end{tabular}}}}%
    \put(0.16088867,0.60333333){\makebox(0,0)[rt]{\lineheight{1.25}\smash{\begin{tabular}[t]{r}50\end{tabular}}}}%
    \put(0.04866667,0.37000033){\rotatebox{90}{\makebox(0,0)[t]{\lineheight{1.25}\smash{\begin{tabular}[t]{c}$\mu_\mathrm{a}$ (deg)\end{tabular}}}}}%
    \put(0,0){\includegraphics[width=\unitlength,page=3]{mu_a_ref_wpred.pdf}}%
  \end{picture}%
\endgroup%

%% file: alpha_a_ref_wpred.pdf_tex
\begingroup%
  \makeatletter%
  \providecommand\color[2][]{%
    \errmessage{(Inkscape) Color is used for the text in Inkscape, but the package 'color.sty' is not loaded}%
    \renewcommand\color[2][]{}%
  }%
  \providecommand\transparent[1]{%
    \errmessage{(Inkscape) Transparency is used (non-zero) for the text in Inkscape, but the package 'transparent.sty' is not loaded}%
    \renewcommand\transparent[1]{}%
  }%
  \providecommand\rotatebox[2]{#2}%
  \newcommand*\fsize{\dimexpr\f@size pt\relax}%
  \newcommand*\lineheight[1]{\fontsize{\fsize}{#1\fsize}\selectfont}%
  \ifx\svgwidth\undefined%
    \setlength{\unitlength}{225bp}%
    \ifx\svgscale\undefined%
      \relax%
    \else%
      \setlength{\unitlength}{\unitlength * \real{\svgscale}}%
    \fi%
  \else%
    \setlength{\unitlength}{\svgwidth}%
  \fi%
  \global\let\svgwidth\undefined%
  \global\let\svgscale\undefined%
  \makeatother%
  \begin{picture}(1,0.66666667)%
    \lineheight{1}%
    \setlength\tabcolsep{0pt}%
    \put(0,0){\includegraphics[width=\unitlength,page=1]{alpha_a_ref_wpred.pdf}}%
    \put(0.17333333,0.07622233){\makebox(0,0)[t]{\lineheight{1.25}\smash{\begin{tabular}[t]{c}65\end{tabular}}}}%
    \put(0.54,0.07622233){\makebox(0,0)[t]{\lineheight{1.25}\smash{\begin{tabular}[t]{c}70\end{tabular}}}}%
    \put(0.90666667,0.07622233){\makebox(0,0)[t]{\lineheight{1.25}\smash{\begin{tabular}[t]{c}75\end{tabular}}}}%
    \put(0.54000067,0.01933333){\makebox(0,0)[t]{\lineheight{1.25}\smash{\begin{tabular}[t]{c}Time (s)\end{tabular}}}}%
    \put(0,0){\includegraphics[width=\unitlength,page=2]{alpha_a_ref_wpred.pdf}}%
    \put(0.16088867,0.13740733){\makebox(0,0)[rt]{\lineheight{1.25}\smash{\begin{tabular}[t]{r}-5\end{tabular}}}}%
    \put(0.16088867,0.27444433){\makebox(0,0)[rt]{\lineheight{1.25}\smash{\begin{tabular}[t]{r}0\end{tabular}}}}%
    \put(0.16088867,0.41148133){\makebox(0,0)[rt]{\lineheight{1.25}\smash{\begin{tabular}[t]{r}5\end{tabular}}}}%
    \put(0.16088867,0.54851867){\makebox(0,0)[rt]{\lineheight{1.25}\smash{\begin{tabular}[t]{r}10\end{tabular}}}}%
    \put(0.04866667,0.37000033){\rotatebox{90}{\makebox(0,0)[t]{\lineheight{1.25}\smash{\begin{tabular}[t]{c}$\alpha$ (deg)\end{tabular}}}}}%
    \put(0,0){\includegraphics[width=\unitlength,page=3]{alpha_a_ref_wpred.pdf}}%
  \end{picture}%
\endgroup%

%% file: phi_tau_upset.pdf_tex
\begingroup%
  \makeatletter%
  \providecommand\color[2][]{%
    \errmessage{(Inkscape) Color is used for the text in Inkscape, but the package 'color.sty' is not loaded}%
    \renewcommand\color[2][]{}%
  }%
  \providecommand\transparent[1]{%
    \errmessage{(Inkscape) Transparency is used (non-zero) for the text in Inkscape, but the package 'transparent.sty' is not loaded}%
    \renewcommand\transparent[1]{}%
  }%
  \providecommand\rotatebox[2]{#2}%
  \newcommand*\fsize{\dimexpr\f@size pt\relax}%
  \newcommand*\lineheight[1]{\fontsize{\fsize}{#1\fsize}\selectfont}%
  \ifx\svgwidth\undefined%
    \setlength{\unitlength}{225bp}%
    \ifx\svgscale\undefined%
      \relax%
    \else%
      \setlength{\unitlength}{\unitlength * \real{\svgscale}}%
    \fi%
  \else%
    \setlength{\unitlength}{\svgwidth}%
  \fi%
  \global\let\svgwidth\undefined%
  \global\let\svgscale\undefined%
  \makeatother%
  \begin{picture}(1,0.66666667)%
    \lineheight{1}%
    \setlength\tabcolsep{0pt}%
    \put(0,0){\includegraphics[width=\unitlength,page=1]{phi_tau_upset.pdf}}%
    \put(0.17333333,0.07622233){\makebox(0,0)[t]{\lineheight{1.25}\smash{\begin{tabular}[t]{c}65\end{tabular}}}}%
    \put(0.54,0.07622233){\makebox(0,0)[t]{\lineheight{1.25}\smash{\begin{tabular}[t]{c}70\end{tabular}}}}%
    \put(0.90666667,0.07622233){\makebox(0,0)[t]{\lineheight{1.25}\smash{\begin{tabular}[t]{c}75\end{tabular}}}}%
    \put(0.54000067,0.01933333){\makebox(0,0)[t]{\lineheight{1.25}\smash{\begin{tabular}[t]{c}Time (s)\end{tabular}}}}%
    \put(0,0){\includegraphics[width=\unitlength,page=2]{phi_tau_upset.pdf}}%
    \put(0.16088867,0.11){\makebox(0,0)[rt]{\lineheight{1.25}\smash{\begin{tabular}[t]{r}-60\end{tabular}}}}%
    \put(0.16088867,0.23333333){\makebox(0,0)[rt]{\lineheight{1.25}\smash{\begin{tabular}[t]{r}-30\end{tabular}}}}%
    \put(0.16088867,0.35666667){\makebox(0,0)[rt]{\lineheight{1.25}\smash{\begin{tabular}[t]{r}0\end{tabular}}}}%
    \put(0.16088867,0.48){\makebox(0,0)[rt]{\lineheight{1.25}\smash{\begin{tabular}[t]{r}30\end{tabular}}}}%
    \put(0.16088867,0.60333333){\makebox(0,0)[rt]{\lineheight{1.25}\smash{\begin{tabular}[t]{r}60\end{tabular}}}}%
    \put(0.04866667,0.37000033){\rotatebox{90}{\makebox(0,0)[t]{\lineheight{1.25}\smash{\begin{tabular}[t]{c}$\Phi_\mathrm{\tau}$ (deg)\end{tabular}}}}}%
    \put(0,0){\includegraphics[width=\unitlength,page=3]{phi_tau_upset.pdf}}%
  \end{picture}%
\endgroup%

%% file: theta_tau_upset.pdf_tex
\begingroup%
  \makeatletter%
  \providecommand\color[2][]{%
    \errmessage{(Inkscape) Color is used for the text in Inkscape, but the package 'color.sty' is not loaded}%
    \renewcommand\color[2][]{}%
  }%
  \providecommand\transparent[1]{%
    \errmessage{(Inkscape) Transparency is used (non-zero) for the text in Inkscape, but the package 'transparent.sty' is not loaded}%
    \renewcommand\transparent[1]{}%
  }%
  \providecommand\rotatebox[2]{#2}%
  \newcommand*\fsize{\dimexpr\f@size pt\relax}%
  \newcommand*\lineheight[1]{\fontsize{\fsize}{#1\fsize}\selectfont}%
  \ifx\svgwidth\undefined%
    \setlength{\unitlength}{225bp}%
    \ifx\svgscale\undefined%
      \relax%
    \else%
      \setlength{\unitlength}{\unitlength * \real{\svgscale}}%
    \fi%
  \else%
    \setlength{\unitlength}{\svgwidth}%
  \fi%
  \global\let\svgwidth\undefined%
  \global\let\svgscale\undefined%
  \makeatother%
  \begin{picture}(1,0.66666667)%
    \lineheight{1}%
    \setlength\tabcolsep{0pt}%
    \put(0,0){\includegraphics[width=\unitlength,page=1]{theta_tau_upset.pdf}}%
    \put(0.17333333,0.07622233){\makebox(0,0)[t]{\lineheight{1.25}\smash{\begin{tabular}[t]{c}65\end{tabular}}}}%
    \put(0.54,0.07622233){\makebox(0,0)[t]{\lineheight{1.25}\smash{\begin{tabular}[t]{c}70\end{tabular}}}}%
    \put(0.90666667,0.07622233){\makebox(0,0)[t]{\lineheight{1.25}\smash{\begin{tabular}[t]{c}75\end{tabular}}}}%
    \put(0.54000067,0.01933333){\makebox(0,0)[t]{\lineheight{1.25}\smash{\begin{tabular}[t]{c}Time (s)\end{tabular}}}}%
    \put(0,0){\includegraphics[width=\unitlength,page=2]{theta_tau_upset.pdf}}%
    \put(0.16088867,0.11){\makebox(0,0)[rt]{\lineheight{1.25}\smash{\begin{tabular}[t]{r}-40\end{tabular}}}}%
    \put(0.16088867,0.23333333){\makebox(0,0)[rt]{\lineheight{1.25}\smash{\begin{tabular}[t]{r}-30\end{tabular}}}}%
    \put(0.16088867,0.35666667){\makebox(0,0)[rt]{\lineheight{1.25}\smash{\begin{tabular}[t]{r}-20\end{tabular}}}}%
    \put(0.16088867,0.48){\makebox(0,0)[rt]{\lineheight{1.25}\smash{\begin{tabular}[t]{r}-10\end{tabular}}}}%
    \put(0.16088867,0.60333333){\makebox(0,0)[rt]{\lineheight{1.25}\smash{\begin{tabular}[t]{r}0\end{tabular}}}}%
    \put(0.04866667,0.37000033){\rotatebox{90}{\makebox(0,0)[t]{\lineheight{1.25}\smash{\begin{tabular}[t]{c}$\Theta_\mathrm{\tau}$ (deg)\end{tabular}}}}}%
    \put(0,0){\includegraphics[width=\unitlength,page=3]{theta_tau_upset.pdf}}%
  \end{picture}%
\endgroup%

%% file: upset_aircraft_visual.pdf_tex
\begingroup%
  \makeatletter%
  \providecommand\color[2][]{%
    \errmessage{(Inkscape) Color is used for the text in Inkscape, but the package 'color.sty' is not loaded}%
    \renewcommand\color[2][]{}%
  }%
  \providecommand\transparent[1]{%
    \errmessage{(Inkscape) Transparency is used (non-zero) for the text in Inkscape, but the package 'transparent.sty' is not loaded}%
    \renewcommand\transparent[1]{}%
  }%
  \providecommand\rotatebox[2]{#2}%
  \newcommand*\fsize{\dimexpr\f@size pt\relax}%
  \newcommand*\lineheight[1]{\fontsize{\fsize}{#1\fsize}\selectfont}%
  \ifx\svgwidth\undefined%
    \setlength{\unitlength}{225bp}%
    \ifx\svgscale\undefined%
      \relax%
    \else%
      \setlength{\unitlength}{\unitlength * \real{\svgscale}}%
    \fi%
  \else%
    \setlength{\unitlength}{\svgwidth}%
  \fi%
  \global\let\svgwidth\undefined%
  \global\let\svgscale\undefined%
  \makeatother%
  \begin{picture}(1,0.66666667)%
    \lineheight{1}%
    \setlength\tabcolsep{0pt}%
    \put(0,0){\includegraphics[width=\unitlength,page=1]{upset_aircraft_visual.pdf}}%
    \put(0.183924,0.202441){\makebox(0,0)[rt]{\lineheight{1.25}\smash{\begin{tabular}[t]{r}0\end{tabular}}}}%
    \put(0,0){\includegraphics[width=\unitlength,page=2]{upset_aircraft_visual.pdf}}%
    \put(0.182999,0.166837){\makebox(0,0)[t]{\lineheight{1.25}\smash{\begin{tabular}[t]{c}0\end{tabular}}}}%
    \put(0,0){\includegraphics[width=\unitlength,page=3]{upset_aircraft_visual.pdf}}%
    \put(0.183924,0.326425){\makebox(0,0)[rt]{\lineheight{1.25}\smash{\begin{tabular}[t]{r}200\end{tabular}}}}%
    \put(0,0){\includegraphics[width=\unitlength,page=4]{upset_aircraft_visual.pdf}}%
    \put(0.06865433,0.36229033){\rotatebox{90}{\makebox(0,0)[t]{\lineheight{1.25}\smash{\begin{tabular}[t]{c}$\mathrm{z}_\mathrm{W}$ (m)\end{tabular}}}}}%
    \put(0,0){\includegraphics[width=\unitlength,page=5]{upset_aircraft_visual.pdf}}%
    \put(0.183924,0.450409){\makebox(0,0)[rt]{\lineheight{1.25}\smash{\begin{tabular}[t]{r}400\end{tabular}}}}%
    \put(0,0){\includegraphics[width=\unitlength,page=6]{upset_aircraft_visual.pdf}}%
    \put(0.77583967,0.12829967){\makebox(0,0)[rt]{\lineheight{1.25}\smash{\begin{tabular}[t]{r}200\end{tabular}}}}%
    \put(0,0){\includegraphics[width=\unitlength,page=7]{upset_aircraft_visual.pdf}}%
    \put(0.252381,0.021851){\makebox(0,0)[t]{\lineheight{1.25}\smash{\begin{tabular}[t]{c}$\mathrm{x}_\mathrm{W}$ (m)\end{tabular}}}}%
    \put(0,0){\includegraphics[width=\unitlength,page=8]{upset_aircraft_visual.pdf}}%
    \put(0.71751833,0.018206){\makebox(0,0)[t]{\lineheight{1.25}\smash{\begin{tabular}[t]{c}$\mathrm{y}_\mathrm{W}$ (m)\end{tabular}}}}%
    \put(0,0){\includegraphics[width=\unitlength,page=9]{upset_aircraft_visual.pdf}}%
    \put(0.67330233,0.09990067){\makebox(0,0)[rt]{\lineheight{1.25}\smash{\begin{tabular}[t]{r}0\end{tabular}}}}%
    \put(0,0){\includegraphics[width=\unitlength,page=10]{upset_aircraft_visual.pdf}}%
    \put(0.390582,0.07916267){\makebox(0,0)[t]{\lineheight{1.25}\smash{\begin{tabular}[t]{c}500\end{tabular}}}}%
    \put(0,0){\includegraphics[width=\unitlength,page=11]{upset_aircraft_visual.pdf}}%
    \put(0.570765,0.07150167){\makebox(0,0)[rt]{\lineheight{1.25}\smash{\begin{tabular}[t]{r}-200\end{tabular}}}}%
    \put(0,0){\includegraphics[width=\unitlength,page=12]{upset_aircraft_visual.pdf}}%
  \end{picture}%
\endgroup%

%% file: upset_aircraft_visual_magnified.pdf_tex
\begingroup%
  \makeatletter%
  \providecommand\color[2][]{%
    \errmessage{(Inkscape) Color is used for the text in Inkscape, but the package 'color.sty' is not loaded}%
    \renewcommand\color[2][]{}%
  }%
  \providecommand\transparent[1]{%
    \errmessage{(Inkscape) Transparency is used (non-zero) for the text in Inkscape, but the package 'transparent.sty' is not loaded}%
    \renewcommand\transparent[1]{}%
  }%
  \providecommand\rotatebox[2]{#2}%
  \newcommand*\fsize{\dimexpr\f@size pt\relax}%
  \newcommand*\lineheight[1]{\fontsize{\fsize}{#1\fsize}\selectfont}%
  \ifx\svgwidth\undefined%
    \setlength{\unitlength}{225bp}%
    \ifx\svgscale\undefined%
      \relax%
    \else%
      \setlength{\unitlength}{\unitlength * \real{\svgscale}}%
    \fi%
  \else%
    \setlength{\unitlength}{\svgwidth}%
  \fi%
  \global\let\svgwidth\undefined%
  \global\let\svgscale\undefined%
  \makeatother%
  \begin{picture}(1,0.66666667)%
    \lineheight{1}%
    \setlength\tabcolsep{0pt}%
    \put(0,0){\includegraphics[width=\unitlength,page=1]{upset_aircraft_visual_magnified.pdf}}%
    \put(0.17333333,0.07344433){\makebox(0,0)[rt]{\lineheight{1.25}\smash{\begin{tabular}[t]{r}-300\end{tabular}}}}%
    \put(0.29555567,0.07344433){\makebox(0,0)[rt]{\lineheight{1.25}\smash{\begin{tabular}[t]{r}-200\end{tabular}}}}%
    \put(0.41777767,0.07344433){\makebox(0,0)[rt]{\lineheight{1.25}\smash{\begin{tabular}[t]{r}-100\end{tabular}}}}%
    \put(0.54,0.07344433){\makebox(0,0)[rt]{\lineheight{1.25}\smash{\begin{tabular}[t]{r}0\end{tabular}}}}%
    \put(0.66222233,0.07344433){\makebox(0,0)[rt]{\lineheight{1.25}\smash{\begin{tabular}[t]{r}100\end{tabular}}}}%
    \put(0.78444433,0.07344433){\makebox(0,0)[rt]{\lineheight{1.25}\smash{\begin{tabular}[t]{r}200\end{tabular}}}}%
    \put(0.90666667,0.07344433){\makebox(0,0)[rt]{\lineheight{1.25}\smash{\begin{tabular}[t]{r}300\end{tabular}}}}%
    \put(0.54000033,0.01522233){\makebox(0,0)[t]{\lineheight{1.25}\smash{\begin{tabular}[t]{c}$\mathrm{y}_\mathrm{W}$ (m)\end{tabular}}}}%
    \put(0,0){\includegraphics[width=\unitlength,page=2]{upset_aircraft_visual_magnified.pdf}}%
    \put(0.160889,0.10722233){\makebox(0,0)[rt]{\lineheight{1.25}\smash{\begin{tabular}[t]{r}0\end{tabular}}}}%
    \put(0.160889,0.22944433){\makebox(0,0)[rt]{\lineheight{1.25}\smash{\begin{tabular}[t]{r}100\end{tabular}}}}%
    \put(0.160889,0.35166667){\makebox(0,0)[rt]{\lineheight{1.25}\smash{\begin{tabular}[t]{r}200\end{tabular}}}}%
    \put(0.160889,0.473889){\makebox(0,0)[rt]{\lineheight{1.25}\smash{\begin{tabular}[t]{r}300\end{tabular}}}}%
    \put(0.160889,0.596111){\makebox(0,0)[rt]{\lineheight{1.25}\smash{\begin{tabular}[t]{r}400\end{tabular}}}}%
    \put(0.04733333,0.36500033){\rotatebox{90}{\makebox(0,0)[t]{\lineheight{1.25}\smash{\begin{tabular}[t]{c}$\mathrm{z}_\mathrm{W}$ (m)\end{tabular}}}}}%
    \put(0,0){\includegraphics[width=\unitlength,page=3]{upset_aircraft_visual_magnified.pdf}}%
    \put(0.70830755,0.49186046){\color[rgb]{0.14901961,0.14901961,0.14901961}\rotatebox{-19.5}{\makebox(0,0)[rt]{\lineheight{1.25}\smash{\begin{tabular}[t]{r}Flight direction\end{tabular}}}}}%
    \put(0,0){\includegraphics[width=\unitlength,page=4]{upset_aircraft_visual_magnified.pdf}}%
  \end{picture}%
\endgroup%

%% file: w_acc_without_pred.pdf_tex
\begingroup%
  \makeatletter%
  \providecommand\color[2][]{%
    \errmessage{(Inkscape) Color is used for the text in Inkscape, but the package 'color.sty' is not loaded}%
    \renewcommand\color[2][]{}%
  }%
  \providecommand\transparent[1]{%
    \errmessage{(Inkscape) Transparency is used (non-zero) for the text in Inkscape, but the package 'transparent.sty' is not loaded}%
    \renewcommand\transparent[1]{}%
  }%
  \providecommand\rotatebox[2]{#2}%
  \newcommand*\fsize{\dimexpr\f@size pt\relax}%
  \newcommand*\lineheight[1]{\fontsize{\fsize}{#1\fsize}\selectfont}%
  \ifx\svgwidth\undefined%
    \setlength{\unitlength}{225bp}%
    \ifx\svgscale\undefined%
      \relax%
    \else%
      \setlength{\unitlength}{\unitlength * \real{\svgscale}}%
    \fi%
  \else%
    \setlength{\unitlength}{\svgwidth}%
  \fi%
  \global\let\svgwidth\undefined%
  \global\let\svgscale\undefined%
  \makeatother%
  \begin{picture}(1,0.66666667)%
    \lineheight{1}%
    \setlength\tabcolsep{0pt}%
    \put(0,0){\includegraphics[width=\unitlength,page=1]{w_acc_without_pred.pdf}}%
    \put(0.17333333,0.07622233){\makebox(0,0)[t]{\lineheight{1.25}\smash{\begin{tabular}[t]{c}65\end{tabular}}}}%
    \put(0.54,0.07622233){\makebox(0,0)[t]{\lineheight{1.25}\smash{\begin{tabular}[t]{c}70\end{tabular}}}}%
    \put(0.90666667,0.07622233){\makebox(0,0)[t]{\lineheight{1.25}\smash{\begin{tabular}[t]{c}75\end{tabular}}}}%
    \put(0.54000067,0.01933333){\makebox(0,0)[t]{\lineheight{1.25}\smash{\begin{tabular}[t]{c}Time (s)\end{tabular}}}}%
    \put(0,0){\includegraphics[width=\unitlength,page=2]{w_acc_without_pred.pdf}}%
    \put(0.16088867,0.11){\makebox(0,0)[rt]{\lineheight{1.25}\smash{\begin{tabular}[t]{r}-6\end{tabular}}}}%
    \put(0.16088867,0.19222233){\makebox(0,0)[rt]{\lineheight{1.25}\smash{\begin{tabular}[t]{r}-4\end{tabular}}}}%
    \put(0.16088867,0.27444433){\makebox(0,0)[rt]{\lineheight{1.25}\smash{\begin{tabular}[t]{r}-2\end{tabular}}}}%
    \put(0.16088867,0.35666667){\makebox(0,0)[rt]{\lineheight{1.25}\smash{\begin{tabular}[t]{r}0\end{tabular}}}}%
    \put(0.16088867,0.438889){\makebox(0,0)[rt]{\lineheight{1.25}\smash{\begin{tabular}[t]{r}2\end{tabular}}}}%
    \put(0.16088867,0.521111){\makebox(0,0)[rt]{\lineheight{1.25}\smash{\begin{tabular}[t]{r}4\end{tabular}}}}%
    \put(0.16088867,0.60333333){\makebox(0,0)[rt]{\lineheight{1.25}\smash{\begin{tabular}[t]{r}6\end{tabular}}}}%
    \put(0.04866667,0.37000033){\rotatebox{90}{\makebox(0,0)[t]{\lineheight{1.25}\smash{\begin{tabular}[t]{c}$\mathrm{a}_\mathrm{winch}$ $\left(\mathrm{m/s}^2\right)$\end{tabular}}}}}%
    \put(0,0){\includegraphics[width=\unitlength,page=3]{w_acc_without_pred.pdf}}%
  \end{picture}%
\endgroup%

%% file: w_acc_with_pred.pdf_tex
\begingroup%
  \makeatletter%
  \providecommand\color[2][]{%
    \errmessage{(Inkscape) Color is used for the text in Inkscape, but the package 'color.sty' is not loaded}%
    \renewcommand\color[2][]{}%
  }%
  \providecommand\transparent[1]{%
    \errmessage{(Inkscape) Transparency is used (non-zero) for the text in Inkscape, but the package 'transparent.sty' is not loaded}%
    \renewcommand\transparent[1]{}%
  }%
  \providecommand\rotatebox[2]{#2}%
  \newcommand*\fsize{\dimexpr\f@size pt\relax}%
  \newcommand*\lineheight[1]{\fontsize{\fsize}{#1\fsize}\selectfont}%
  \ifx\svgwidth\undefined%
    \setlength{\unitlength}{225bp}%
    \ifx\svgscale\undefined%
      \relax%
    \else%
      \setlength{\unitlength}{\unitlength * \real{\svgscale}}%
    \fi%
  \else%
    \setlength{\unitlength}{\svgwidth}%
  \fi%
  \global\let\svgwidth\undefined%
  \global\let\svgscale\undefined%
  \makeatother%
  \begin{picture}(1,0.66666667)%
    \lineheight{1}%
    \setlength\tabcolsep{0pt}%
    \put(0,0){\includegraphics[width=\unitlength,page=1]{w_acc_with_pred.pdf}}%
    \put(0.17333333,0.07622233){\makebox(0,0)[t]{\lineheight{1.25}\smash{\begin{tabular}[t]{c}65\end{tabular}}}}%
    \put(0.54,0.07622233){\makebox(0,0)[t]{\lineheight{1.25}\smash{\begin{tabular}[t]{c}70\end{tabular}}}}%
    \put(0.90666667,0.07622233){\makebox(0,0)[t]{\lineheight{1.25}\smash{\begin{tabular}[t]{c}75\end{tabular}}}}%
    \put(0.54000067,0.01933333){\makebox(0,0)[t]{\lineheight{1.25}\smash{\begin{tabular}[t]{c}Time (s)\end{tabular}}}}%
    \put(0,0){\includegraphics[width=\unitlength,page=2]{w_acc_with_pred.pdf}}%
    \put(0.16088867,0.11){\makebox(0,0)[rt]{\lineheight{1.25}\smash{\begin{tabular}[t]{r}-6\end{tabular}}}}%
    \put(0.16088867,0.19222233){\makebox(0,0)[rt]{\lineheight{1.25}\smash{\begin{tabular}[t]{r}-4\end{tabular}}}}%
    \put(0.16088867,0.27444433){\makebox(0,0)[rt]{\lineheight{1.25}\smash{\begin{tabular}[t]{r}-2\end{tabular}}}}%
    \put(0.16088867,0.35666667){\makebox(0,0)[rt]{\lineheight{1.25}\smash{\begin{tabular}[t]{r}0\end{tabular}}}}%
    \put(0.16088867,0.438889){\makebox(0,0)[rt]{\lineheight{1.25}\smash{\begin{tabular}[t]{r}2\end{tabular}}}}%
    \put(0.16088867,0.521111){\makebox(0,0)[rt]{\lineheight{1.25}\smash{\begin{tabular}[t]{r}4\end{tabular}}}}%
    \put(0.16088867,0.60333333){\makebox(0,0)[rt]{\lineheight{1.25}\smash{\begin{tabular}[t]{r}6\end{tabular}}}}%
    \put(0.04866667,0.37000033){\rotatebox{90}{\makebox(0,0)[t]{\lineheight{1.25}\smash{\begin{tabular}[t]{c}$\mathrm{a}_\mathrm{winch}$ $\left(\mathrm{m/s}^2\right)$\end{tabular}}}}}%
    \put(0,0){\includegraphics[width=\unitlength,page=3]{w_acc_with_pred.pdf}}%
  \end{picture}%
\endgroup%

%% file: w_v_ro_without_pred.pdf_tex
\begingroup%
  \makeatletter%
  \providecommand\color[2][]{%
    \errmessage{(Inkscape) Color is used for the text in Inkscape, but the package 'color.sty' is not loaded}%
    \renewcommand\color[2][]{}%
  }%
  \providecommand\transparent[1]{%
    \errmessage{(Inkscape) Transparency is used (non-zero) for the text in Inkscape, but the package 'transparent.sty' is not loaded}%
    \renewcommand\transparent[1]{}%
  }%
  \providecommand\rotatebox[2]{#2}%
  \newcommand*\fsize{\dimexpr\f@size pt\relax}%
  \newcommand*\lineheight[1]{\fontsize{\fsize}{#1\fsize}\selectfont}%
  \ifx\svgwidth\undefined%
    \setlength{\unitlength}{225bp}%
    \ifx\svgscale\undefined%
      \relax%
    \else%
      \setlength{\unitlength}{\unitlength * \real{\svgscale}}%
    \fi%
  \else%
    \setlength{\unitlength}{\svgwidth}%
  \fi%
  \global\let\svgwidth\undefined%
  \global\let\svgscale\undefined%
  \makeatother%
  \begin{picture}(1,0.66666667)%
    \lineheight{1}%
    \setlength\tabcolsep{0pt}%
    \put(0,0){\includegraphics[width=\unitlength,page=1]{w_v_ro_without_pred.pdf}}%
    \put(0.17333333,0.07622233){\makebox(0,0)[t]{\lineheight{1.25}\smash{\begin{tabular}[t]{c}65\end{tabular}}}}%
    \put(0.54,0.07622233){\makebox(0,0)[t]{\lineheight{1.25}\smash{\begin{tabular}[t]{c}70\end{tabular}}}}%
    \put(0.90666667,0.07622233){\makebox(0,0)[t]{\lineheight{1.25}\smash{\begin{tabular}[t]{c}75\end{tabular}}}}%
    \put(0.54000067,0.01933333){\makebox(0,0)[t]{\lineheight{1.25}\smash{\begin{tabular}[t]{c}Time (s)\end{tabular}}}}%
    \put(0,0){\includegraphics[width=\unitlength,page=2]{w_v_ro_without_pred.pdf}}%
    \put(0.16088867,0.14083333){\makebox(0,0)[rt]{\lineheight{1.25}\smash{\begin{tabular}[t]{r}-5\end{tabular}}}}%
    \put(0.16088867,0.295){\makebox(0,0)[rt]{\lineheight{1.25}\smash{\begin{tabular}[t]{r}0\end{tabular}}}}%
    \put(0.16088867,0.44916667){\makebox(0,0)[rt]{\lineheight{1.25}\smash{\begin{tabular}[t]{r}5\end{tabular}}}}%
    \put(0.16088867,0.60333333){\makebox(0,0)[rt]{\lineheight{1.25}\smash{\begin{tabular}[t]{r}10\end{tabular}}}}%
    \put(0.04866667,0.37000033){\rotatebox{90}{\makebox(0,0)[t]{\lineheight{1.25}\smash{\begin{tabular}[t]{c}$\mathrm{v}_\mathrm{winch}$ $\left(\mathrm{m/s}\right)$\end{tabular}}}}}%
    \put(0,0){\includegraphics[width=\unitlength,page=3]{w_v_ro_without_pred.pdf}}%
  \end{picture}%
\endgroup%

%% file: w_v_ro_with_pred.pdf_tex
\begingroup%
  \makeatletter%
  \providecommand\color[2][]{%
    \errmessage{(Inkscape) Color is used for the text in Inkscape, but the package 'color.sty' is not loaded}%
    \renewcommand\color[2][]{}%
  }%
  \providecommand\transparent[1]{%
    \errmessage{(Inkscape) Transparency is used (non-zero) for the text in Inkscape, but the package 'transparent.sty' is not loaded}%
    \renewcommand\transparent[1]{}%
  }%
  \providecommand\rotatebox[2]{#2}%
  \newcommand*\fsize{\dimexpr\f@size pt\relax}%
  \newcommand*\lineheight[1]{\fontsize{\fsize}{#1\fsize}\selectfont}%
  \ifx\svgwidth\undefined%
    \setlength{\unitlength}{225bp}%
    \ifx\svgscale\undefined%
      \relax%
    \else%
      \setlength{\unitlength}{\unitlength * \real{\svgscale}}%
    \fi%
  \else%
    \setlength{\unitlength}{\svgwidth}%
  \fi%
  \global\let\svgwidth\undefined%
  \global\let\svgscale\undefined%
  \makeatother%
  \begin{picture}(1,0.66666667)%
    \lineheight{1}%
    \setlength\tabcolsep{0pt}%
    \put(0,0){\includegraphics[width=\unitlength,page=1]{w_v_ro_with_pred.pdf}}%
    \put(0.17333333,0.07622233){\makebox(0,0)[t]{\lineheight{1.25}\smash{\begin{tabular}[t]{c}65\end{tabular}}}}%
    \put(0.54,0.07622233){\makebox(0,0)[t]{\lineheight{1.25}\smash{\begin{tabular}[t]{c}70\end{tabular}}}}%
    \put(0.90666667,0.07622233){\makebox(0,0)[t]{\lineheight{1.25}\smash{\begin{tabular}[t]{c}75\end{tabular}}}}%
    \put(0.54000067,0.01933333){\makebox(0,0)[t]{\lineheight{1.25}\smash{\begin{tabular}[t]{c}Time (s)\end{tabular}}}}%
    \put(0,0){\includegraphics[width=\unitlength,page=2]{w_v_ro_with_pred.pdf}}%
    \put(0.16088867,0.14083333){\makebox(0,0)[rt]{\lineheight{1.25}\smash{\begin{tabular}[t]{r}-5\end{tabular}}}}%
    \put(0.16088867,0.295){\makebox(0,0)[rt]{\lineheight{1.25}\smash{\begin{tabular}[t]{r}0\end{tabular}}}}%
    \put(0.16088867,0.44916667){\makebox(0,0)[rt]{\lineheight{1.25}\smash{\begin{tabular}[t]{r}5\end{tabular}}}}%
    \put(0.16088867,0.60333333){\makebox(0,0)[rt]{\lineheight{1.25}\smash{\begin{tabular}[t]{r}10\end{tabular}}}}%
    \put(0.04866667,0.37000033){\rotatebox{90}{\makebox(0,0)[t]{\lineheight{1.25}\smash{\begin{tabular}[t]{c}$\mathrm{v}_\mathrm{winch}$ $\left(\mathrm{m/s}\right)$\end{tabular}}}}}%
    \put(0,0){\includegraphics[width=\unitlength,page=3]{w_v_ro_with_pred.pdf}}%
  \end{picture}%
\endgroup%

%% file: v_w_x_rupture_zoom_noavoid.pdf_tex
\begingroup%
  \makeatletter%
  \providecommand\color[2][]{%
    \errmessage{(Inkscape) Color is used for the text in Inkscape, but the package 'color.sty' is not loaded}%
    \renewcommand\color[2][]{}%
  }%
  \providecommand\transparent[1]{%
    \errmessage{(Inkscape) Transparency is used (non-zero) for the text in Inkscape, but the package 'transparent.sty' is not loaded}%
    \renewcommand\transparent[1]{}%
  }%
  \providecommand\rotatebox[2]{#2}%
  \ifx\svgwidth\undefined%
    \setlength{\unitlength}{240bp}%
    \ifx\svgscale\undefined%
      \relax%
    \else%
      \setlength{\unitlength}{\unitlength * \real{\svgscale}}%
    \fi%
  \else%
    \setlength{\unitlength}{\svgwidth}%
  \fi%
  \global\let\svgwidth\undefined%
  \global\let\svgscale\undefined%
  \makeatother%
  \begin{picture}(1,0.66666667)%
    \put(0,0){\includegraphics[width=\unitlength,page=1]{v_w_x_rupture_zoom_noavoid.pdf}}%
    \put(0.14333333,0.04833333){\makebox(0,0)[b]{\smash{65}}}%
    \put(0.52416667,0.04833333){\makebox(0,0)[b]{\smash{70}}}%
    \put(0.905,0.04833333){\makebox(0,0)[b]{\smash{75}}}%
    \put(0.52416733,-0.00066667){\makebox(0,0)[b]{\smash{Time (s)}}}%
    \put(0,0){\includegraphics[width=\unitlength,page=2]{v_w_x_rupture_zoom_noavoid.pdf}}%
    \put(0.13,0.08766667){\makebox(0,0)[rb]{\smash{14}}}%
    \put(0.13,0.21641667){\makebox(0,0)[rb]{\smash{15}}}%
    \put(0.13,0.34516667){\makebox(0,0)[rb]{\smash{16}}}%
    \put(0.13,0.47391667){\makebox(0,0)[rb]{\smash{17}}}%
    \put(0.13,0.60266667){\makebox(0,0)[rb]{\smash{18}}}%
    \put(0.03133333,0.359167){\rotatebox{90}{\makebox(0,0)[b]{\smash{$\mathrm{v}_\mathrm{w,x,W}$ (m/s)}}}}%
    \put(0,0){\includegraphics[width=\unitlength,page=3]{v_w_x_rupture_zoom_noavoid.pdf}}%
  \end{picture}%
\endgroup%

%% file: v_w_x_rupture_zoom.pdf_tex
\begingroup%
  \makeatletter%
  \providecommand\color[2][]{%
    \errmessage{(Inkscape) Color is used for the text in Inkscape, but the package 'color.sty' is not loaded}%
    \renewcommand\color[2][]{}%
  }%
  \providecommand\transparent[1]{%
    \errmessage{(Inkscape) Transparency is used (non-zero) for the text in Inkscape, but the package 'transparent.sty' is not loaded}%
    \renewcommand\transparent[1]{}%
  }%
  \providecommand\rotatebox[2]{#2}%
  \ifx\svgwidth\undefined%
    \setlength{\unitlength}{240bp}%
    \ifx\svgscale\undefined%
      \relax%
    \else%
      \setlength{\unitlength}{\unitlength * \real{\svgscale}}%
    \fi%
  \else%
    \setlength{\unitlength}{\svgwidth}%
  \fi%
  \global\let\svgwidth\undefined%
  \global\let\svgscale\undefined%
  \makeatother%
  \begin{picture}(1,0.66666667)%
    \put(0,0){\includegraphics[width=\unitlength,page=1]{v_w_x_rupture_zoom.pdf}}%
    \put(0.14333333,0.04833333){\makebox(0,0)[b]{\smash{65}}}%
    \put(0.52416667,0.04833333){\makebox(0,0)[b]{\smash{70}}}%
    \put(0.905,0.04833333){\makebox(0,0)[b]{\smash{75}}}%
    \put(0.52416733,-0.00066667){\makebox(0,0)[b]{\smash{Time (s)}}}%
    \put(0,0){\includegraphics[width=\unitlength,page=2]{v_w_x_rupture_zoom.pdf}}%
    \put(0.13,0.08766667){\makebox(0,0)[rb]{\smash{14}}}%
    \put(0.13,0.21641667){\makebox(0,0)[rb]{\smash{15}}}%
    \put(0.13,0.34516667){\makebox(0,0)[rb]{\smash{16}}}%
    \put(0.13,0.47391667){\makebox(0,0)[rb]{\smash{17}}}%
    \put(0.13,0.60266667){\makebox(0,0)[rb]{\smash{18}}}%
    \put(0.03133333,0.359167){\rotatebox{90}{\makebox(0,0)[b]{\smash{$\mathrm{v}_\mathrm{w,x,W}$ (m/s)}}}}%
    \put(0,0){\includegraphics[width=\unitlength,page=3]{v_w_x_rupture_zoom.pdf}}%
  \end{picture}%
\endgroup%

%% file: va_over_alpha_upset.pdf_tex
\begingroup%
  \makeatletter%
  \providecommand\color[2][]{%
    \errmessage{(Inkscape) Color is used for the text in Inkscape, but the package 'color.sty' is not loaded}%
    \renewcommand\color[2][]{}%
  }%
  \providecommand\transparent[1]{%
    \errmessage{(Inkscape) Transparency is used (non-zero) for the text in Inkscape, but the package 'transparent.sty' is not loaded}%
    \renewcommand\transparent[1]{}%
  }%
  \providecommand\rotatebox[2]{#2}%
  \newcommand*\fsize{\dimexpr\f@size pt\relax}%
  \newcommand*\lineheight[1]{\fontsize{\fsize}{#1\fsize}\selectfont}%
  \ifx\svgwidth\undefined%
    \setlength{\unitlength}{203.70698547bp}%
    \ifx\svgscale\undefined%
      \relax%
    \else%
      \setlength{\unitlength}{\unitlength * \real{\svgscale}}%
    \fi%
  \else%
    \setlength{\unitlength}{\svgwidth}%
  \fi%
  \global\let\svgwidth\undefined%
  \global\let\svgscale\undefined%
  \makeatother%
  \begin{picture}(1,0.85546962)%
    \lineheight{1}%
    \setlength\tabcolsep{0pt}%
     \put(0.8,0.7169525){\makebox(0,0)[t]{\lineheight{1.25}\smash{\begin{tabular}[t]{c}Probability\end{tabular}}}}%
    \put(0,0){\includegraphics[width=\unitlength,page=1]{va_over_alpha_upset.pdf}}%
    \put(0.1750964,0.10358112){\makebox(0,0)[t]{\lineheight{1.25}\smash{\begin{tabular}[t]{c}0\end{tabular}}}}%
    \put(0.41287653,0.10358112){\makebox(0,0)[t]{\lineheight{1.25}\smash{\begin{tabular}[t]{c}5\end{tabular}}}}%
    \put(0.65065704,0.10358112){\makebox(0,0)[t]{\lineheight{1.25}\smash{\begin{tabular}[t]{c}10\end{tabular}}}}%
    \put(0.46043308,0.04074565){\makebox(0,0)[t]{\lineheight{1.25}\smash{\begin{tabular}[t]{c}$\alpha_\mathrm{a}$ (deg)\end{tabular}}}}%
    \put(0,0){\includegraphics[width=\unitlength,page=2]{va_over_alpha_upset.pdf}}%
    \put(0.16135129,0.25723564){\makebox(0,0)[rt]{\lineheight{1.25}\smash{\begin{tabular}[t]{r}30\end{tabular}}}}%
    \put(0.16135129,0.41956697){\makebox(0,0)[rt]{\lineheight{1.25}\smash{\begin{tabular}[t]{r}35\end{tabular}}}}%
    \put(0.16135129,0.58189793){\makebox(0,0)[rt]{\lineheight{1.25}\smash{\begin{tabular}[t]{r}40\end{tabular}}}}%
    \put(0.03739862,0.42806705){\rotatebox{90}{\makebox(0,0)[t]{\lineheight{1.25}\smash{\begin{tabular}[t]{c}$\mathrm{v}_\mathrm{a}$ (m/s)\end{tabular}}}}}%
    \put(0,0){\includegraphics[width=\unitlength,page=3]{va_over_alpha_upset.pdf}}%
    \put(0.86,0.13578384){\makebox(0,0)[lt]{\lineheight{1.25}\smash{\begin{tabular}[t]{l}0\end{tabular}}}}%
    \put(0.86,.3670835){\makebox(0,0)[lt]{\lineheight{1.25}\smash{\begin{tabular}[t]{l}0.0025\end{tabular}}}}%
    \put(0.86,0.59769525){\makebox(0,0)[lt]{\lineheight{1.25}\smash{\begin{tabular}[t]{l}0.005\end{tabular}}}}%
    \put(0,0){\includegraphics[width=\unitlength,page=4]{va_over_alpha_upset.pdf}}%
  \end{picture}%
\endgroup%

%% file: va_over_alpha_noupset.pdf_tex
\begingroup%
  \makeatletter%
  \providecommand\color[2][]{%
    \errmessage{(Inkscape) Color is used for the text in Inkscape, but the package 'color.sty' is not loaded}%
    \renewcommand\color[2][]{}%
  }%
  \providecommand\transparent[1]{%
    \errmessage{(Inkscape) Transparency is used (non-zero) for the text in Inkscape, but the package 'transparent.sty' is not loaded}%
    \renewcommand\transparent[1]{}%
  }%
  \providecommand\rotatebox[2]{#2}%
  \newcommand*\fsize{\dimexpr\f@size pt\relax}%
  \newcommand*\lineheight[1]{\fontsize{\fsize}{#1\fsize}\selectfont}%
  \ifx\svgwidth\undefined%
    \setlength{\unitlength}{197.44532776bp}%
    \ifx\svgscale\undefined%
      \relax%
    \else%
      \setlength{\unitlength}{\unitlength * \real{\svgscale}}%
    \fi%
  \else%
    \setlength{\unitlength}{\svgwidth}%
  \fi%
  \global\let\svgwidth\undefined%
  \global\let\svgscale\undefined%
  \makeatother%
  \begin{picture}(1,0.88259945)%
    \lineheight{1}%
    \setlength\tabcolsep{0pt}%
    \put(0,0){\includegraphics[width=\unitlength,page=1]{va_over_alpha_noupset.pdf}}%
         \put(0.83,0.7469525){\makebox(0,0)[t]{\lineheight{1.25}\smash{\begin{tabular}[t]{c}Probability\end{tabular}}}}%
    \put(0.1806493,0.10686603){\makebox(0,0)[t]{\lineheight{1.25}\smash{\begin{tabular}[t]{c}0\end{tabular}}}}%
    \put(0.42597024,0.10686603){\makebox(0,0)[t]{\lineheight{1.25}\smash{\begin{tabular}[t]{c}5\end{tabular}}}}%
    \put(0.67129157,0.10686603){\makebox(0,0)[t]{\lineheight{1.25}\smash{\begin{tabular}[t]{c}10\end{tabular}}}}%
    \put(0.47503496,0.04203783){\makebox(0,0)[t]{\lineheight{1.25}\smash{\begin{tabular}[t]{c}$\alpha_\mathrm{a}$ (deg)\end{tabular}}}}%
    \put(0,0){\includegraphics[width=\unitlength,page=2]{va_over_alpha_noupset.pdf}}%
    \put(0.16646828,0.15316239){\makebox(0,0)[rt]{\lineheight{1.25}\smash{\begin{tabular}[t]{r}25\end{tabular}}}}%
    \put(0.16646828,0.34828399){\makebox(0,0)[rt]{\lineheight{1.25}\smash{\begin{tabular}[t]{r}30\end{tabular}}}}%
    \put(0.16646828,0.5434056){\makebox(0,0)[rt]{\lineheight{1.25}\smash{\begin{tabular}[t]{r}35\end{tabular}}}}%
    \put(0.03858465,0.4416425){\rotatebox{90}{\makebox(0,0)[t]{\lineheight{1.25}\smash{\begin{tabular}[t]{c}$\mathrm{v}_\mathrm{a}$ (m/s)\end{tabular}}}}}%
    \put(0,0){\includegraphics[width=\unitlength,page=3]{va_over_alpha_noupset.pdf}}%
    \put(0.89,0.14079934){\makebox(0,0)[lt]{\lineheight{1.25}\smash{\begin{tabular}[t]{l}0\end{tabular}}}}%
    \put(0.89,0.39568382){\makebox(0,0)[lt]{\lineheight{1.25}\smash{\begin{tabular}[t]{l}0.003\end{tabular}}}}%
    \put(0.89,0.65056869){\makebox(0,0)[lt]{\lineheight{1.25}\smash{\begin{tabular}[t]{l}0.006\end{tabular}}}}%
    \put(0,0){\includegraphics[width=\unitlength,page=4]{va_over_alpha_noupset.pdf}}%
  \end{picture}%
\endgroup%

%% file: fp_fn_curve.pdf_tex
\begingroup%
  \makeatletter%
  \providecommand\color[2][]{%
    \errmessage{(Inkscape) Color is used for the text in Inkscape, but the package 'color.sty' is not loaded}%
    \renewcommand\color[2][]{}%
  }%
  \providecommand\transparent[1]{%
    \errmessage{(Inkscape) Transparency is used (non-zero) for the text in Inkscape, but the package 'transparent.sty' is not loaded}%
    \renewcommand\transparent[1]{}%
  }%
  \providecommand\rotatebox[2]{#2}%
  \newcommand*\fsize{\dimexpr\f@size pt\relax}%
  \newcommand*\lineheight[1]{\fontsize{\fsize}{#1\fsize}\selectfont}%
  \ifx\svgwidth\undefined%
    \setlength{\unitlength}{225bp}%
    \ifx\svgscale\undefined%
      \relax%
    \else%
      \setlength{\unitlength}{\unitlength * \real{\svgscale}}%
    \fi%
  \else%
    \setlength{\unitlength}{\svgwidth}%
  \fi%
  \global\let\svgwidth\undefined%
  \global\let\svgscale\undefined%
  \makeatother%
  \begin{picture}(1,0.66666667)%
    \lineheight{1}%
    \setlength\tabcolsep{0pt}%
    \put(0,0){\includegraphics[width=\unitlength,page=1]{fp_fn_curve.pdf}}%
    \put(0.17333333,0.07622233){\makebox(0,0)[t]{\lineheight{1.25}\smash{\begin{tabular}[t]{c}0\end{tabular}}}}%
    \put(0.32,0.07622233){\makebox(0,0)[t]{\lineheight{1.25}\smash{\begin{tabular}[t]{c}0.1\end{tabular}}}}%
    \put(0.46666667,0.07622233){\makebox(0,0)[t]{\lineheight{1.25}\smash{\begin{tabular}[t]{c}0.2\end{tabular}}}}%
    \put(0.61333333,0.07622233){\makebox(0,0)[t]{\lineheight{1.25}\smash{\begin{tabular}[t]{c}0.3\end{tabular}}}}%
    \put(0.76,0.07622233){\makebox(0,0)[t]{\lineheight{1.25}\smash{\begin{tabular}[t]{c}0.4\end{tabular}}}}%
    \put(0.90666667,0.07622233){\makebox(0,0)[t]{\lineheight{1.25}\smash{\begin{tabular}[t]{c}0.5\end{tabular}}}}%
    \put(0.54000033,0.01933333){\makebox(0,0)[t]{\lineheight{1.25}\smash{\begin{tabular}[t]{c}$\Pr\left(\mathrm{FP}\right)$ $(\%)$\end{tabular}}}}%
    \put(0,0){\includegraphics[width=\unitlength,page=2]{fp_fn_curve.pdf}}%
    \put(0.160889,0.11){\makebox(0,0)[rt]{\lineheight{1.25}\smash{\begin{tabular}[t]{r}0\end{tabular}}}}%
    \put(0.160889,0.23333333){\makebox(0,0)[rt]{\lineheight{1.25}\smash{\begin{tabular}[t]{r}2\end{tabular}}}}%
    \put(0.160889,0.35666667){\makebox(0,0)[rt]{\lineheight{1.25}\smash{\begin{tabular}[t]{r}4\end{tabular}}}}%
    \put(0.160889,0.48){\makebox(0,0)[rt]{\lineheight{1.25}\smash{\begin{tabular}[t]{r}6\end{tabular}}}}%
    \put(0.160889,0.60333333){\makebox(0,0)[rt]{\lineheight{1.25}\smash{\begin{tabular}[t]{r}8\end{tabular}}}}%
    \put(0.04866667,0.37000033){\rotatebox{90}{\makebox(0,0)[t]{\lineheight{1.25}\smash{\begin{tabular}[t]{c}$\Pr\left(\hat{y}=1 \;\middle\vert\;  y=-1 \right)$ $(\%)$\end{tabular}}}}}%
    \put(0,0){\includegraphics[width=\unitlength,page=3]{fp_fn_curve.pdf}}%
    \put(0.78330967,0.1395){\makebox(0,0)[lt]{\lineheight{1.25}\smash{\begin{tabular}[t]{l}$\mathrm{F}_\mathrm{t,set}+8\%$\end{tabular}}}}%
    \put(0.37074067,0.13716667){\makebox(0,0)[lt]{\lineheight{1.25}\smash{\begin{tabular}[t]{l}$\mathrm{F}_\mathrm{t,set}+10\%$\end{tabular}}}}%
    \put(0.23678567,0.175643){\makebox(0,0)[lt]{\lineheight{1.25}\smash{\begin{tabular}[t]{l}$\mathrm{F}_\mathrm{t,set}+12\%$\end{tabular}}}}%
    \put(0.187434,0.228287){\makebox(0,0)[lt]{\lineheight{1.25}\smash{\begin{tabular}[t]{l}$\mathrm{F}_\mathrm{t,set}+14\%$\end{tabular}}}}%
    \put(0.17333333,0.58343533){\makebox(0,0)[lt]{\lineheight{1.25}\smash{\begin{tabular}[t]{l}$\mathrm{F}_\mathrm{t,set}+16\%$\end{tabular}}}}%
  \end{picture}%
\endgroup%

%% file: loss_over_downtime.pdf_tex
\begingroup%
  \makeatletter%
  \providecommand\color[2][]{%
    \errmessage{(Inkscape) Color is used for the text in Inkscape, but the package 'color.sty' is not loaded}%
    \renewcommand\color[2][]{}%
  }%
  \providecommand\transparent[1]{%
    \errmessage{(Inkscape) Transparency is used (non-zero) for the text in Inkscape, but the package 'transparent.sty' is not loaded}%
    \renewcommand\transparent[1]{}%
  }%
  \providecommand\rotatebox[2]{#2}%
  \newcommand*\fsize{\dimexpr\f@size pt\relax}%
  \newcommand*\lineheight[1]{\fontsize{\fsize}{#1\fsize}\selectfont}%
  \ifx\svgwidth\undefined%
    \setlength{\unitlength}{225bp}%
    \ifx\svgscale\undefined%
      \relax%
    \else%
      \setlength{\unitlength}{\unitlength * \real{\svgscale}}%
    \fi%
  \else%
    \setlength{\unitlength}{\svgwidth}%
  \fi%
  \global\let\svgwidth\undefined%
  \global\let\svgscale\undefined%
  \makeatother%
  \begin{picture}(1,0.66666667)%
    \lineheight{1}%
    \setlength\tabcolsep{0pt}%
    \put(0,0){\includegraphics[width=\unitlength,page=1]{loss_over_downtime.pdf}}%
    \put(0.17333333,0.07622233){\makebox(0,0)[t]{\lineheight{1.25}\smash{\begin{tabular}[t]{c}0\end{tabular}}}}%
    \put(0.41777767,0.07622233){\makebox(0,0)[t]{\lineheight{1.25}\smash{\begin{tabular}[t]{c}5\end{tabular}}}}%
    \put(0.66222233,0.07622233){\makebox(0,0)[t]{\lineheight{1.25}\smash{\begin{tabular}[t]{c}10\end{tabular}}}}%
    \put(0.90666667,0.07622233){\makebox(0,0)[t]{\lineheight{1.25}\smash{\begin{tabular}[t]{c}15\end{tabular}}}}%
    \put(0.54000033,0.01933333){\makebox(0,0)[t]{\lineheight{1.25}\smash{\begin{tabular}[t]{c}$\Delta \mathrm{T}_\mathrm{nop}$ (days)\end{tabular}}}}%
    \put(0,0){\includegraphics[width=\unitlength,page=2]{loss_over_downtime.pdf}}%
    \put(0.160889,0.11){\makebox(0,0)[rt]{\lineheight{1.25}\smash{\begin{tabular}[t]{r}0\end{tabular}}}}%
    \put(0.160889,0.27444433){\makebox(0,0)[rt]{\lineheight{1.25}\smash{\begin{tabular}[t]{r}0.02\end{tabular}}}}%
    \put(0.160889,0.438889){\makebox(0,0)[rt]{\lineheight{1.25}\smash{\begin{tabular}[t]{r}0.04\end{tabular}}}}%
    \put(0.160889,0.60333333){\makebox(0,0)[rt]{\lineheight{1.25}\smash{\begin{tabular}[t]{r}0.06\end{tabular}}}}%
    \put(0.04866667,0.37000033){\rotatebox{90}{\makebox(0,0)[t]{\lineheight{1.25}\smash{\begin{tabular}[t]{c}$\mathrm{L/E}_\mathrm{pc}$ $(\%)$\end{tabular}}}}}%
    \put(0,0){\includegraphics[width=\unitlength,page=3]{loss_over_downtime.pdf}}%
  \end{picture}%
\endgroup%